%% file: main.tex
\documentclass[11pt,a4paper]{article}

\usepackage{xcolor}
\definecolor{forestgreen}{rgb}{0.13,0.55,0.13}

\usepackage{jheppub}

\usepackage[utf8]{inputenc}
\usepackage[english]{babel}

\usepackage{cleveref}

\hypersetup{pdfpagemode={UseOutlines},bookmarksopen=true,
   bookmarksopenlevel=0,bookmarksnumbered=true,hypertexnames=false,
   colorlinks,citecolor=green,urlcolor=blue,linkcolor=red,
   pdfstartview={FitV},unicode,breaklinks=true}

\usepackage{graphicx}
\graphicspath{{Figures/}}
\usepackage{epstopdf}
\usepackage[export]{adjustbox}

\usepackage{tabu}
\usepackage{tabularx}
\def\minwd#1#2#3\endminwd{\stackengine{0pt}{#3}{\rule{#2}{0pt}}{O}{#1}{F}{F}{L}}
\newcolumntype{L}[1]{>{\minwd l{#1}}l<{\endminwd}}
\newcolumntype{C}[1]{>{\minwd c{#1}}c<{\endminwd}}
\newcolumntype{R}[1]{>{\minwd r{#1}}r<{\endminwd}}

\usepackage{caption}
\usepackage{subcaption}
\usepackage{float}
\usepackage{stackengine}

\input{macros}

\usepackage{tikz}
\usetikzlibrary{shapes,arrows,positioning,calc,decorations.markings,math,arrows.meta}

\title{Zig-zag deformations of toric quiver gauge theories\\
Part I: reflexive polytopes}

\author{Stefano Cremonesi,}
\author{Jos\'{e} S\'{a}}

\affiliation{
Department of Mathematical Sciences, Durham University, 
Durham DH1 3LE, UK}

\emailAdd{stefano.cremonesi} \emailAdd{jose.a.cerqueira-sa@durham.ac.uk}

\abstract{We study one-parameter families of $U(1)^2$ preserving deformations relating pairs of toric quiver gauge theories on D-branes probing local toric (pseudo) del Pezzo surfaces. The superpotential deformations are defined by zig-zag paths in the brane tiling and are non-trivial in the chiral ring if the geometry has a non-isolated singularity. In the dual $(p,q)$ web, the deformation is realized as a Hanany-Witten move that reverses a semi-infinite fivebrane. We use these deformations to find RG flows between 4d $\mathcal{N}=1$ SCFTs on D3-branes probing local toric (pseudo) del Pezzo surfaces of the same degree, and briefly comment on the interpretation for BPS quivers of rank one 5d SCFTs on $S^1$.
}

\begin{document}

\maketitle

\input{Sections/Intro}
\input{Sections/Tilings}
\input{Sections/Deformations}

\input{Sections/Resolutions}

\acknowledgments
We thank Federico Carta, Cyril Closset and Michele Del Zotto for discussions and collaboration on related projects. 
SC is supported in part by STFC through grant ST/T000708/1.
JS is supported by Funda\c{c}\~{a}o para a Ci\^{e}ncia e a Tecnologia I.P. through grant SFRH/BD/ 136280/2018. 
SC gratefully acknowledges support from the Simons Center for Geometry and Physics, Stony Brook University, where some of the results in this paper were presented at the workshop `5d N=1 SCFTs and Gauge Theories on Brane Webs'.

\appendix
\input{Appendices/ZZflows}

\input{Appendices/QuiverReps}

\bibliographystyle{JHEP}
\bibliography{main}

\end{document}

%% file: macros.tex
\usepackage{amsmath,amsfonts,amssymb,amscd,amsbsy,amsthm,stmaryrd}
\SetSymbolFont{stmry}{bold}{U}{stmry}{m}{n}
\usepackage{mathtools,textcomp,stackrel}
\usepackage{relsize}
\usepackage{xfrac}
\usepackage{bm}
\usepackage{esint}
\usepackage{enumerate}

\usepackage[bbgreekl]{mathbbol}
\DeclareSymbolFontAlphabet{\mathbb}{AMSb}
\DeclareSymbolFontAlphabet{\mathbbl}{bbold}

\usepackage{stackengine}

\renewcommand{\eqref}[1]{(\ref{#1})}

\newcommand{\fracpd}[2][]{\frac{\partial #1}{\partial #2}}

\usepackage{xparse}

\ExplSyntaxOn
\NewDocumentCommand{\shortenmathcapital}{m}{
	\prop_set_from_keyval:Nn \l_tmpa_prop{#1}
	\int_step_inline:nnn{`A}{`Z}{
		\prop_map_inline:Nn \l_tmpa_prop{
			\cs_new_protected:cpx { 
				####1 \char_generate:nn{##1}{11}
			}{
				\exp_not:N ####2{ \char_generate:nn{##1}{11} }
			}
		}
	}
}
\ExplSyntaxOff

\shortenmathcapital{bb=\mathbb,cal=\mathcal,bbl=\mathbbl}

\DeclareMathOperator{\tr}{tr}

\DeclareMathOperator{\Tr}{Tr}
\DeclareMathOperator{\perm}{perm}
\DeclareMathOperator{\im}{im}

\DeclareMathOperator{\Sym}{Sym}
\DeclareMathOperator{\Spec}{Spec}

\DeclareMathOperator{\Proj}{Proj}
\DeclareMathOperator{\Rep}{Rep}
\DeclareMathOperator{\Hom}{Hom}

\renewcommand{\Re}{\mathrm{Re}}

\DeclareDocumentCommand{\shortexact}{O{} O{} mmmm}{
\xymatrix{
	0\ar[r] & #3\ar[r]^-{#1} & #4\ar[r]^-{#2} & #5\ar[r] & 0#6
}}
\DeclareDocumentCommand{\leftexact}{O{} O{} mmmm}{
\xymatrix{
	0\ar[r] & #3\ar[r]^-{#1} & #4\ar[r]^-{#2} & #5 #6
}}
\DeclareDocumentCommand{\rightexact}{O{} O{} mmmm}{
\xymatrix{
	#3\ar[r]^-{#1} & #4\ar[r]^-{#2} & #5\ar[r] & 0#6
}}

\DeclarePairedDelimiterX\angl[1]{\langle}{\rangle}{#1}
\DeclarePairedDelimiterX\abs[1]{\lvert}{\rvert}{#1}
\DeclarePairedDelimiterX\norm[1]{\lVert}{\rVert}{#1}
\DeclarePairedDelimiterX\bkt[1]{\lbrack}{\rbrack}{#1}
\DeclarePairedDelimiterX\poisson[2]{\lbrace}{\rbrace}{#1,#2}
\DeclarePairedDelimiterX\comm[2]{\lbrack}{\rbrack}{#1,#2}
\DeclarePairedDelimiterX\acomm[2]{\lbrace}{\rbrace}{#1,#2}
\DeclarePairedDelimiterX\set[1]{\lbrace}{\rbrace}{\mkern1mu#1\mkern1mu}
\DeclarePairedDelimiterX\setst[2]{\lbrace}{\rbrace}{\mkern1mu#1\mkern3mu\delimsize\vert\mkern3mu#2\mkern1mu}
\DeclarePairedDelimiterX\privcoset[2]{.}{.}{\raisebox{.04em}{$#1$}\delimsize/\raisebox{-.08em}{$#2$}}
\makeatletter
\def\coset{\privcoset*}
\makeatother
\DeclarePairedDelimiterX\privccoset[3]{.}{.}{\raisebox{.04em}{$#1$}\mkern1mu\delimsize/\mkern-5mu\delimsize/_{\mkern-3mu#2}\mkern3mu\raisebox{-.06em}{$#3$}}
\makeatletter
\def\ccoset{\privccoset*}
\makeatother

\DeclarePairedDelimiterX\bra[1]{\langle}{\rvert}{#1}
\DeclarePairedDelimiterX\ket[1]{\lvert}{\rangle}{#1}
\DeclarePairedDelimiterX\braket[2]{\langle}{\rangle}{#1\delimsize\vert#2}
\DeclarePairedDelimiterX\braOket[3]{\langle}{\rangle}{#1\delimsize\vert#2\delimsize\vert#3}

\theoremstyle{definition}

\newtheorem*{theorem*}{Theorem}

\newtheorem*{lemma*}{Lemma}

\theoremstyle{remark}

\numberwithin{theorem}{section}
\numberwithin{lemma}{section}
\numberwithin{corollary}{section}
\numberwithin{proposition}{section}
\numberwithin{exercise}{section}
\numberwithin{example}{section}
\numberwithin{definition}{section}

%% file: Sections/Intro.tex

\section{Introduction}
\label{sec:Intro}

Quiver gauge theories describing the low energy dynamics on D-branes probing singularities have been extensively studied for more than 25 years, in the context of the AdS/CFT correspondence (starting from \cite{Douglas:1996sw,Klebanov:1998hh,Acharya:1998db,Morrison:1998cs}) and for model building in string theory (see \emph{e.g.} \cite{Verlinde:2005jr,Wijnholt:2007vn}).
Engineering strongly coupled field theories in this way allows to geometrize many of their properties, which are harder to access directly in field theory.
Conversely, understanding the field theories from first principles can shed light on aspects of the associated geometry. 

The correspondence is best understood when the geometry has restricted holonomy so that the worldvolume theory preserves supersymmetry \cite{Acharya:1998db}, ensuring greater theoretical control.
In this paper we focus on conical Gorenstein 3-fold singularities $\calY$ which admit a Ricci-flat K\"ahler metric with $SU(3)$ holonomy, which we refer to as Calabi-Yau threefold (CY$_3$) cones with a common abuse of terminology.
The low energy worldvolume theory on D$p$-branes $(p\le 3)$ probing $\mathcal{Y}\times\mathbb{R}^{3-p}$ enjoys $4d$ $\mathcal{N}=1$ SUSY if $p=3$ (in which case the superalgebra extends to the superconformal algebra $SU(2,2|1)$) or its dimensional reductions 3d $\mathcal{N}=2$, 2d $\mathcal{N}=(2,2)$, \emph{etc}, if $p < 3$.
The $p=3$ case provides $\text{AdS}_5/\text{CFT}_4$ dual pairs, and is the language in which the paper is written.
However, most of the analysis is $p$-independent, and an important motivation for revisiting this setup comes from the $p=0$ case, which provides BPS quivers for 5d $\mathcal{N}=1$ SCFTs compactified on $S^1$ (4d Kaluza-Klein theories) \cite{Closset:2019juk}.
A key aspect of the correspondence is that the moduli space of supersymmetric vacua of the field theory on $N$ regular D-branes probing a conical $CY_3$ $\mathcal{Y}$ contains a component, the so called \emph{geometric branch}, which is isomorphic to the configuration space of $N$ points on $\mathcal{Y}$. 

The situation is under even better control if the singularity is toric, in which case the field theory on regular D-branes enjoys (at least) a mesonic $U(1)^2$ symmetry in addition to the $U(1)_R$ symmetry, and is known as a toric quiver gauge theory \cite{Morrison:1998cs,Feng:2000mi}.
The maps between toric geometry and toric quiver gauge theories are completely understood by exploiting the graph-theoretic properties of bipartite graphs which describe twice T-dual brane configurations to D-branes at toric singularities, known as brane tilings \cite{Hanany:2005ve,Franco:2005rj,Franco:2005sm,Feng:2005gw}.  

In this paper we consider toric geometries with non-isolated singularities, which are necessarily of type $A$.
When a regular D-brane reaches an $A_{k-1}$ singularity, it can marginally decay into $k$ fractional D-branes of so called $\mathcal{N}=2$ type, which can separately probe this one-complex-dimensional singular locus \cite{Douglas:1996sw,Bertolini:2003iv,Franco:2005zu}. In the field theory, this is reflected in an additional branch of the moduli space of vacua that intersects the geometric branch at each of its non-isolated singularities.
In this work we dub this branch an \emph{$\mathcal{N}=2$ Coulomb branch} for its analogy to the Coulomb branch of 4d $\mathcal{N}=2$ theories. 

The non-isolated singularity of $\mathcal{Y}$ has an interesting and sometimes underappreciated role in the geometric engineering of 5d SCFTs via M-theory on $\mathbb{R}^{1,4}\times \mathcal{Y}$ \cite{Witten:1996qb,Morrison:1996xf,Intriligator:1997pq}, or equivalently via webs of fivebranes in type IIB string theory \cite{Aharony:1997bh}, where an $A_{k-1}$ singularity translates into $k$ parallel semi-infinite fivebranes (see \cite{Leung:1997tw} for the relation between the two constructions).
M-theory on an $A_{k-1}$ singularity engineers 7d SYM with gauge Lie algebra $\mathfrak{su}(k)$ \cite{Witten:1995ex,Sen:1997js}.
On the other hand, M-theory on a CY$_3$ cone $\mathcal{Y}$ with an isolated singularity engineers a 5d SCFT \cite{Morrison:1996xf}.
If $\mathcal{Y}$  contains both isolated and non-isolated singularities, M-theory on $\mathcal{Y}$ engineers a 5d SCFT \emph{coupled} to a 7d SYM theory as a codimension two half-BPS defect: in particular, an $SU(k)$ global symmetry of the 5d SCFT couples to the $su(k)$ gauge field in the bulk.
Further compactifying on a circle, so that the background is dual to type IIA string theory on $\mathbb{R}^{1,3}\times \mathcal{Y}$, engineers a Kaluza-Klein theory.
The BPS quiver of this Kaluza-Klein theory is nothing but the worldline theory on regular or fractional D0-branes ($p=0$ above) \cite{Closset:2019juk}.
Crucially, the BPS quiver with superpotential is sensitive not only to the singularity at the apex of $\mathcal{Y}$, but also to any non-isolated singularity that intersects it.
Therefore it is a BPS quiver for the circle compactification of the full coupled 5d-7d system, not just the 5d SCFT.
The Hilbert space of BPS particles in a given charge sector, which is a quantization of the moduli space of semi-stable quiver representations with given dimension vector, knows about the higher-dimensional degrees of freedom through quiver representations that correspond to $\mathcal{N}=2$ Coulomb branch vacua. 

For the purpose of engineering purely 5d SCFTs not coupled to higher dimensional degrees of freedom, it is therefore desirable to deform the geometry $\mathcal{Y}$ and lift any non-isolated singularities.
Such deformations break the toric $U(1)^3$ symmetry to a proper subgroup, which should contain $U(1)$ for the field theory on D-brane probes to have a continuous R-symmetry (a necessary condition for superconformal invariance if $p=3$).
Such general non-toric deformations of toric quiver gauge theories were studied in  \cite{Butti:2006nk}. 

In this paper we are interested in non-toric deformations of a special type.
We construct one-parameter families of deformations of toric quiver gauge theories, which preserve $U(1)^2$ symmetry for generic values of the deformation parameter $\mu \in \mathbb{C}\cup \{\infty\}\cong \mathbb{P}^1$, and interpolate between two toric models at $\mu=0$ and $\mu=\infty$ respectively.
We then study how the deformation of the quiver gauge theory impacts its moduli space of vacua and is reflected in the geometry probed by regular D-branes.
For definiteness, in this paper we will focus on toric geometries which are affine cones over weak toric del Pezzo surfaces, or complex cones over pseudo-del Pezzo surfaces in physics parlance \cite{Feng:2002fv}, corresponding to reflexive toric diagrams.%
\footnote{The ideas apply more generally and will be  explored \cite{Sa2023} and developed \cite{CCS2023} further elsewhere.}
We turn on a superpotential deformation in one-to-one correspondence with a \emph{zig-zag path} $\eta$, a path in the brane tiling which follows the edges and turns maximally left (resp. right) at white (resp. left) nodes \cite{kenyon2003introduction,Hanany:2005ss}.
This corresponds to an outward pointing normal vector to an edge of the toric diagram, and to an external leg in the dual $(p,q)$-web. 
The superpotential is deformed by $\delta W=\mu \mathcal{O}_\eta$, where $\mathcal{O}_\eta$ is the difference of the two mesonic chiral operators represented by the loops which wind oppositely to the chosen zig-zag path in the tiling, on either side of it.
We dub this a \emph{zig-zag deformation}. 

The zig-zag deformation is non-trivial in the chiral ring if and only if the chosen zig-zag path is parallel to other zig-zag paths in the tiling, signalling a non-isolated singularity in the geometry.
In such cases, the two mesonic operators entering $\mathcal{O}_\eta$ are $\mathcal{N}=2$ Coulomb moduli of the undeformed toric quiver gauge theory, which are lifted by the deformation.
At the same time, the geometric branch of the abelian theory on the worldvolume of a regular D-brane is deformed, lifting (partially or fully) the non-isolated singularity. 

We study the resulting deformed geometry and show that a new toric geometry arises as $\mu\to \infty$, whose toric diagram is a mutation of the toric diagram of the $\mu=0$ model.
If the chosen zig-zag operator has length 4, the $\mu=\infty$ model is described by a new brane tiling obtained from brane tiling of the $\mu=0$ model by a \emph{zig-zag move} that reverses the chosen zig-zag, generalizing the observation of \cite{Bianchi:2014qma} for mass deformations.
Equivalently, the $\mu=\infty$ brane tiling can be obtained from the $\mu=0$ brane tiling by applying specular duality, followed by a toric Seiberg duality on the face dual to the chosen zig-zag, and then applying specular duality again. This is in agreement with  \cite{Franco:2023flw}, which related mutations of generalized toric polytopes (GTPs) \cite{Benini:2009gi, vanBeest:2020kou} to mutations of twin quivers.
We show that this matches precisely the field theory analysis, where we turn on the zig-zag deformation of the original toric model, integrate out any massive fields, and find field redefinitions to recover the quiver and toric superpotential of a new toric model as $\mu\to \infty$.
For $\mu$ large but finite, the $\mu\to \infty$ model is also deformed by a (possibly trivial) zig-zag deformation $\frac{1}{\mu} \mathcal{O}'_{\eta'}$, where $\eta'$ is the reversed zig-zag.
See \cref{fig:mirror-argument} for the general idea. 
For the toric del Pezzo theories that we study in this paper, zig-zag deformations define RG flows between 4d $\mathcal{N}=1$ toric SCFTs, where the IR model has fewer $\mathcal{N}=2$ Coulomb moduli than the IR model and the geometry has less non-isolated singularity.
In terms of the BPS quivers of 5d rank one $E_{n\le 5}$ SCFTs (coupled to 7d SYM), each flow removes the coupling to some higher dimensional degrees of freedom.
See \cref{fig:defsSummary} for a summary of the flows.
\input{Sections/SummaryFig}

Finally, we observe (and test for arbitrary resolution parameters) that the triangulated toric diagrams of the $\mu=0$ and $\mu=\infty$ geometries are related by a mutation of a triangulated polytope (the untriangulated version was considered in \cite{Franco:2023flw}). As already pointed out in \cite{Franco:2023flw,Franco:2023mkw}, this mutation is dual to reversing a fivebrane in the $(p,q)$-web, by making it terminate on a 7-brane, sliding the 7-brane along the line of the five-brane keeping track of Hanany-Witten transitions, and finally rotating the $SL(2,\mathbb{Z})$ monodromy cut so that it disappears in the limit.
In this paper we have tested this observation in detail against purely field theoretical results for arbitrary FI parameters, and keeping track of the superpotential. We will derive this fact from string dualities in \cite{CCS2023}, where the ideas of this paper will be generalized and put in a broader mathematical context.%
\footnote{Several of the results of this paper and of the forthcoming work \cite{CCS2023} have been presented in a number of talks over the last two years \cite{talks1,talks2}.}

We note that upon adding orientifolds, some of the non-toric mass deformations of toric models discussed in \cite{Bianchi:2014qma} have been shown to lead to  $\mathcal{N}=1$ conformal dualities \cite{Antinucci:2020yki, Antinucci:2021edv, Amariti:2021lhk, Amariti:2022dui, Amariti:2022dyi}. It would be interesting to study the interplay of orientifolds and the more general zig-zag deformations considered in this paper to search for new families of conformal dualities.%
\footnote{SC thanks Salvo Mancani for discussions.}

The paper is organized as follows.
\Cref{sec:Tilings} reviews generalities of toric quiver gauge theories and brane tilings.
In \Cref{sec:ToricDefs} we introduce zig-zag deformations and study them for toric quiver gauge theories on a regular D-brane probing local toric pseudo del Pezzo surfaces.
We analyze the chiral ring of the zig-zag deformed theory, identify the new toric model in the $\mu\to \infty$ limit, and describe how the toric endpoints of zig-zag deformations are related in terms of brane tilings and specular and toric duality.
Finally, in \Cref{sec:Resolutions} we study the interplay of the zig-zag deformation and FI parameters, showing that the field theory and quiver representation theory analysis are in precise agreement with the mutation of (triangulated) lattice polytopes and the dual Hanany-Witten move for the 5-brane web with a 7-brane.
We also include two appendices for self-containedness: \Cref{sec:RGflows} lists the details of all the zig-zag deformations of toric del Pezzo models and the field redefinitions needed to manifest the toric symmetry in the $\mu\to\infty$ limit; \cref{sec:quiver-reps-moduli} reviews generalities of quiver representation theory used in \Cref{sec:Resolutions}. An ancillary file detailing the interplay of zig-zag deformations and resolutions is available upon request.

{\bf{Note added:}} After this manuscript was completed, an interesting paper appeared which uses a local piece of our \emph{zig-zag move} in the brane tiling discussed in this paper (in fact, of the move of  \cite{Bianchi:2014qma}) to describe how the insertion of certain defects in the 4d $\mathcal{N}=2$ KK theory modifies its BPS quiver and superpotential. It would be interesting to work out the relation between the two constructions.

%% file: Sections/SummaryFig.tex

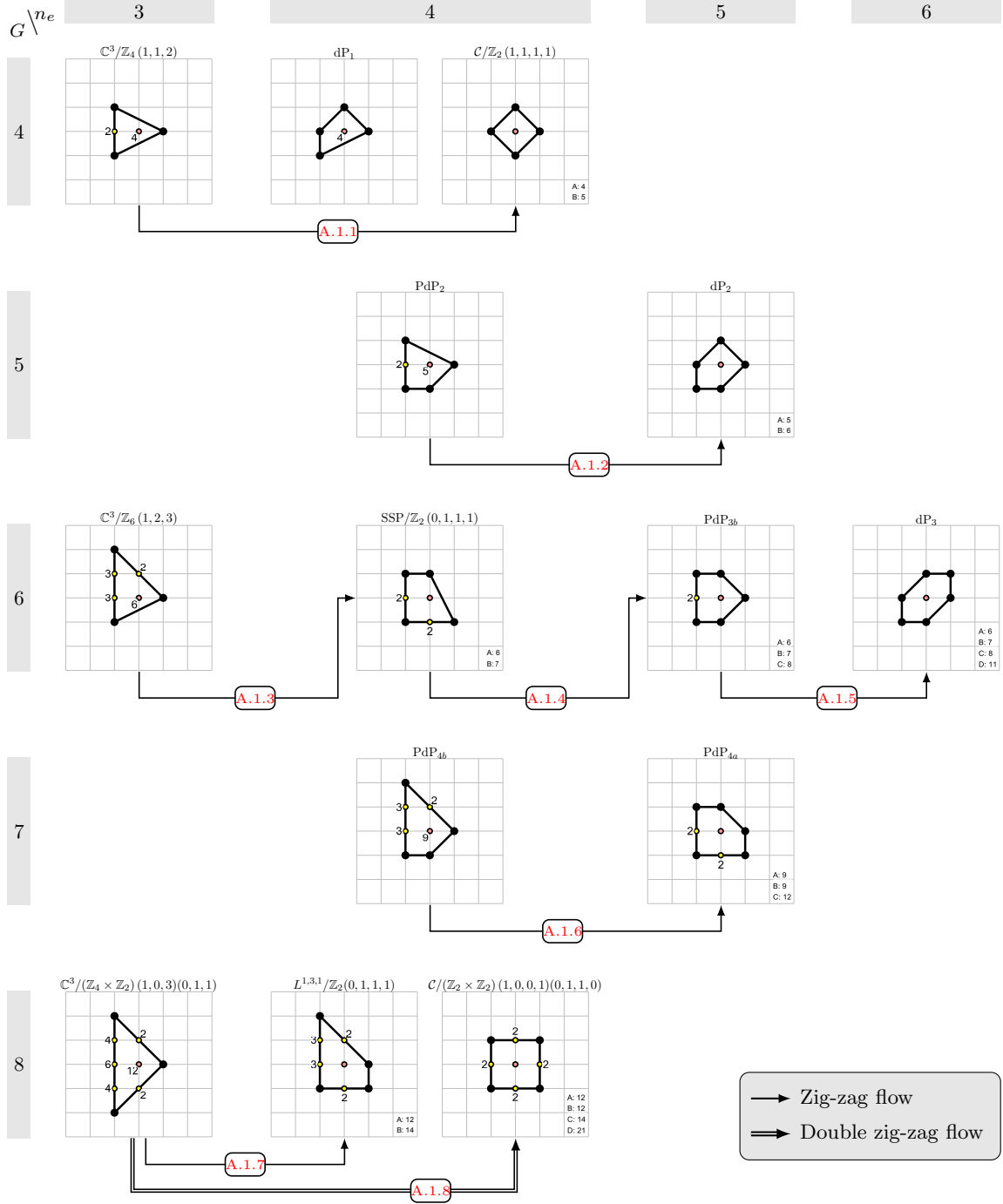
\begin{figure}[ht]
   \begin{tikzpicture}
      \newlength{\sizeTD}
      \newlength{\sepX}
      \newlength{\sepXsmall}
      \newlength{\sepY}
      \newlength{\hX}
      \newlength{\hsX}
      \newlength{\hY}
      \newlength{\startX}
      \newlength{\startY}
      \newlength{\posGx}
      \newlength{\posNPy}
      
      \setlength{\sizeTD}{2.25cm}
      \setlength{\sepX}{0.85cm}
      \setlength{\sepXsmall}{0.333cm}
      \setlength{\sepY}{1.3cm}
      \setlength{\hX}{\dimexpr(\sepX + \sizeTD)\relax}
      \setlength{\hsX}{\dimexpr(\sepXsmall + \sizeTD)\relax}
      \setlength{\hY}{\dimexpr(-\sepY -\sizeTD)\relax}

      \setlength{\posGx}{\dimexpr(-0.25\sepX -1.6cm)\relax}
      \setlength{\posNPy}{\dimexpr(+0.25\sepY +1.5cm)\relax}
      \setlength{\startY}{\dimexpr(-1\hY)\relax}
      
      \newcommand{\includetd}[1]{%
         \includegraphics[width=0.99\sizeTD]{#1-td}%
      }%
      \newcommand{\tdtexlabel}[2]{%
         \node[TDlabel, yshift=-0.2mm] at (#1.north) {\scalebox{0.5}{#2}};%
      }%
      \newcommand{\mySlash}[2]{\ensuremath{%
         \!\sideset{_#1}{\!\!^#2}{\mathop\backslash}}
      }%

      \tikzstyle{Gnelabel}=[inner sep=0pt]
      \tikzstyle{TDimg}=[inner sep=0pt]
      \tikzstyle{Grect}=[rectangle, fill=black!10, 
         minimum height=\sizeTD, minimum width=0.35cm, inner sep=0pt]
      \tikzstyle{NPrect}=[rectangle, fill=black!10, 
         minimum height=0.35cm, minimum width=\sizeTD, inner sep=0pt]
      \tikzstyle{NPrectlong}=[rectangle, fill=black!10, 
         minimum height=0.35cm, minimum width=\dimexpr(\sizeTD+\hsX)\relax, inner sep=0pt]
      \tikzstyle{TDlabel}=[inner sep=0pt, anchor=south]
      \tikzstyle{rgArrow}=[-latex, semithick]
      \tikzstyle{defDetails}=[rectangle, semithick, fill=white, draw=black, font=\tiny, 
         rounded corners=0.1cm, yshift=0cm, inner sep=0pt,
         minimum height=0.35cm, minimum width=0.6cm]
      \tikzstyle{blacknode}=[shape=circle, draw=black, line width=2, minimum width=2mm]

      \node[Gnelabel] (GNP)  at (\posGx+0.07\hX, \posNPy+0.05\hY) {\mySlash{G}{{n_{e}}}};

      \node[Grect] (G4) at (\posGx, 1*\hY+\startY) {\scalebox{0.8}{$4$}};
      \node[Grect] (G5) at (\posGx, 2*\hY+\startY) {\scalebox{0.8}{$5$}};
      \node[Grect] (G6) at (\posGx, 3*\hY+\startY) {\scalebox{0.8}{$6$}};
      \node[Grect] (G7) at (\posGx, 4*\hY+\startY) {\scalebox{0.8}{$7$}};
      \node[Grect] (G8) at (\posGx, 5*\hY+\startY) {\scalebox{0.8}{$8$}};

      \node[NPrect]     (NP3)  at (0*\hX,            \posNPy) {\scalebox{0.8}{$3$}};
      \coordinate       (NP4a) at (1*\hX,            \posNPy);
      \coordinate       (NP4b) at (1*\hX + 1*\hsX,   \posNPy);
      \node[NPrectlong] (NP4)  at (1*\hX + 0.5*\hsX, \posNPy) {\scalebox{0.8}{$4$}};
      \node[NPrect]     (NP5)  at (2*\hX + 1*\hsX,   \posNPy) {\scalebox{0.8}{$5$}};
      \node[NPrect]     (NP6)  at (3*\hX + 1*\hsX,   \posNPy)  {\scalebox{0.8}{$6$}};

      \draw let 
         \p{G8}=(G8), 
         \p{G7}=(G7), 
         \p{G6}=(G6), 
         \p{G5}=(G5), 
         \p{G4}=(G4), 
         \p{NP3}=(NP3), 
         \p{NP4}=(NP4), 
         \p{NP4a}=(NP4a),
         \p{NP4b}=(NP4b),
         \p{NP5}=(NP5),
         \p{NP6}=(NP6)
      in
         node[TDimg] (PdP1td) at (\x{NP3},\y{G4}) {\includetd{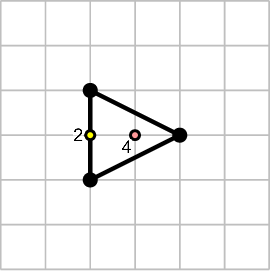}}
         node[TDimg] (dP1td) at (\x{NP4a},\y{G4}) {\includetd{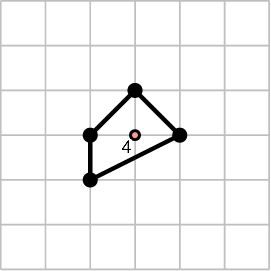}}
         node[TDimg] (F0td) at (\x{NP4b},\y{G4}) {\includetd{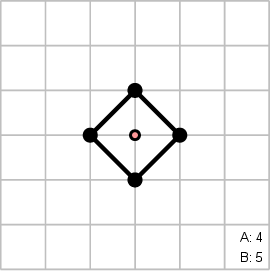}}
         node[TDimg] (PdP2td) at (\x{NP4},\y{G5}) {\includetd{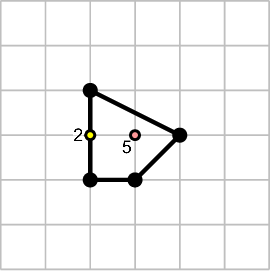}}
         node[TDimg] (dP2td) at (\x{NP5},\y{G5}) {\includetd{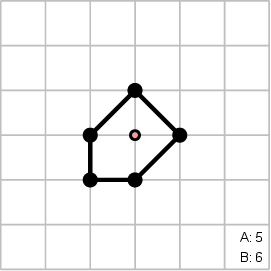}}
         node[TDimg] (PdP3atd) at (\x{NP3},\y{G6}) {\includetd{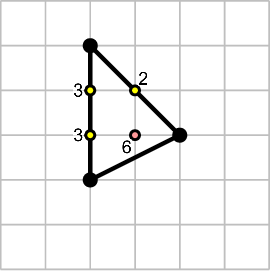}}
         node[TDimg] (PdP3ctd) at (\x{NP4},\y{G6}) {\includetd{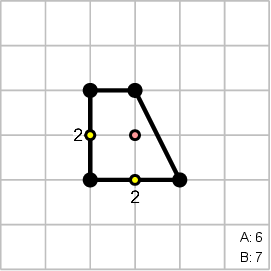}}
         node[TDimg] (PdP3btd) at (\x{NP5},\y{G6}) {\includetd{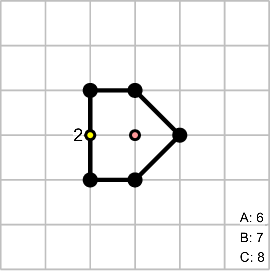}}
         node[TDimg] (dP3td) at (\x{NP6},\y{G6}) {\includetd{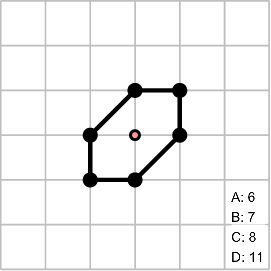}}
         node[TDimg] (PdP4btd) at (\x{NP4},\y{G7}) {\includetd{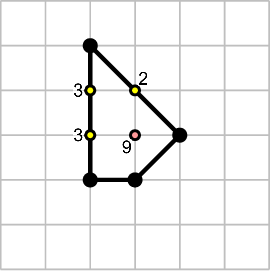}}
         node[TDimg] (PdP4atd) at (\x{NP5},\y{G7}) {\includetd{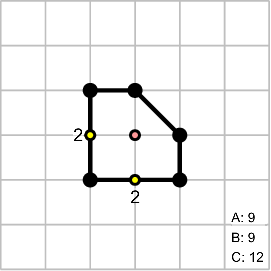}}
         node[TDimg] (C3Z4Z2td) at (\x{NP3},\y{G8}) {\includetd{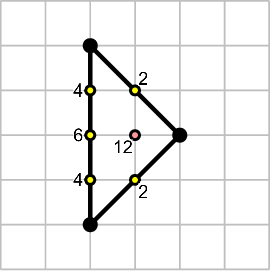}}
         node[TDimg] (L131Z2td) at (\x{NP4a},\y{G8}) {\includetd{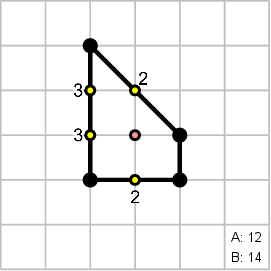}}
         node[TDimg] (PdP5td) at (\x{NP4b},\y{G8}) {\includetd{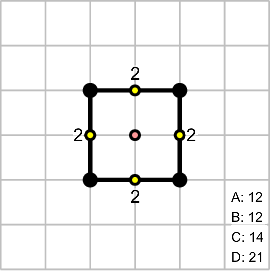}}

         node[rectangle] (endpnt) at ($(\x{NP3}, \y{G8}) + (0, 0.7\hY)$) {}
      ;
      
      \tdtexlabel{PdP1td}{$\bbC^3/\bbZ_4 \,(1,1,2)$}
      \tdtexlabel{dP1td}{$\text{dP}_1$}
      \tdtexlabel{F0td}{$\calC/\bbZ_2 \,(1,1,1,1)$}
      \tdtexlabel{PdP2td}{$\text{PdP}_{2}$}
      \tdtexlabel{dP2td}{$\text{dP}_{2}$}
      \tdtexlabel{PdP3atd}{$\bbC^3/\bbZ_6 \,(1,2,3)$}
      \tdtexlabel{PdP3ctd}{$\text{SSP}/\bbZ_2 \,(0,1,1,1)$}
      \tdtexlabel{PdP3btd}{$\text{PdP}_{3b}$}
      \tdtexlabel{dP3td}{$\text{dP}_{3}$}
      \tdtexlabel{PdP4btd}{$\text{PdP}_{4b}$}
      \tdtexlabel{PdP4atd}{$\text{PdP}_{4a}$}
      \tdtexlabel{C3Z4Z2td}{$\bbC^3/(\bbZ_4 \times \bbZ_2) \,(1,0,3)(0,1,1)$}
      \tdtexlabel{L131Z2td}{$L^{1,3,1}/\bbZ_2 (0,1,1,1)$}
      \tdtexlabel{PdP5td}{$\calC/(\bbZ_2 \times \bbZ_2) \,(1,0,0,1)(0,1,1,0)$}

      \draw[rgArrow] (PdP1td.south) |- ++(2,-0.4) -- 
         node[defDetails]{\ref{sec:PdP1todP1}} 
         ++(2,0) -|  (F0td.south);

      \draw[rgArrow] (PdP2td.south) |- ++(1.4,-0.4) -- 
         node[defDetails]{\ref{sec:PdP2todP2}} 
         ++(2,0) -| (dP2td.south);

      \draw[rgArrow] (PdP3atd.south) |- ++(0.5,-0.4) -- 
         node[defDetails]{\ref{sec:PdP3atoPdP3c}} 
         ++(2.5,0) -| +(0,1) |- (PdP3ctd.west);
      
      \draw[rgArrow] (PdP3ctd.south) |- ++(0.5,-0.4) -- 
         node[defDetails]{\ref{sec:PdP3ctoPdP3b}} 
         ++(2.5,0) -| +(0,1) |- (PdP3btd.west);
      
      \draw[rgArrow] (PdP3btd.south) |- ++(1,-0.4) -- 
         node[defDetails]{\ref{sec:PdP3btodP3}} 
         ++(1.5,0) -| (dP3td.south);
      
      \draw[rgArrow] (PdP4btd.south) |- ++(1,-0.4) -- 
         node[defDetails]{\ref{sec:PdP4btoPdP4a}} 
         ++(2,0) -| (PdP4atd.south);

      \draw[rgArrow] ($(C3Z4Z2td.south) + (0.1cm, 0.cm)$) |- ++(1,-0.4) -- 
         node[defDetails]{\ref{sec:C3Z4Z2toL131Z2}} 
         ++(1,0) -| (L131Z2td.south);

      \draw[rgArrow, double, semithick] ($(C3Z4Z2td.south) + (-0.1cm, 0.cm)$) |- ++(4,-0.8) -- 
         node[defDetails]{\ref{sec:C3Z4Z2toPdP5}} 
         ++(1,0) -| (PdP5td.south);

      \matrix (Leg) [draw, below left, rounded corners, anchor=south east, fill=black!10] at ($ (current bounding box.south east) + (0em, 3em) $) {
         \draw[rgArrow, thick, anchor=west] (0.2,0) -- (0.8,0) node{\footnotesize Zig-zag flow}; \\
         \draw[rgArrow, double, thick, anchor=west] (0.2,0) -- (0.8,0) node{\footnotesize Double zig-zag flow}; \\
      };
   \end{tikzpicture}
   \caption{Toric-to-toric 4d $\mathcal{N}=1$ RG flows connecting geometries with reflexive toric diagrams via zig-zag deformations. Rows and columns are organized by the number of nodes $G$ in the quiver and the number of extremal points $n_e$ in the polytope, following \cite{Hanany:2012hi}. Links in the arrows point to details in \cref{sec:RGflows}.}
   \label{fig:defsSummary}
\end{figure}

%% file: Sections/Tilings.tex

\newcommand{\tikzcircle}[2][black,fill=white]{\raisebox{1pt}{\tikz[baseline=-0.5ex]\draw[#1,radius=#2] (0,0) circle;}}%

\section{Toric Quiver Gauge Theories and Brane Tilings}
\label{sec:Tilings}

\subsection{Brane tiling dictionary}
\label{sec:brane-tiling-dict}

A \emph{brane tiling} \cite{Franco:2005rj,Hanany:2005ve,Kennaway:2007tq} is a bipartite graph on a 2-torus $\bbT^2$ which encodes quiver and superpotential of a so \emph{toric} supersymmetric gauge theory with four supercharges (4d $\calN=1$ and dimensional reductions thereof). This is the low energy field theory on the worldvolume of $N$ D-branes probing a CY$_3$ cone $\calY$ with $U(1)^3$ isometry.
The graph consists of 0-dimensional vertices (or nodes), 1-dimensional edges connecting a pair of vertices, and 2-dimensional faces bounded by edges.
The graph is bipartite, meaning that vertices are coloured black or white and edges connect vertices of different colours. The colouring encodes a counterclockwise/clockwise ordering of the edges incident to a vertex.

The data of a brane tiling and its field theory interpretation are given by:
\begin{itemize}
    \item \textbf{Faces} correspond to $U(N)_i$ group factors,%
\footnote{In four dimensions the low energy gauge groups are special unitary, since central $U(1)$ factors are massive or decoupled. Considering unitary gauge groups is nevertheless useful to study mesonic moduli space of vacua, which we focus on in this paper. Baryonic branches can also be studied, by relating baryonic VEVs to FI parameters (see section \ref{sec:Resolutions}). See \cite{Martelli:2008cm} and references therein for a comprehensive discussion.} for $i\in\left\{1,\dots,G\right\}$, where $G$ denotes the number of unitary gauge group factors (sometimes $F$ for the number of faces). 
    \item \textbf{Edges} between two faces $i$ and $j$ represent chiral superfields $X_{ij}^a$, which transform in the bifundamental representation 
    of $U(N)_i \times U(N)_j$. If $i=j$, the chiral superfield is in the adjoint representation. 
    The direction of the arrow in the quiver diagram is determined by the orientation of the distinct-colored vertices that the corresponding edge connects to in the tiling. We write $E$ for the number of chiral superfields/edges.
    \item \textbf{Vertices} \tikzcircle{4pt}/\tikzcircle[fill=black]{4pt}~ represent toric superpotential terms.
    Each superpotential term is the single trace of the product of bifundamentals chiral superfields associated to the incoming edges of the corresponding vertex, ordered clockwise/counter-clockwise and with a $+$/$-$  sign according to the white/black colour of the vertex.
\end{itemize}
The bipartite nature of the graph guarantees that each bifundamental chiral superfield appears exactly once in exactly two superpotential monomials with opposite signs.
This is known as the \emph{toric condition} and leads to F-term equations of the form
\begin{align}
    \fracpd[W]{X_{ij}} = 0 \quad\Rightarrow\quad  X_{j, c_1} \cdots X_{c_{r}, i}  = X_{j, d_1} \cdots X_{d_{s}, i} ~,
\end{align}
for two specific connected paths $(c_1,\dots,c_r)$ and $(d_1,\dots,d_s)$ in the brane tiling/quiver. 

The dual graph of the brane tiling is the \emph{periodic quiver diagram} \cite{Franco:2005rj}: it consists of 0-dimensional vertices (or nodes) representing unitary gauge groups, 1-dimensional directed edges (or arrows) representing bifundamental chiral superfields, and 2-dimensional oriented faces whose boundaries represent (positive or negative) superpotential terms. 
The usual quiver graph refers only to the 1-skeleton structure and encodes the gauge group and matter content of the gauge theory.
The quiver data is encoded in the \emph{incidence matrix} $d \in \textbf{M}_{G\times E}(\bbZ)$ of the graph:
\begin{equation}
    d_{ie}=\begin{cases}
        +1 & t(e)=i\\
        -1 & s(e)=i\\
        0 & \text{else}
    \end{cases}
    \label{eq:incidence}
\end{equation}
where $t(e)$ and $s(e)$ denote the nodes at the tail and head of the arrow (or directed edge) $e$.
In the quiver gauge theory, we associate to each edge $e$ a chiral superfield $X_e$ in the bifundamental representation of $U(N)_{t(e)}\times U(N)_{s(e)}$.
By an abuse of notation, we will interchangeably denote a chiral superfield as $X_e$ according to the corresponding edge in the graph or as $X_{ij}^{a}$ according to the nodes $i=t(e)$ and $j=s(e)$ it connects. 
The superscript $a$ labels the $A_{ij}$ bifundamentals between nodes $i$ and $j$ in the quiver.

Toric quiver gauge theories are generically chiral, hence there is a non-trivial condition for cancellation of gauge anomalies.
In the most general case, where the gauge group is
\begin{align}
    \calG = \prod_{i=1}^G U(N_i) ~,
\end{align}
the four-dimensional $SU(N_i)$ gauge anomaly cancellation\footnote{The central $U(1)_i \subset U(N_i)$ generically also have (mixed) gauge anomalies, which are cancelled in string theory by St\"{u}ckelberg terms that make the corresponding gauge bosons massive \cite{Buican:2006sn,Martelli:2008cm}. Anomaly-free central $U(1)_i$ factors decouple in the infrared and become non-anomalous baryonic symmetries.} condition reads \cite{Benvenuti:2004dw}
\begin{align}
    \sum_{j} (A_{ij} - A_{ji}) N_{j} = 0 ~,
    \label{eq:gauge_anomaly}
\end{align}
where $A \in \textbf{M}_{G\times G}(\bbZ)$ is the adjacency matrix of the quiver, with entry $A_{ij}$ counting the number of arrows from node $i$ to node $j$.

\subsection{The moduli space of vacua and its components}

The moduli space of supersymmetric vacua $\calM$ of a 4d $\calN=1$ low energy quiver gauge theory on the worldvolume of regular D3-branes consists of various components (or \emph{branches}).

The first go-to tool for studying the moduli of quiver gauge theories is the master space $\calF^\flat$ \cite{Hanany:2010zz, Forcella:2008bb,Forcella:2008eh,Forcella:2008ng}.
The master space is a toric variety of dimension $3N + G - 1$ and is defined as the space of solutions to the F-term equations, given the chiral superfields as matrices, quotiented by $SU(N)^{G}$, since only the non-abelian part of $\calG = U(N)^G$ couples in the IR.
In particular, for the abelian theory on the worldvolume of a single D3-brane, $\calF^\flat$ is the same as the full moduli space of the low energy quiver theory, of dimension $G+2$.
The master space $\calF^\flat$ typically decomposes into multiple irreducible components: the non-trivial top component, referred as the \emph{coherent component} ${}^{\text{Irr}}\calF^\flat$, and multiple lower-dimensional pieces, typically freely-generated.

In general, we can write the master space as
\begin{align}
    \calF^\flat = \Spec\left( \coset{ \bbC[X_{e_1}, \dots, X_{e_E}] }{ \angl{\partial_X W} } \right)^{SU(N)^{G}_{\bbC}} ~.
    \label{eq:master-space-nonabelian}
\end{align}
By $\calR^{SU(N)^{G}_{\bbC}}$, we mean the algebraic invariants of the ring $\calR$ under the complexified non-abelian gauge symmetry $SU(N)^{G}_{\bbC}=SL(N,\bbC)^G$, \emph{i.e.} traces (mesons) and determinants (baryons).
These invariants form the spectrum of chiral BPS operators and their vacuum expectation values (VEVs) parametrize two components of the moduli spaces, the \emph{mesonic branch} and the \emph{baryonic branch}. 

If the CY$_3$ cone $\calY$ only has an isolated singularity at the tip of the cone, the two branches are:
\begin{itemize}
    \item \emph{Mesonic branch = Geometric branch} $\calM^\text{mes}=\calM^\text{geom}$\cite{Forcella:2008bb,Forcella:2008eh, Berenstein:2002ge, Martelli:2008cm}: it describes $N$ regular D3-branes probing the CY$_3$ cone $\calY$.
    It is an affine variety parametrised by VEVs of mesonic operators (henceforth often referred to as mesons), \emph{i.e.} traces of products of chiral superfields which are associated to cycles in the quiver:
    \begin{align}
        M_i = \tr\left( X_{e_1} \cdots X_{e_\ell} \right)~.
    \end{align}
    In general, we have that $\calM^\text{geom} = \Sym^N(\calY)$ with dimension $3N$. If $N=1$,  $\calM^\text{geom} = \calY$ reproduces the affine Calabi-Yau geometry, hence the name geometric branch.
    \item \emph{Baryonic branch} $\calM^\text{bar}$ \cite{Forcella:2007wk, Butti:2006au}: it describes $N$ regular D3-branes probing the moduli space of (partial) resolutions $\tilde{\calY}_{\xi}$ of the singular CY$_3$ cone $\calY$. It is parametrised by VEVs of dibaryonic operators (or simply dibaryons) of the schematic form
    \begin{align}
        B_j = \det\left( X_{e_1} \dots X_{e_m} \right) ~,
    \end{align}
    where $( e_1, \dots, e_m )$ are open paths in the quiver.     Geometrically, the full moduli space of vacua ${}^{\text{Irr}}\calF^\flat$ can be thought as the total space of the fibration $\tilde{\calY}_{\xi}$ over a complexified $(G-1)$-dimensiontal K\"{a}hler moduli space of $\calY$ parametrised by a vector $\xi$.
    The fibre directions are mesonic, while the base directions are baryonic.
    In the gauged linear sigma model (GLSM) that describes the toric geometry of $\tilde{\calY}_{\xi}$, the K\"{a}hler moduli $\Re(\xi)$ are realized by Fayet-Iliopoulos (FI) parameters.
\end{itemize}
If we gauge the central $U(1)^G \subset U(N)^G$, the baryonic branch disappears and the moduli space of vacua consists of the mesonic branch only.

In this paper we are interested in Calabi-Yau cones $\calY$ which have lines of non-isolated singularities, which are of $A$-type since we assume that $\calY$ is toric. Then there are additional mesonic components of the moduli space of vacua on top of the geometric branch:
\begin{itemize}
    \item \emph{$\calN=2$ Coulomb branch} $\calM^{\calN=2}$: for each non-isolated singularity  of $\calY$, locally of $A_{k-1}$ type, this extra mesonic component of the moduli space of vacua describes $k N$ $\calN=2$ fractional D3-branes \cite{Franco:2005zu} probing the locus of the non-isolated $A_{k-1}$ singularity.%
    \footnote{If $\calY=\bbC^2/\Gamma \times \bbC$, with $\Gamma\subset SU(2)$ a finite subgroup, this is literally the Coulomb branch of an $\calN=2$ theory. The physics of the fractional branes and the geometry of this component of the moduli space of vacua is largely determined by the local geometry of $\calY$ near the singularity, hence the name for the type of fractional branes and for the branch of the moduli space. The latter is a misnomer, as there is no Coulomb branch for $\calN=1$ gauge theories, but we stick to it for the sake of brevity.}     For $N=1$ and for each $A_{k-1}$ singularity, this branch is parametrised by VEVs of $k$ independent chiral mesons, which take identical VEVs on $\calM^\text{geom}$. The intersection of an $\calN=2$ Coulomb branch and the geometric branch is a locus of non-isolated singularity of the latter. A regular D3-brane on a non-isolated singularity of $\calY$ can marginally decay into $k$ different $\calN=2$ fractional branes which separately probe the singular locus. 
    Conversely, $k$ coincident  $\calN=2$ fractional branes of different type can recombine into a regular D3-brane, which is free to explore the whole Calabi-Yau $\calY$.
\end{itemize}

The full mesonic moduli space of vacua is then the union of the geometric branch and the $\calN=2$ Coulomb branches: $\calM^\text{mes}=\calM^\text{geom}\cup \calM^{\calN=2}$.
We will use $f(M_i, \dots) \simeq 0$ to indicate relations in the full chiral ring, while we will use $\sim$ to denote relations which only hold on the geometric component, but not on $\calN=2$ Coulomb branches.\footnote{Note that the presence of additional components in $\calF^\flat$ does not immediately imply $\calN=2$ Coulomb branches, since isolated singularities can still produce reducible master spaces.}

In this work we focus on the worldvolume theories on a stack of $N$ regular D3-branes, meaning that the resulting quiver theory has $N_i = N$. The gauge anomaly cancellation is thus enforced by the bipartite nature of brane tilings.
We will initially be interested in the conical geometry probed by D3-branes, which is the geometric moduli space of their worldvolume quiver gauge theory, as well as in the $\calN=2$ Coulomb branches associated to non-isolated singularities of this geometry.
We will use brane tilings to construct and study non-toric deformations of Klebanov-Witten type \cite{Klebanov:1998hh}, which interpolate via deformation of complex structure between different toric Calabi-Yau geometries $\calY$ (hence geometric branches), and lift non-isolated singularities (hence $\calN=2$ Coulomb branches).
We will also observe the consequences for the baryonic branch by studying the effects of the deformations on resolutions $\tilde{\calY}_{\xi}$ and volumes of holomorphic 2-cycles therein.
All of this can be studied in the abelian case $N_i=N=1$, which we thus restrict to in the following.
The case with $N>1$ can be recovered by appropriate symmetric products.

\subsection{Fast-forward algorithm and toric diagrams}

The brane tiling data can be encoded in the \emph{Kasteleyn matrix} \cite{Hanany:2005ve,Franco:2005rj}. 
This is the node-node connectivity matrix of the bipartite graph, which also keeps track of the winding number
\begin{align}
    h(X_{e}) = \Big( h_z(X_{e}), h_w(X_{e}) \Big)
\end{align}
of the edges.
To define it we need to pick a fundamental domain and choose two primitive $1$-cycles $(\gamma_z, \gamma_w)$ on $\bbT^2$.
By fixing the bipartite orientation of the edges to be directed from the white to black nodes, we have that $h_{z,w}$ equals $+1$/$-1$ if the edge $X_e$ crosses the cycle $\gamma_{z,w}$ in the positive/negative direction, and $0$ if no crossing occurred.
As a convention, we index the row and columns of the Kasteleyn matrix as black and white nodes respectively.
Therefore, the Kasteleyn matrix takes the form
\begin{align}
    K_{ij}(z,w) = \sum_{e \in B_i \cap W_j} 
    a_e \; z^{h_z(X_e)} \; w^{h_w(X_e)} ~,
\end{align}
where $B_i \cap W_j$ denotes the set of tiling edges that connect to both vertices $B_i$ and $W_j$, and $a_e$ is an \emph{edge weight}.

A \emph{perfect matching} $p_\alpha$ \cite{Hanany:2005ve,Franco:2005rj}  (or dimer collection) is a set of edges (``dimers'') which touch each vertex of the brane tiling exactly once.
Obtaining all perfect matchings is a combinatorial problem that can be solved via the Kasteleyn matrix: each term in the expansion of its permanent is associated to a unique perfect matching,
\begin{align}
    \perm K(z,w) = \sum_{\alpha=1}^{c} 
    z^{h_z(p_\alpha)} \; w^{h_w(p_\alpha)} \prod_{X_e \in p_\alpha} a_e ~,
    \label{eq:permK}
\end{align}
with the winding number given by
\begin{align}
    h(p_\alpha) = \sum_{X_e \in p_\alpha} h(X_e) ~.
    \label{eq:windingPalpha}
\end{align}
Keeping the $a_e$ edge weight general we can keep track of which edge $e$ belongs to a particular perfect matching $p_\alpha$ (if we take all $a_e = 1$, the permanent lists the multiplicities of the perfect matchings for each monomial $z^a w^b$.)
The collection of perfect matchings is summarised in the \emph{perfect matching matrix} $P \in \textbf{M}_{E\times c}(\bbZ)$ \cite{Hanany:2005ve,Franco:2005rj}, with entries
\begin{align}
    P_{e\alpha} = \begin{cases}
        1 ~, & X_e \in p_\alpha \\
        0 ~, & X_e \notin p_\alpha
    \end{cases} ~.
\end{align}

The \emph{toric diagram} $\Delta$ \cite{Leung:1997tw,Franco:2005rj,Hanany:2005ve} of the CY$_3$ cone $\calY=\calM^{\rm{geom}}$ can be read off directly from \cref{eq:permK}: to each monomial $z^a w^b$ in the expansion we associate a lattice point $(a,b)$.
Alternatively, the coordinates of the lattice points can also be obtained from the winding numbers of differences of perfect matchings and a reference perfect matching.
These form the lattice points of a convex lattice polygon $\Delta \subset \bbZ^2$, which we consider modulo $SL(2,\bbZ)$ transformations and lattice translations.
Alternatively, the perfect matching technology gives the toric diagram directly via the \emph{fast-forward} algorithm \cite{Franco:2005rj} (which improves the  \emph{forward} algorithm of \cite{Feng:2000mi,Feng:2001xr}):
\begin{itemize}
    \item 
To each perfect matching $p_\alpha$ we associate a field in an abelian gauged linear sigma model (GLSM) \cite{Witten:1993yc,Feng:2000mi}, which we call a \emph{perfect matching variable} and also denote by $p_\alpha$, with a slight abuse of notation.
    Expressing the bifundamentals in terms of perfect matching variables as
    \begin{align}
    	X_e = \prod_{\alpha=1}^{c} \left( p_\alpha \right)^{P_{e\alpha}} 
    	\label{eq:bifund_from_pm}
    \end{align}
   solves the F-term equations automatically
    \begin{align}
    	\bar F_e = \fracpd[W]{X_e} = 0~,
    \end{align}    
    at the expense of introducing an abelian gauge symmetry that leaves each bifundamental in \eqref{eq:bifund_from_pm} invariant.   
    The charges of perfect matching variables under this gauge symmetry are encoded in a charge matrix $Q_F \in\mathbf{M}_{(c-G-2)\times c}(\bbZ)$, defined as
    \begin{align}
        Q_F P^T = 0 \quad\Rightarrow\quad Q_F=(\ker P)^T ~.
        \label{eq:Q_F}
    \end{align}
    We remark that we can bypass the brane tiling construction step and still obtain $P$ by grouping superpotential terms.
    While we lose the edge-winding information of the fundamental domain ($z=w=1$ in \eqref{eq:permK}), the same combinatorics that allows us to obtain the perfect matchings is still present.
    \item There are D-terms for each of the $U(1)$ factors in the quiver, which take the form
    \begin{align}
        D_i = - g_i^2 \left( \sum_e d_{ie} \abs{ X_e }^{2} - \xi_{i} \right) ~,
    \end{align}
    where $\xi_{i}$ are Fayet-Iliopoulos (FI) parameters.
    The D-term charge matrix $Q_D \in \mathbf{M}_{(G-1)\times c}(\bbZ)$ is obtained by solving the matrix equation
    \begin{align}
        Q_D P^T = \hat d ~,
        \label{eq:Q_D}
    \end{align}
    where we define the \emph{reduced incidence matrix} $\hat d \in\mathbf{M}_{(G-1)\times E}(\bbZ)$  by subtracting all remaining rows of $d$ by one of the rows, e.g. $\hat d_{ie} = d_{ie} - d_{Ge} ~,~ i\in\set{1,\dots,G-1}$.
    This matrix encodes the charges of the bifundamentals under the faithfully acting gauge group of the quiver $U(1)^G/U(1)$.

    \item We can combine $Q_F$ and $Q_D$ in a total charge matrix $Q_t \in\mathbf{M}_{(c-3)\times c}(\bbZ)$
    \begin{align}
        Q_t = \begin{pmatrix}
            Q_F \\ Q_D
        \end{pmatrix} ~.
    \end{align}
    Then, the $3\times c$ matrix
    \begin{align}
        G_t = (\ker Q_t )^T
    \end{align}
    gives a collection of $c$ lattice points in $\bbZ^3$ representing the  charges of the perfect matching variables under the $U(1)^3$ toric symmetries.
    For the model to describe D3-branes probing a toric Calabi-Yau the lattice points must be coplanar, so that it can be projectivized to obtain the toric diagram $\Delta \subset \bbZ^2$. 
\end{itemize}

The geometric branch of the moduli space $\calM^{\rm{geom}}$ of the 4d $\calN = 1$ supersymmetric gauge theory with abelian gauge group $U(1)^G$ coincides with the classical moduli space of the GLSM introduced above. 
This is the K\"{a}hler quotient by the abelian gauge group with the total charge matrix $Q_t$
\begin{align}
    \calM^\text{geom} = \ccoset{\bbC^c}{}{U(1)^{c-3}_{Q_t}}
\end{align}
at level $\xi = 0$, which results in a toric CY$_3$ cone $\calY$.
We stress that while the perfect matching technology is very useful to extract the geometric branch, it does not provide a general solution of the $F$-term equations and does not capture $\calN=2$ Coulomb branches.

\subsection{Zig-zag paths}

A \emph{zig-zag path} $\eta$ is a special type of closed oriented path in a brane tiling, which forms a homology cycle on $\bbT^2$ \cite{Hanany:2005ss}.
By definition, a zig-zag path follows the edges of the brane tiling, making a maximal left turn at each white node and a maximal right turn at each black node,  until the path closes in $T^2$.
A convenient way to depict a zig-zag path is to deform it slightly so that it crosses in the middle each of the edges that it follows, keeping the black vertex on the left and the white vertex on the right as we do when we go from the tiling to the dual periodic quiver.
From the description above, we can work out how a brane tiling edge is oriented regarding a given zig-zag $\eta$: we define a \emph{zig} to be an edge that goes from a black to white node along $\eta$, otherwise we call it a \emph{zag}.
A given edge is exactly a zig and a zag of two zig-zag paths.
\begin{figure}[!ht]
    \centering
    \hspace*{\fill}%
    \begin{subfigure}[c]{0.72\columnwidth}
        \includegraphics[width=\columnwidth]{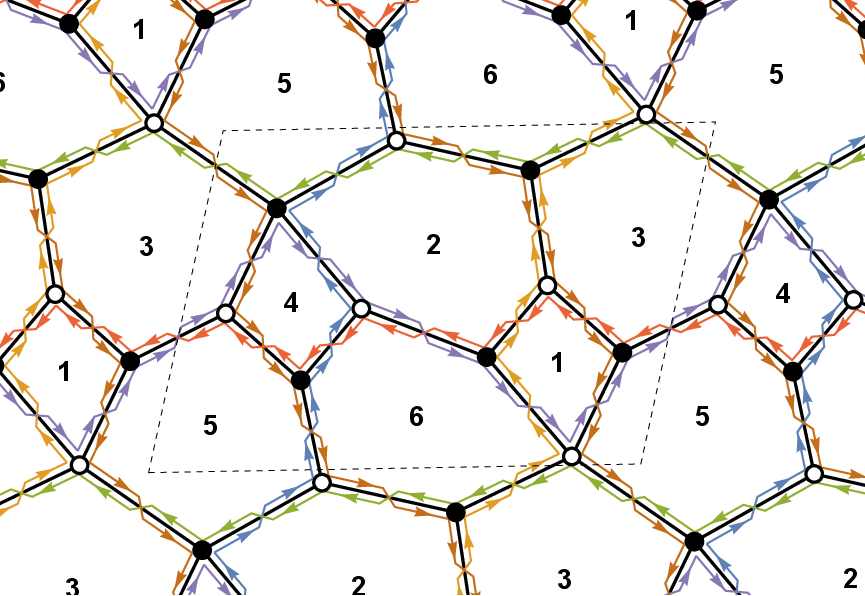}
        \caption{}
        \label{fig:PdP3cB-til-zz}
    \end{subfigure}%
    \hspace*{\fill}%
    \hspace*{\fill}%
    \begin{minipage}[c]{0.22\columnwidth}
        \begin{subfigure}[c]{\columnwidth}
            \includegraphics[width=\columnwidth]{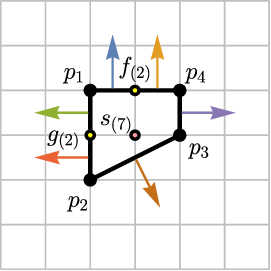}
            \caption{}
            \label{fig:PdP3cB-td-zz}
        \end{subfigure}
        \\[0.8em]
        \begin{subfigure}[c]{\columnwidth}
            \includegraphics[width=\columnwidth]{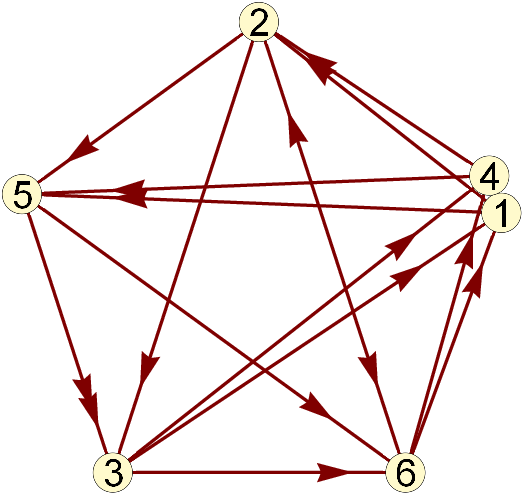}
            \caption{}
            \label{fig:PdP3cB-quiver}
        \end{subfigure}
    \end{minipage}
    \hspace*{\fill}%
    \caption{Model $\text{PdP}_{3c}$ Phase B \cite{Hanany:2012hi}: (a) brane tiling decorated with zig-zag paths; (b) toric diagram with normal vectors $(p,q)$ associated to zig-zags; (c) quiver diagram.}
    \label{fig:PdP3cB-til-quiver-td}
\end{figure}

The zig-zag paths form a collection of loops in $T^2$, with winding numbers $h(\eta_i)$ in terms of the homology basis of the reference fundamental domain, similar to \cref{eq:windingPalpha}.
From these we can reconstruct the data of the singular toric geometry: winding numbers of zig-zag paths form a set of vectors in $\bbZ^2$, which represent the outward pointing normals of edges in the toric diagram (compare \cref{fig:PdP3cB-td-zz} with \cref{fig:PdP3cB-til-zz}).

\subsection{Toric (Seiberg) duality and specular duality of brane tilings}

\begin{figure}[!ht]
    \centering
    \begin{tikzpicture}
        \node at (0,0) {\includegraphics[width=0.225\textwidth]{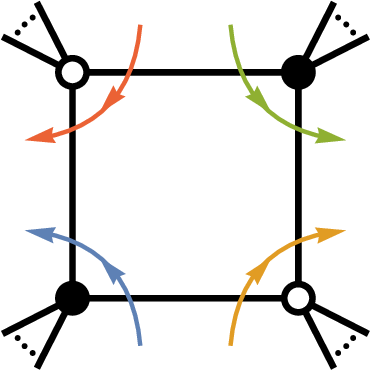}};
        \node at (3.5,0) {$\longleftrightarrow$};
        \node at (7,-0.02) {\includegraphics[width=0.225\textwidth]{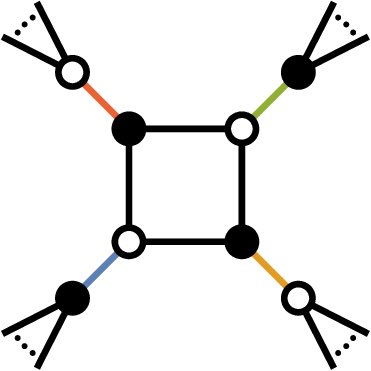}};
    \end{tikzpicture}
    \caption{Toric (Seiberg) duality as an operation on the brane tiling. Coloured arrows represent chiral superfield products. Often, one can integrate out massive fields by shrinking pairs of edges which share a bivalent vertex.}
    \label{fig:ToricDuality}
\end{figure}

Often, different quiver gauge theories, represented by different brane tilings, are related by a so called \emph{toric duality} \cite{Feng:2001bn, Beasley:2001zp}.
This is just a manifestation of $\calN=1$ Seiberg duality \cite{Seiberg:1994pq} for theories with toric moduli spaces. 
A Seiberg duality on a quiver node of rank $N_c$ defines a mutation to a new (dual) node of rank $N_f - N_c$, with $N_f = \sum_{i\ne c} N_i A_{ic} = \sum_{j\ne c} A_{cj} N_j$ guaranteed by the gauge anomaly cancellation.
In general, Seiberg duality relates an infinite tree of dual quiver gauge theories by allowing the mutation of any sequence of nodes in the quiver \cite{Franco:2003ja}.
To ensure that the Seiberg dual model is also toric, the ranks $N_i = N$ of the gauge groups in the worldvolume theory of $N$ regular D3-branes probing a toric CY 3-fold singularity must be unchanged.
The mutated node $i$ that obeys this condition has $N_f = 2 N$ and corresponds to a square in the brane tiling.
The result of Seiberg duality on square faces is graphically represented in \cref{fig:ToricDuality}.

Toric duality relates different brane tilings with identical mesonic moduli space $\calM^\text{mes}$. The associated toric diagrams are $GL(2,\bbZ)$ equivalent. 
However, the multiplicities of internal points are not the same and this leads to a different ${}^{\text{Irr}}\calF^\flat$ due to anomalous $U(1)$ baryonic symmetries.
These become equivalent if we restrict to non-anomalous charges \cite{Forcella:2008ng}.
Toric dual brane tilings are referred to as \emph{toric phases} of the same geometry $\calY$.

\emph{Specular duality} \cite{Feng:2005gw,Hanany:2012vc} is instead an application of mirror symmetry to brane tilings.
It is a correspondence between a brane tiling on $\bbT^2$ and a brane tiling defined on the mirror curve $\Sigma$ of the toric 3-fold $\calY$.
A toric $\calY$, which is specified by a convex lattice polytope $\Delta\subset \bbZ^2$, has a mirror geometry $\calW$ defined by the double fibration over the $Z$ plane,
\begin{align}
    \begin{aligned}
        Z &= P(z,w) := \sum_{(p,q)\in\Delta} c_{p,q} z^p w^q \\
        Z &= u v~,
    \end{aligned}
\end{align}
for $z,w\in\bbC^\times$ and $u,v\in\bbC$.
The complex coefficients $c_{p,q}$ parameterize the complex structure deformations of $\calW$ and are mirror dual to the K\"{a}hler moduli of $\calY$ \cite{Hanany:2001py}.
The mirror curve $\Sigma_Z$ is defined by the first equation $P(z,w)-Z=0$ and encodes all the toric geometry information of $\calY$ through the \emph{Newton polynomial} $P(z,w)$ of $\Delta$.

\begin{figure}[!ht]
    \centering
    \begin{tikzpicture}
        \node at (0,0) {\includegraphics[width=0.33\textwidth]{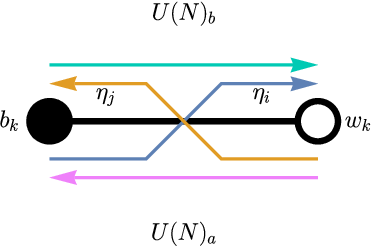}};
        \node at (4,0) {$\longleftrightarrow$};
        \node at (8,-0.02) {\includegraphics[width=0.33\textwidth]{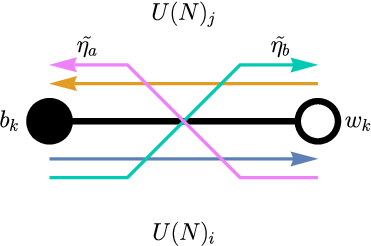}};
    \end{tikzpicture}
    \caption{Untwisting map, followed by the identification of punctures with gauge groups.}
    \label{fig:UntwistingMap}
\end{figure}

The fibre at the origin, $P(z,w) = 0$, which we denote simply by $\Sigma$, is of particular relevance as it can be related to the brane tiling on $\bbT^2$ via the \emph{untwisting map} \cite{Feng:2005gw}, exemplified graphically by \cref{fig:UntwistingMap}.

In the untwisting map, the number of edges and vertices in the tiling remain the same. A given edge $X_{ab}$ crossed by a zig $\eta_{i}$ and a zag $\eta_{j}$ is relabeled as $X_{ij}$.
The map acts on the boundary of a face (gauge group $U(N)_a$), turning it into a zig-zag path $\tilde{\eta}_a$, which has non-trivial homology on the new tiling.
On the other hand, the original zig-zag paths $\eta_i$ become closed polygon cycles that wind around punctures of $\gamma_i$ of the Riemann surface $\Sigma$.
The number of punctures of $\Sigma_{Z}$ is given by the number of external points $B$ of $\Delta$, which is the same as the number of zig-zag paths.
The untwisting map leads to the union of closed polygon cycles winding around punctures $\gamma_i$ of $\Sigma$, usually referred as a \emph{shiver}.
By mapping $\gamma_i$ to gauge groups $U(N)_{i}$ we obtain a consistent  tiling.
This is summarized in \cref{tb:specDuality}.
\begin{table}[!ht]
    \centering
    \tabulinesep=1mm
    \begin{tabu}{l|l}
        brane tiling on $\bbT^2$ & brane tiling on $\Sigma$ \\
        \hline\hline
        zig-zag path $\eta_i$ & face/gauge group $U(N)_i$  \\
        \hline
        face/gauge group $U(N)_a$ & zig-zag path $\tilde{\eta}_a$ \\
        \hline
        node/term $w_k,~b_k$ & node/term $w_k,~b_k$ \\
        \hline
        edge/field $X_{ab}$ & edge/field $X_{ij}$ \\
        \hline
    \end{tabu}
    \caption{Specular duality mapping.}
    \label{tb:specDuality}
\end{table}

The specular duality exchanges $B$ zig-zag paths with $G$ face boundaries and vice versa, while keeping the number of nodes and edges in the tiling.
From Pick's theorem, $G = 2\, \mathrm{Area}(\Delta) = B + 2 I - 2$, it is easy to show that
\begin{align}
    g(\Sigma) = g(\bbT^2) - \frac{B - G}{2} = I ~,
\end{align}
\emph{i.e.} the new consistent tiling is embedded in a Riemann surface of genus equal to the number of internal points $I$ in the toric diagram $\Delta$ of the original model.
In this paper we restrict to models with reflexive toric diagrams, with $g(\bbT^2) = g(\Sigma) = 1$. This set is closed under specular duality.
See \cite{Hanany:2012vc} for the full web of duals for quiver gauge theories describing (pseudo-)del Pezzo geometries. 

%% file: Sections/Deformations.tex

\section{Zig-zag deformations of toric (pseudo) del Pezzo theories}
\label{sec:ToricDefs}

The focus of this work is to classify and study a special class of superpotential deformations which relate worldvolume theories of D3-branes probing local toric (pseudo) del Pezzo geometries \cite{Feng:2002fv}, which have reflexive toric diagrams \cite{Hanany:2012hi}.
To do so, we perform a series of operations on the brane tiling, which encodes the effect of relevant (or marginal, in one case) superpotential deformations.
The deformation violates the toric condition, breaking the $U(1)^3$ mesonic symmetry to a $U(1)^2$ subgroup.
The extra term $\delta W$ in the deformed superpotential
\begin{align}
   W_\text{def} = W + \delta W ~
\end{align}
have UV superconformal R-charge $R_{\mathrm{sc}}\left[ \delta W \right] \le 2$ so that the deformation is relevant or marginal. We are mostly interested in special deformations which have an extra emergent $U(1)$ symmetry in the infrared, restoring the full toric $U(1)^3$ symmetry.

We present a general framework for deformations of brane tiling models by an operator $\calO_{\eta}$ fully defined by a zig-zag path $\eta$. We call $\calO_{\eta}$ a \emph{zig-zag operator} and the superpotential deformation a \emph{zig-zag deformation}. 
Firstly, we will work through some examples of relevant deformations of this type by manipulating the gauge theory, integrating out any fields which become massive after the deformation and finding field redefinitions which lead to a new toric superpotential.
We will also give an interpretation of the zig-zag deformation in terms of brane tilings, extending \cite{Bianchi:2014qma}, as well as analyze it from the perspective of the chiral ring and the moduli space of vacua.
Finally, we will present the main argument, which holds for all $\calO_{\eta}$ of zig-zag paths $\eta$ of length 4.
This is verified for all reflexive geometries. The full details can be found in \cref{sec:RGflows}. 

\subsection{Zig-zag operator}

The chiral ring of mesonic operators, and the mesonic branch of the moduli space of vacua which is obtained by replacing chiral operators by their VEVs, are important for finding the operators that trigger deformations that lead to new toric models. We reviewed in the previous section the notions of geometric branch of the moduli space, which is isomorphic to the geometry $\calY$ probed by a regular brane, and of $\calN=2$ Coulomb branches emanating from loci of non-isolated singularities of $\calY$, which are probed by $\calN=2$ fractional branes.

\begin{figure}[!ht]
   \centering
   \includegraphics[width=0.6\textwidth]{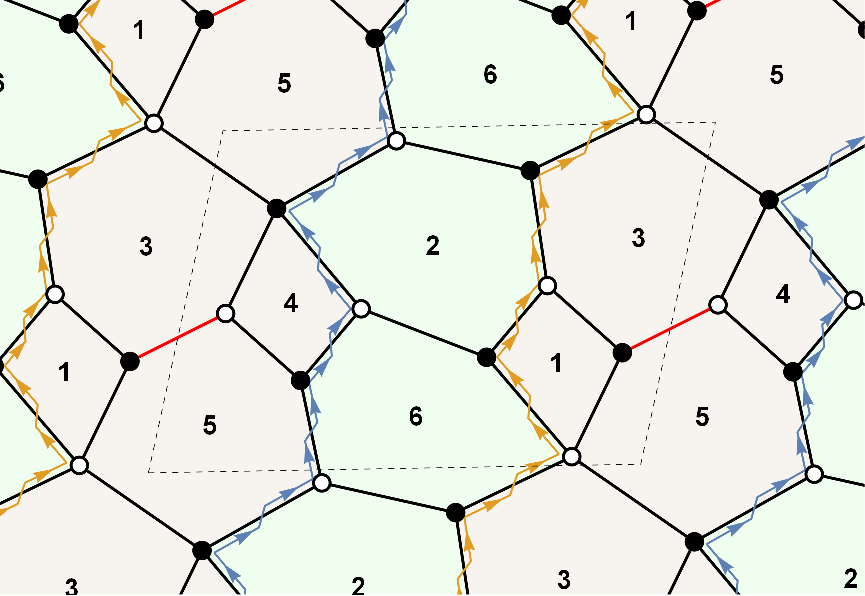}
   \caption{Tiling of $\text{PdP}_{3c}$ phase B, focusing on the two parallel zig-zag paths and $\calN=2$ fractional  branes strips associated to the most relevant (lowest $R_{\mathrm{sc}}$) zigzag deformation. The red edge, $X_{53}^{2}$, indicates the F-term equation that relates the two non-zero winding mesons of the strip in the chiral ring, i.e. $X_{31} X_{15} X_{53}^{1} \simeq X_{34} X_{45} X_{53}^{1}$.}
   \label{fig:PdP3cBFracBands}
\end{figure}

In the brane tiling, $\calN=2$ fractional branes are associated to strips of faces in the tiling bounded by zig-zag paths with the same homology class \cite{Franco:2005zu}.
If $\calY$ has a non-isolated $A_{k}$ singularity ($k \ge 1$), there is a subset of zig-zags $\set{\eta_0, \eta_1, \dots, \eta_k}$ in the same homology class, which divide the brane tiling into $k+1$ strips, corresponding to $k+1$ $\calN=2$ fractional branes that a regular brane can split into.
For each strip, we have a mesonic operator in the chiral ring, which is a loop in the tiling with opposite winding numbers to the zig-zag paths at the boundary of the strip. In all the tilings we consider (and presumably in general) it is always possible to find a place along the strip where it is exactly one face wide. Taking that face as a start and end point of a path winding along the strip, and using that homotopic open paths are F-term equivalent \cite{Hanany:2006nm}, it follows that for each strip there is exactly one meson generator in the chiral ring with mesonic charges equal and opposite to the winding numbers of the boundary zig-zags.
Thus, for each $A_{k}$ singularity, there are $k+1$ inequivalent mesons $\set{M_0, M_1, \dots, M_k}$ in the chiral ring, since by the zig-zag path construction there are no F-terms connecting them.
This is exemplified in \cref{fig:PdP3cBFracBands}.

For each zig-zag path separating two strips, say $\eta$, there are two inequivalent mesonic generators $\calO_{\eta}^{L} \not\simeq \calO_{\eta}^{R}$ in the chiral ring which have the same charges: 
\begin{align}
   \calO_{\eta}^{L} = X_{e_{i_1}} \cdots X_{e_{i_r}} ~, \qquad\quad \calO_{\eta}^{R} = X_{e_{j_1}} \cdots X_{e_{j_s}}~,
\end{align}
where $c^L = e_{i_1} \cdots e_{i_r}$ and $c^R = e_{j_1} \cdots e_{j_s}$ correspond to the 1-cycles in the tiling running through all the edges on the immediate left and right sides of the oriented zig-zag path $\eta$.
The mesonic operators $\calO_{\eta_{j}}^{L}$ and $\calO_{\eta_{j}}^{R}$ take equal VEV in the geometric branch $\calM^\text{geom}$ (we write $\calO_{\eta_{j}}^{L} \sim \calO_{\eta_{j}}^{R}$), but different VEVs in an $\calN=2$ Coulomb branch $\calM^{\calN=2}$ (hence $\calO_{\eta_{j}}^{L} \not\simeq \calO_{\eta_{j}}^{R}$).
Additionally, for a set of zig-zag paths in the same homology class $\set{\eta_0, \eta_1, \dots, \eta_k}$, ordered by adjacency in the brane tiling, we have $\calO_{\eta_{j}}^{R} \simeq \calO_{\eta_{j+1}}^{L} ~(\simeq M_{j+1} )$.
These two chiral mesons can be forced to become equivalent in the chiral ring by turning on a superpotential deformation 
\begin{align}
   \delta W_\eta = \mu \Big( \calO_{\eta}^{L}  - \calO_{\eta}^{R} \Big) \equiv \mu \,\calO_{\eta}~.
   \label{eq:deltaWzigzag}
\end{align}
We remark that the \emph{zig-zag operator} $\calO_{\eta}$ can be written as
\begin{align}
   \calO_{\eta} \equiv
   \prod_{k=n}^{1} \frac{\partial^2 W}{\partial \eta_{2k} \partial \eta_{2k-1}} 
   - (-1)^n \prod_{k=n}^{1} \frac{\partial^2 W}{\partial \eta_{2k+1} \partial \eta_{2k}}
   ~,
   \label{eq:zig-zag-operator}
\end{align}
where $\eta_i = \eta_{(i\;\mathrm{mod}\,2n)}$ are the $2n$ chiral fields that represent the zig-zag closed path.
In this definition we assume that $\eta_1$ is an edge that goes from a black to white node along $\eta$.
The operation $\frac{\partial}{\partial \eta_{i}}$ is a \emph{cyclic derivative} with respect to paths in the quiver, defined as
\begin{align}
\begin{aligned}
   \frac{\partial}{\partial X_{e_i}} \Tr(X_{e_1} \dots X_{e_n}) &= X_{e_{i+1}} \dots X_{e_n} X_{e_1} \dots X_{e_{i-1}} ~,\\
   \frac{\partial}{\partial X_{e_i}} X_{e_1} \dots X_{e_n} &= X_{e_{i+1}} \dots X_{e_{i-1}} \qquad\text{if}\quad i=1,n ~, \\
\end{aligned}
\end{align}
otherwise undefined.
The $(-1)^n$ factor on \cref{eq:zig-zag-operator} takes into account that the right-side second derivative acts on black nodes in the tiling.
Note that the chiral mesons defining $\calO_{\eta}$ are oriented in the opposite direction to $\eta$.

In a theory with four supercharges, a chiral operator $\calO$ which acquires a VEV in the full supersymmetric moduli space cannot be nilpotent in the chiral ring, because VEVs of chiral operators factorize.
For a theory with a $U(1)_R$ symmetry and a superpotential $W_\text{def} \equiv W + \delta W \equiv W + \mu\, \calO$, where $\fracpd{\mu} W=0$, one can show that 
\begin{align}
   \angl*{\fracpd{\mu} W_\text{def}} = \angl*{\calO}= 0 ~,
\end{align}
where $\angl*{\cdot}$ is the expectation value in a supersymmetric vacuum.
In the chiral ring of the deformed theory the operator $\calO$ must be nilpotent, i.e. $\calO^n \simeq 0$ for some nilpotency index $n$.
This argument was made for SCFTs with marginal deformations in \cite{Komargodski:2020ved}, but it extends to any theory with a $U(1)_R$ preserving deformation, marginal or not.

In our case, we have that $\angl*{\calO_{\eta}^{L}} = \angl*{\calO_{\eta}^{R}}$ after the deformation, some $\calN=2$ fractional brane moduli are lifted, and the order of non-isolated $A_k$ singularity decreases.
For the toric theories that we study in this paper, the deformed F-terms force $\calO_{\eta}^{L} \simeq \calO_{\eta}^{R}$ in the chiral ring, which means that the nilpotency index is 1.

\subsection{Mesonic moduli from chiral rings}
 
In this section we consider the mesonic branch of the moduli space $\calM^\text{mes}$, hence we take abelian gauge group $U(1)^G$ with vanishing FI parameters, and we focus in particular on its geometric component $\calM^\text{geom}$, which describes the Calabi-Yau cone $\calY$. Since we are interested in deformations that violate toricity, we will not use the fast-forward algorithm or plethystics \cite{Benvenuti:2006qr,Forcella:2008eh}, and instead derive the algebraic description explicitly.
For this description it is convenient to study the parent space obtained by relaxing the D-term constraints, \emph{i.e.} the master space $\calF^\flat$ of the abelian theory \cite{Hanany:2010zz, Forcella:2008bb,Forcella:2008eh,Forcella:2008ng}.
Geometrically, this is a non-compact $CY_{G+2}$ cone, which consists of 3-dimensional resolved Calabi-Yau cones $\tilde\calY_\xi$ fibered over a ($G-1$)-dimensional base space parameterized by FI parameters $\xi$ (with $\sum_{i=1}^{G} \xi_i = 0$).
The fiber at $\xi = 0$ is the singular cone $\calY$.
Algebraically, the master space is described by the zero locus of a set of homogeneous polynomials in an ambient affine space. For $N=1$, \cref{eq:master-space-nonabelian} is simply
\begin{align}
   \calF^\flat = \Spec \Big( k[X_{e_1}, \ldots, X_{e_E}] / \angl{\partial_X W} \Big) ~,
\end{align}
where $k = \bbC$ or $k = \bbC[\mu^{\pm}]$, for a deformation parameter $\mu$.

To obtain the mesonic moduli we still need to quotient by $\calG = U(1)^G$.
We will do so by constructing the \emph{chiral ring} of the abelian theory, i.e. the quotient $k[M_{c_1}, \ldots, M_{c_s}]/I_{\text{chiral}}$ of the polynomial ring of a finite set of mesonic generators $\set{M_c}$, modulo the chiral ideal $I_{\text{chiral}}$.
The ideal $I_{\text{chiral}}$ captures F-term equations in the $\calG_\bbC$-invariant sector of the master space coordinate ring $k[X_{e_1}, \ldots, X_{e_E}]/\angl{\partial_X W}$.
The chiral ring can be obtain via
\begin{align}
   I_\text{chiral} = \ker\Big( \Phi_W : k[M_{c_1}, \ldots, M_{c_s}] \to 
   k[X_{e_1}, \ldots, X_{e_E}] / \angl{\partial_X W} \Big) ~,
   \label{eq:IchiralKernel}
\end{align}
where the homomorphism $\Phi_W$ is trivially constructed by assigning a mesonic generator $M_c$ (no relations) to a cycle $c = e_1 e_2 \cdots e_n$ in the quiver
\begin{align}
   \Phi_W(M_c) = X_{e_1} X_{e_2} \dots X_{e_n} ~,
\end{align}
If $W = 0$, the image $\im(\Phi_W)$ is just the ring of $\calG_\bbC$-invariants $k[X_{e_1}, \ldots, X_{e_E}]^{\calG_\bbC}$ and $\ker(\Phi_W) = I_{\text{chiral}}$ consists of relations due to compositions of cycles, which follow purely from gauge invariance.
In the general case, $\Phi_W$ is trivially defined by the inclusion $k[X_{e_1}, \ldots, X_{e_E}]^{\calG_\bbC} \hookrightarrow k[X_{e_1}, \ldots, X_{e_E}]$, mapping generators to quiver cycles that descend to equivalence classes in $\im(\Phi_W) \subsetneq k[X_{e_1}, \ldots, X_{e_E}] / \angl{\partial_X W}$.
The kernel%
\footnote{This can be easily obtain via computational algebraic geometry software, e.g. Macaulay2 \cite{M2}.}
of $\Phi_W$ is an ideal $I_\text{chiral}$ that contains all the relations between cycles in the abelian quiver up to F-terms.
Alternatively, this can be computed via the elimination of chiral superfield variables $\set{X_e}$ in an ideal composed of the F-term relations and the map $\Phi_W$ that encodes the complexified gauge group orbits, sitting in the larger base ring $k[M_{c_1}, \ldots, M_{c_s}, X_{e_1}, \ldots, X_{e_E}]$, i.e.
\begin{align}
   I_\text{chiral} = \angl{\partial_{X} W, M_{c_1} - \Phi_W(M_{c_1}), \ldots, M_{c_s} - \Phi_W(M_{c_s})} \cap k[M_{c_1}, \ldots, M_{c_s}] ~.
   \label{eq:IchiralElim}
\end{align}

This construction is particularly useful if the toric $CY_3$ cone $\calY$ has lines of non-isolated $A_{k-1}$ singularities. 
As we reviewed, in this case the moduli space includes additional $\calN=2$ Coulomb branches, which account for the regular D3-brane splitting into $\calN=2$ fractional D3-branes which separately probe the locus of non-isolated singularities.
For this reason, the variety associated to the chiral ring $k[M_i] / I_\text{chiral}$ may be the union of several irreducible components, including the three-fold $\calM^\text{geom} = \calY$ and several other varieties.
These additional components can be detected via the \emph{primary decomposition}
\begin{align}
   I_\text{chiral} = I_\text{geom} \cap J_1 \cap \dots \cap J_\ell ~,
\end{align}
where the top non-trivial component $I_\text{geom}$ is denoted as the \emph{geometric ideal}.
Generally,%
\footnote{One brane tiling, $\text{PdP}_5$ phase B \cite{Hanany:2012hi}, is found to have an additional branch, parametrized by chiral mesons with opposite winding in the tiling. We conjecture that this is related to the fact that this tiling contains two oppositely oriented strips, one of which contains the other.}
each component $J_i$ is associated to a single non-isolated $A_{k-1}$ singularity, which is freely generated by $k$ chiral mesons with no relations. These give rise to affine varieties $\bbC^k$, which intersect $\calY$ at 1-dimensional singular loci.
We can identify the geometric branch with the quotient ring associated to the 3-dimensional non-trivial component of the primary decomposition
\begin{align}
   \calM^\text{geom} = \Spec \Big( k[M_{c_1}, \ldots, M_{c_s}] / I_\text{geom} \Big) ~.
\end{align}
This algorithm is a distilled version of the affine GIT quotient \cite{Martelli:2008cm,Closset:2012ep,Luty:1995sd}, adapted to quiver gauge theories.
We crucially used the fact that for a quiver with $\calG = U(1)^G$ the typical unruliness of the chiral ring is kept in check by the commutativity and finiteness of the mesonic generators.

\subsection{RG flows between toric del Pezzos}
\label{sec:deformation-delPezzo}

The superpotential deformed by relevant terms violates the toric condition and breaks the mesonic and $R$-symmetries down to a $U(1)^2$ subgroup.
We will be mostly interested in RG flows with toric endpoints, namely with an emergent $U(1)^3$ symmetry in the IR (we will discuss an IR endpoint which is a marginal deformation of a toric model in \cref{sec:deformations-marginal}).
By construction, the deformation does not change the number of non-anomalous and anomalous baryonic symmetries (of the $SU(N)^G$ theory).
If the IR endpoint of the RG flow is toric, its toric diagram must have the same number of internal points $I$ and external points $E$ as the toric diagram of the undeformed UV theory, since the rank of the anomalous baryonic symmetry is $2I$ and the rank of the non-anomalous baryonic symmetry is $E-3$ for toric models \cite{Butti:2007jv,Butti:2006au,Butti:2005vn,Forcella:2008eh}.
In geometric terms, the deformation does not change the degree of the del Pezzo surface.

We first integrate out any massive bifundamentals appearing in  superpotential terms of the form $\mu X_{ab} X_{ba}$, by imposing the F-term equations
\begin{align}
   \fracpd[W_\text{def}]{ (X_{ab}, X_{ba}) } = 0 ~,
   \label{eq:integrate-out-fterms}
\end{align}
which modifies the superpotential as follows:
\begin{align}
   \mu X_{ab} X_{ba} + X_{ab} g_{ba}(X) - f_{ab}(X) X_{ba} + \,\cdots \,\mapsto\, \frac{1}{\mu} f_{ab}(X) g_{ba}(X) + \,\cdots
   \label{eq:integrating-out}
\end{align}

The resulting superpotential does not usually make manifest the toric symmetry, but a particular set of field redefinitions may restore the toric condition.
For the case of gauge theories resulting from reflexive polytopes, we were able to restore the desired form by field redefinitions of degree up to 2,
\begin{align}
   X_{ij}^{k} = \sum_{m} \alpha_{ij}^{m} \tilde{X}_{ij}^{m} +  \sum_{l,m,n} \beta_{ilj}^{knm} \tilde{X}_{il}^{m} \tilde{X}_{lj}^{n} ~,
   \label{eq:field_redef}
\end{align}
for some coefficients $\alpha_{ij}^{m}$, $\beta_{ilj}^{knm}$ such that $\alpha_{ij}^{k} \ne 0$.
The coefficients in the non-trivial field redefinitions are proportional to $1/\mu$, and the Jacobian of this change of variables obeys (up to an overall sign)
\begin{align}
   \det\left(\fracpd[X]{\tilde{X}}\right) = \mu^{(E_\mathrm{i} - E_\mathrm{f} - V_\mathrm{f})/2} ~,
\end{align}
where $E_\mathrm{i}$ is the number of bifundamentals of the initial model and $V_\mathrm{f}, E_\mathrm{f}$ denote respectively the number of superpotential terms and bifundamentals of the final model after integrating out massive degrees of freedom.
The resulting low-energy superpotential can be written as 
\begin{align}
   W_\text{low}(X) = \frac{1}{\mu}\, W'(\tilde{X}) ~,
   \label{eq:Wfinal}
\end{align}
with $W'$ toric.
The remaining power of $\mu$ can be cancelled by a complexified $U(1)_R$ transformation $\tilde X_e \mapsto \mu^{R[\tilde X_e]/2} \, \tilde X_e$  on all bifundamentals,  which cancels the $\mu^{-V_\mathrm{f}/2}$ factor in the Jacobian.
Note that the power of $\mu$ in the combined Jacobian is given by $(E_\mathrm{i} - E_\mathrm{f})/2$.

\subsubsection{\texorpdfstring{$\text{PdP}_{3c}$}{PdP3c} to \texorpdfstring{$\text{PdP}_{3b}$}{PdP3b}}

Take for example one of the toric phases of the \emph{Pseudo del Pezzo 3c} model (phase B in \cite{Hanany:2012hi}), with superpotential
\begin{align}
   \begin{aligned}
      W_{\text{PdP}_{3c}}^{(B)} &= X_{12} X_{23} X_{31} + X_{25} X_{56} X_{62} + X_{26} X_{64} X_{42} + X_{34} X_{45} X_{53}^2 \\ 
      & \qquad + X_{15} X_{53}^1 X_{36} X_{61} - X_{12} X_{26} X_{61} - X_{15} X_{53}^2 X_{31} - X_{23} X_{36} X_{62} \\ 
      & \qquad - X_{45} X_{56} X_{64} - X_{25} X_{53}^1 X_{34} X_{42} ~,
   \end{aligned}
   \label{eq:WPdP3cB}
\end{align}
and with tiling, quiver and toric diagram in \cref{fig:PdP3cB-til-quiver-td}.
It is useful to list the generators of the mesonic branch by enumerating all the chiral mesons%
\footnote{For $N=1$, this corresponds to all indecomposable paths. For $N>1$ these are single trace operators.} in the quiver, along with their mesonic $U(1)^3$ charges, and organize them by their GSLM decomposition using \cref{eq:bifund_from_pm}.
This leads to \cref{tb:generatorsPdP3cB}.
\begin{table}[!ht]
   \centering
   \tabulinesep=1mm
   \begin{tabu}{@{\hskip 0.1cm}l|l|c|c@{\hskip 0.1cm}}
       Gen. ($p_\alpha$)  & Generator ($X_e$) & $U(1)^2$ & $U(1)_{R}$ \\
       \hline\hline
       $p_2^2 p_3 g s$ & \parbox[c]{9.2cm}{$X_{26} X_{62} \sim X_{31} X_{15} X_{53}^{1} \simeq X_{34} X_{45} X_{53}^{1}$} & $(0,\text{-}1)$ & $1.577...$ \\
       \hline
       $p_3^2 p_4^2 f s$ & \parbox[c]{9.2cm}{$X_{15} X_{56} X_{61} \sim X_{23} X_{34} X_{42}$} & $(1,0)$ & $1.690...$ \\
       \hline
       $p_1 p_2 p_3 p_4 f g s$ & \parbox[c]{9.2cm}{$X_{12} X_{23} X_{31} \simeq X_{25} X_{56} X_{62} \simeq  X_{25} X_{53}^1 X_{34} X_{42} \simeq$ \\ $X_{34} X_{45} X_{53}^2 \simeq X_{12} X_{26} X_{61} \simeq X_{15} X_{53}^1 X_{36} X_{61} \simeq$ \\ $ X_{15} X_{53}^2 X_{31} \simeq X_{23} X_{36} X_{62} \simeq X_{45} X_{56} X_{64} \simeq X_{26} X_{64} X_{42}$} & $(0,0)$ & $2$ \\
       \hline
       $p_1^2 p_2^2 f g^2 s$ & \parbox[c]{9.2cm}{$X_{36} X_{64} X_{45} X_{53}^1 \simeq X_{36} X_{62} X_{25} X_{53}^1 \simeq X_{31} X_{12} X_{25} X_{53}^1$} & $(\text{-}1,0)$ & $2.309...$ \\
       \hline
       $p_1^2 p_3 p_4^2 f^2 g s$ & \parbox[c]{9.2cm}{$X_{23} X_{36} X_{64} X_{42} \simeq X_{25} X_{56} X_{64} X_{42} \simeq X_{25} X_{53}^2 X_{34} X_{42} \simeq$ \\ $X_{15} X_{53}^2 X_{36} X_{61} \simeq X_{12} X_{23} X_{36} X_{61} \simeq X_{12} X_{25} X_{56} X_{61}$} & $(0,1)$ & $2.422...$ \\
       \hline
       $p_1^3 p_2 p_4 f^2 g^2 s$ & \parbox[c]{9.2cm}{$X_{36} X_{64} X_{45} X_{53}^2 \simeq X_{25} X_{53}^2 X_{36} X_{62} \simeq X_{12} X_{25} X_{53}^2 X_{31} \simeq
       X_{25} X_{53}^1 X_{36} X_{64} X_{42} \simeq X_{12} X_{25} X_{53}^1 X_{36} X_{61}$} & $(\text{-}1,1)$ & $2.732...$ \\
       \hline
       $p_1^4 p_4^2 f^3 g^2 s$ & \parbox[c]{9.2cm}{$X_{25} X_{53}^2 X_{36} X_{64} X_{42} \simeq X_{12} X_{25} X_{53}^2 X_{36} X_{61}$} & $(\text{-}1,2)$ & $3.154...$
   \end{tabu}
   \caption{Table of generators of the mesonic moduli space for $\text{PdP}_{3c}$ phase B, where $f=\prod_{i=1}^{2}f_i$, $g=\prod_{i=1}^{2}g_i$ and $s=\prod_{i=1}^{7}s_i$ (perfect matchings on \cref{fig:PdP3cB-til-quiver-td}). The superconformal R-charge $U(1)_{R}$ can be obtained via a-maximization \cite{Intriligator:2003jj,Butti:2005ps,Butti:2005vn}.}
   \label{tb:generatorsPdP3cB}
\end{table}
Many chiral mesons are F-term equivalent, so we can pick one representative of each equivalence class and define the map
 $\Phi_W$ as 
\begin{align}
   \begin{aligned}
      &A_1 :~ X_{26} X_{62} &&\qquad B_2 :~ X_{31} X_{15} X_{53}^{1} &&\qquad B_8 :~ X_{15} X_{56} X_{61} \\
      &B_5 :~ X_{23} X_{34} X_{42} &&\qquad B_1 :~ X_{45} X_{56} X_{64} &&\qquad C_1 :~ X_{36} X_{64} X_{45} X_{53}^1 \\
      &C_4 :~ X_{25} X_{56} X_{64} X_{42} &&\qquad C_2 :~ X_{36} X_{64} X_{45} X_{53}^2 &&\qquad D_3 :~ X_{25} X_{53}^2 X_{36} X_{64} X_{42}
   \end{aligned}
   \label{eq:PdP3c-chiral-gens}
\end{align}

The affine cone over $\text{PdP}_{3c}$ contains two non-isolated $A_1$ singularities.
The corresponding zig-zag deformations are triggered by differences of relevant chiral mesons in the first two rows of \cref{tb:generatorsPdP3cB}, which can be further corroborated by the primary decomposition of the chiral ideal $I_\text{chiral} = I_\text{geom} \cap J_1 \cap J_2$,
\begin{align}
   \begin{aligned}
      I_\text{geom} &= \langle B_5-B_8, A_1-B_2, C_2^2-C_1 D_3, C_2 C_4-B_1 D_3, B_1 C_2-B_2 D_3, C_4^2-B_8 D_3, \\
      &~\qquad C_1 C_4-B_2 D_3, B_1 C_4-B_8 C_2, B_8 C_1-B_2 C_4, B_1 C_1-B_2 C_2,B_1^2-B_2 C_4 \rangle \\
      J_1 &= \langle D_3, C_2, C_4, C_1, B_1, B_5, B_8\rangle \\
      J_2 &= \langle D_3, C_2, C_4, C_1, B_1, A_1, B_2 \rangle~.
   \end{aligned}
   \label{eq:PdP3cB-chiral-pd}
\end{align}
As expected, the two components associated to the ideals $J_1$ and $J_2$ are parameterized by the chiral mesons that define the $\calN = 2$ fractional brane strips, which intersect the geometric branch at $B_5 = B_8$ and $A_1 = B_2$, respectively.
\begin{figure}[!ht]
   \centering
   \includegraphics[width={0.3\textwidth}]{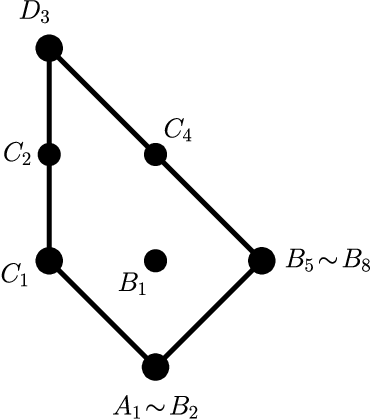}
   \caption{Lattice of the $U(1)^2$ mesonic flavour charges of the generators of $\text{PdP}_{3c}$ phase B (see \cref{tb:generatorsPdP3cB}).}
   \label{fig:PdP3cB-gen-lattice}
\end{figure}

Of the two relevant deformations only the most relevant (smaller $R_\mathrm{sc}$) leads to a new toric model in the IR.%
\footnote{The deformation by the next relevant zig-zag operator $\calO_\eta = X_{15} X_{56} X_{61} - X_{23} X_{34} X_{42}$ does not lead to a toric IR fixed point. Nevertheless it fits into a more general picture that relates zig-zag deformations of two toric models, see \cref{sec:main-argument}.}
A quick inspection of the tiling in \cref{fig:PdP3cBFracBands} or \cref{tb:generatorsPdP3cB} reveals that we have two zig-zag deformations
\begin{align}
   \begin{aligned}
      \calO_{\eta_5} &= X_{26} X_{62} - X_{31} X_{15} X_{53}^{1} \\
      \calO_{\eta_6} &= X_{34} X_{45} X_{53}^{1} - X_{26} X_{62}~,
   \end{aligned}
   \label{eq:PdP3cB-ZZ-deformations}
\end{align}
which are equivalent due to chiral relations, $\calO_{\eta_5} + \calO_{\eta_6} \simeq 0$.
Therefore, we can just pick the RG flow triggered by the superpotential deformation
\begin{align}
   \delta W &= \mu \calO_{\eta_5} = \mu \left(X_{26} X_{62} - X_{31} X_{15} X_{53}^{1}\right) ~.
   \label{eq:PdP3cBdeltaW}
\end{align}
Adding this term modifies the F-term equations, forcing the relation $A_1 \simeq B_2$ in the chiral ring.
The deformation lifts the component $J_1$ and removes the $A_1$ singularity which is intersected by it.
The chiral ring is now given by the relations $I_{\text{chiral}}' = I_{\text{geom}}' \cap J_2'$, where
\begin{align}
   \begin{aligned}
      J_\text{mes}' &= \langle B_5-B_8,C_2 C_4-B_1 D_3,C_1 C_4-B_2 D_3,C_2^2-C_1 D_3,B_1 C_2-B_2 D_3, \\
      &~\qquad B_1 C_1-B_2 C_2, B_1^2 - B_2 C_4, C_4^2 - B_8 D_3 + \mu B_1 C_4, \\
      &~\qquad B_8 C_2 - B_1 C_4 -\mu B_2 C_4, B_2 C_4 - B_8 C_1 + \mu B_1 B_2 \rangle \\
      J_2' &= \langle D_3, C_4, C_2, C_1, B_2, B_1 \rangle~.
   \end{aligned}
\end{align}
There is a slight abuse of notation here.
While the relations hold for the choice of representatives in \cref{eq:PdP3c-chiral-gens}, there is a splitting in the F-term equivalence classes and the generators of the chiral ring are now given by
\begin{align}
   \begin{aligned}
      B_2 &:~ X_{26} X_{62}\simeq X_{53}^1 X_{34} X_{45}\simeq X_{15} X_{53}^1 X_{31} \\
      B_1 &:~ X_{56} X_{64} X_{45}\simeq X_{53}^2 X_{34} X_{45}\simeq X_{26} X_{64} X_{42}\simeq X_{25} X_{56} X_{62}\simeq \\
      &\qquad \simeq X_{15} X_{53}^2 X_{31}\simeq X_{25} X_{53}^1 X_{34} X_{42} \\
      C_4 &:~ X_{25} X_{56} X_{64} X_{42}\simeq X_{25} X_{53}^2 X_{34} X_{42} \\
      B_1 + \mu B_2 &:~ X_{23} X_{36} X_{62}\simeq X_{12} X_{26} X_{61}\simeq X_{12} X_{23} X_{31}\simeq X_{15} X_{53}^1 X_{36} X_{61} \\
      C_4 + \mu B_1 &:~ X_{23} X_{36} X_{64} X_{42}\simeq X_{15} X_{53}^2 X_{36} X_{61}\simeq X_{12} X_{25} X_{56} X_{61} \\
      C_4 +2\mu B_1 + \mu^2 B_2 &:~ X_{12} X_{23} X_{36} X_{61}
   \end{aligned}
\end{align}
and 
\begin{align}
   \begin{aligned}
      C_1 &:~ X_{53}^1 X_{36} X_{64} X_{45}\simeq X_{25} X_{53}^1 X_{36} X_{62}\simeq X_{12} X_{25} X_{53}^1 X_{31} \\
      C_2 &:~ X_{53}^2 X_{36} X_{64} X_{45}\simeq X_{25} X_{53}^2 X_{36} X_{62}\simeq X_{12} X_{25} X_{53}^2 X_{31}\simeq \\
      &\qquad\simeq X_{25} X_{53}^1 X_{36} X_{64} X_{42} \\
      D_3 &:~ X_{25} X_{53}^2 X_{36} X_{64} X_{42} \\
      C_2 + \mu C_1 &:~ X_{12} X_{25} X_{53}^1 X_{36} X_{61} \\
      D_3 + \mu C_2 &:~ X_{12} X_{25} X_{53}^2 X_{36} X_{61}~.
   \end{aligned}
\end{align}
The generators $B_5$ and $B_8$ remain represented by the same quiver cycles in \eqref{eq:PdP3c-chiral-gens}.
Note that, from the point of view of the $\text{PdP}_{3c}$ theory, the mixing of mesonic generators upon deformation occurs in the direction of the winding number $(1,0)$ of the zig-zag $\eta_5$. This matches the mesonic charges of the spurionic parameter $\mu$ in the superpotential deformation $\mu (A_1 - B_2)$, which determine the pattern of global symmetry breaking.%
\footnote{Indeed $\mu$ has $U(1)^3$ charges $(1,0,0)$, where the first two charges are mesonic and the last is the $R$-charge under the $U(1)_R$ symmetry that assigns charge $2$ to all mesonic generators. The $R$-charge is forgotten when the lattice of generators is projected from $\bbZ^3$ to $\bbZ^2$ as in \cref{fig:PdP3cB-gen-lattice}.} 

If we integrate out the massive fields $X_{26}$ and $X_{62}$, by imposing their F-term equations
\begin{align}
   X_{26} = \frac{1}{\mu} \left( X_{23} X_{36} - X_{25} X_{56} \right)
   ~,\qquad
   X_{62} = \frac{1}{\mu} \left( X_{61} X_{12} - X_{64} X_{42} \right)
   ~,
   \label{eq:PdP3c-def-mass-fterms}
\end{align}
the result is not explicitly toric, but we can make it so (up to an overall $1/\mu$) by the field redefinitions
\begin{align}
   \begin{aligned}
      &\quad X_{31} \mapsto -\frac{1}{\mu} X_{31} + \frac{1}{\mu} X_{36} X_{61}
      \qquad X_{53}^1 \mapsto \frac{1}{\mu} X_{53}^1 - \frac{1}{\mu} X_{53}^2 \\ 
      &\quad X_{45} \mapsto \frac{1}{\mu} X_{45} - \frac{1}{\mu} X_{42} X_{25}
   \end{aligned}
   \label{eq:PdP3c-def-field-refef}
\end{align}
as described in \eqref{eq:field_redef}.
Using the methods described previously, we can identify the final result as toric phase B  of the cone over the \emph{Pseudo del Pezzo 3b} \cite{Hanany:2012hi}, with superpotential 
\begin{align}
   \begin{aligned}
      W_{\text{PdP}_{3b}}^{(B)} &= X_{15} X_{53}^1 X_{31} + X_{34} X_{45} X_{53}^2 + X_{12} X_{25} X_{56} X_{61} + X_{23} X_{36} X_{64} X_{42} \\ 
      & \qquad - X_{12} X_{23} X_{31} - X_{45} X_{56} X_{64} - X_{15} X_{53}^2 X_{36} X_{61} - X_{25} X_{53}^1 X_{34} X_{42}~.
   \end{aligned}
   \label{eq:WPdP3bB-fromWPdP3cB}
\end{align}
We can use the fast-forward method and zig-zag paths to read off the changes in the toric diagram.
The removal/reduction of the order of the non-isolated singularity is reflected in the fact that the number of points in the side of the toric diagram orthogonal to the zig-zag's external $(p,q)$-leg is reduced by one.
Consequently, the number of extremal points in the toric diagram increases by one.

\begin{figure}[!ht]
   \centering
   \hspace*{\fill}
   \begin{subfigure}{0.28\textwidth}
      \includegraphics[width=\textwidth]{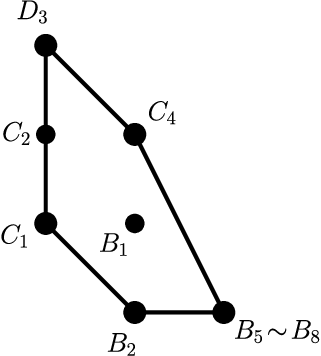}
      \caption{}
      \label{fig:PdP3bB-gen-lattice}
   \end{subfigure}
   \hspace*{\fill}
   \begin{subfigure}{0.28\textwidth}
      \includegraphics[width=\textwidth]{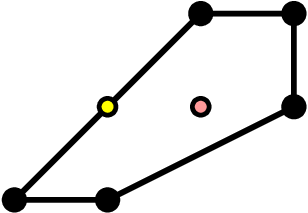}\vspace*{2em}
      \caption{}
      \label{fig:PdP3bB-td-polar-lattice}
   \end{subfigure}
   \hspace*{\fill}
   \caption{(a) Lattice of $U(1)^2$ mesonic flavour charges of the generators and (b) toric diagram  of $\text{PdP}_{3b}$.}
   \label{fig:PdP3bB-lattices}
\end{figure}
The top coherent primary component $I_\text{mes}'$ does not manifest a $U(1)^3$ toric symmetry like in \eqref{eq:PdP3cB-chiral-pd} but, similarly to above, we can find a redefinition that makes the toric symmetry manifest.
We can use \eqref{eq:PdP3c-def-field-refef} to construct the ring isomorphism
\begin{align}
   \begin{aligned}
      &B_1 \mapsto\frac{B_1 -C_4}{\mu} \qquad\qquad & B_2 \mapsto\frac{B_2 -2 B_1 + C_4}{\mu^2} \\ 
      &C_2 \mapsto\frac{C_2 -D_3}{\mu} \qquad\qquad & C_1 \mapsto\frac{C_1 -2 C_2 + D_3}{\mu^2}
   \end{aligned} ~
\end{align}
which is defined for $\mu\neq 0$. Then the relations among mesonic moduli take the toric form
\begin{align}
\hspace{-3pt}  
\begin{aligned}
      I_{\text{geom}}' &= \langle B_8-B_5, B_1^2-B_5 C_2, B_1 C_4-B_5 D_3, B_1 B_2-B_5 C_1,B_2 C_4-B_5 C_2, \\
      &~\quad C_1 C_4- B_1 C_2,C_2 C_4-B_1 D_3,B_2 D_3-B_1 C_2,B_2 C_2-B_1 C_1,C_2^2-C_1 D_3 \rangle ~.
   \end{aligned}
\end{align}
From the superpotential \eqref{eq:WPdP3bB-fromWPdP3cB}, we can find the geometric component of the moduli space of the complex cone over $\text{PdP}_{3b}$, which is isomorphic to $I_{\text{geom}}'$ (as ring quotients).
We can assign $U(1)^2$ charges to each generator consistently with the toric relations in $I_{\text{geom}}'$, obtaining the lattice diagram in \cref{fig:PdP3bB-gen-lattice}.
Note that for geometries described by reflexive polytopes the lattice of generators is polar dual to the toric diagram.

\subsubsection*{Zig-zag move in the brane tiling}

It is also possible to obtain the final brane tiling by a graph deformation of the initial brane tiling.
This builds on \cite{Bianchi:2014qma}, where it was understood that a specific degenerate move of the vertices and edges of the tiling along a zig-zag path described the effect of a mass deformation of Klebanov-Witten type \cite{Klebanov:1998hh}.
The definition of that move required all the vertices on the zig-zag path associated to the deformation to be trivalent.
In order to apply the same 
move to the more general deformations that we consider in this paper, we need to resolve the vertices on the zig-zag path so that they become trivalent, which makes the tiling locally similar to that of $\bbC^2/\bbZ_k \times \bbC$, in a neighbourhood of the zig-zag path.
This is possible by \emph{integrating in}  fields \cite{Intriligator:1994uk}, \emph{i.e.} introducing pairs of bifundamental fields with a superpotential mass term, which upon imposing their F-term equations (or ``integrating out'' as in \eqref{eq:integrating-out}) leads back to the original model. 
In the brane tiling, this corresponds to replacing a white/black node with a total of $k+l>3$ incident edges by two white/black nodes with $k+1$ and $l+1$ incident edges, connected by a 2-valent black/white node in between.
We can always arrange to have $k=2$ (or $l=2$) for the vertices that remain on the zig-zag path after this process, after which the brane tiling move of \cite{Bianchi:2014qma} can be applied.

\begin{figure}[!ht]
   \centering
   \includegraphics[width={0.5\textwidth}, trim={2.2cm 0.3cm 2.2cm 0.3cm}, clip]{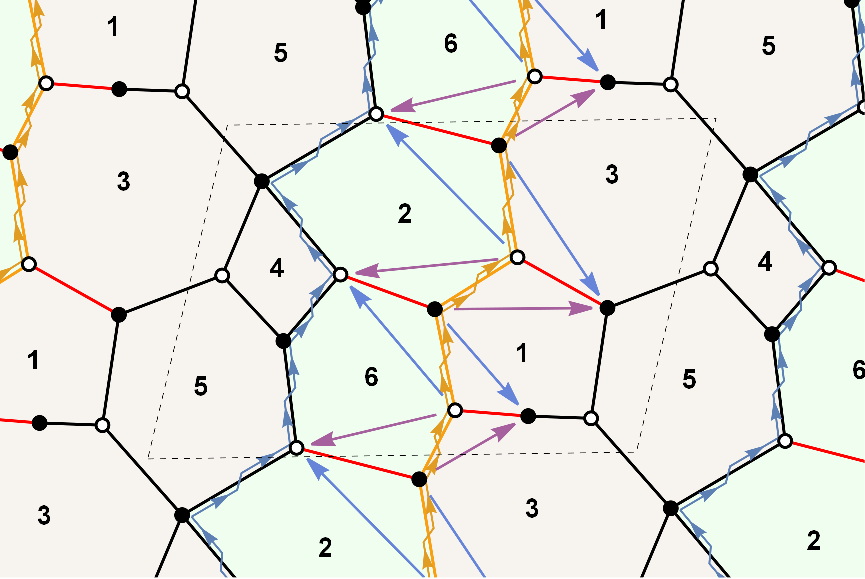}
   \caption{Tiling of $\text{PdP}_{3c}$ phase B with integrated-in node along the deformation zig-zag (yellow), highlighting the two possible (purple/blue) zig-zag moves that realize the deformation.}
   \label{fig:PdP3cB-integrated-in}
\end{figure}

For example, in the case of $\text{PdP}_{3c}$ phase B in \cref{fig:PdP3cBFracBands} and the zig-zag path associated to the deformation \eqref{eq:PdP3cBdeltaW}, we only need to resolve one of the nodes in the zig-zag path, which has the following local effect on the superpotential
\begin{align}
   X_{15} X_{53}^1 X_{36} X_{61} \,\mapsto\, - X_{13}' X_{31}' + X_{15} X_{53}^1 X_{31}' + X_{13}' X_{36} X_{61} ~~,
\end{align}
where primed fields $X_{13}'$, $X_{31}'$ have been integrated in.
In the new tiling (\cref{fig:PdP3cB-integrated-in}), we see that this introduces a new  meson $X_{13}' X_{31}$ in the ``$\calN=2$ fractional brane strip'', which is made of massive chirals and is F-term equivalent to $X_{31} X_{15} X_{53}^{1} \simeq X_{34} X_{45} X_{53}^{1}$ (see \cref{tb:generatorsPdP3cB-intIn}).
\begin{table}[H]
   \centering
   \tabulinesep=1mm
   \begin{tabu}{@{\hskip 0.1cm}l|l|c|c@{\hskip 0.1cm}}
       Gen. ($p_\alpha$)  & Generator ($X_e$) & $U(1)^2$ & $U(1)_{R}$ \\
       \hline\hline
       $p_2^2 p_3 g s$ & \parbox[c]{9.2cm}{$X_{26} X_{62} \sim X_{13}' X_{31} \simeq X_{31} X_{15} X_{53}^{1} \simeq X_{34} X_{45} X_{53}^{1}$} & $(1,0)$ & $1.577...$ 
   \end{tabu}
   \caption{Partial table of generators of the mesonic moduli space for the integrated in $\text{PdP}_{3c}$ phase B model.}
   \label{tb:generatorsPdP3cB-intIn}
\end{table}
This allows us to rewrite the deformation \eqref{eq:PdP3cBdeltaW} as the superpotential mass deformation
\begin{align}
   \delta W' &= \mu \left(X_{26} X_{62} - X_{13}' X_{31}\right) ~.
   \label{eq:PdP3cBdeltaW-intIn}
\end{align}
We can follow the same procedure as before and restore the $U(1)^3$ toric symmetry by the field redefinitions
\begin{align}
   \begin{aligned}
      &\quad X_{45} \mapsto \frac{1}{\mu} X_{45} - \frac{1}{\mu} X_{42} X_{25} ~,
      \qquad X_{53}^1 \mapsto \frac{1}{\mu} X_{53}^1 - \frac{1}{\mu} X_{53}^2
   \end{aligned} ~,
\end{align}
which leads to the same superpotential as \eqref{eq:WPdP3bB-fromWPdP3cB}, with the relabelling $X_{31}' \leftrightarrow X_{31}$.

The result of the deformation and field redefinitions can be visualized in the brane tiling as the folding of the edges involved in the deformation onto the zig-zag path $\eta = X_{12} X_{23} X_{36} X_{61}$.
To perform the move that leads to the endpoint tiling we select alternating edges of a zig-zag path (after integrating in if necessary to ensure that all vertices are trivalent).
Then, every pair of edges directly connected to the zig-zag edge on each side are folded onto it by identifying nodes of the same color, and consequently edges, as indicated by the arrows in figure \ref{fig:PdP3cB-integrated-in}. We call this operation on the brane tiling a \emph{zig-zag move}.
\begin{figure}[H]
   \centering
   \begin{subfigure}[c]{0.39\columnwidth}
      \stackinset{r}{-35pt}{b}{-17pt}{%
         \includegraphics[width={0.333\textwidth}, clip]{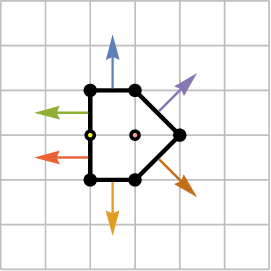}
      }{%
         \includegraphics[width={\textwidth}, trim={2.2cm 0.3cm 2.2cm 0.3cm}, clip]{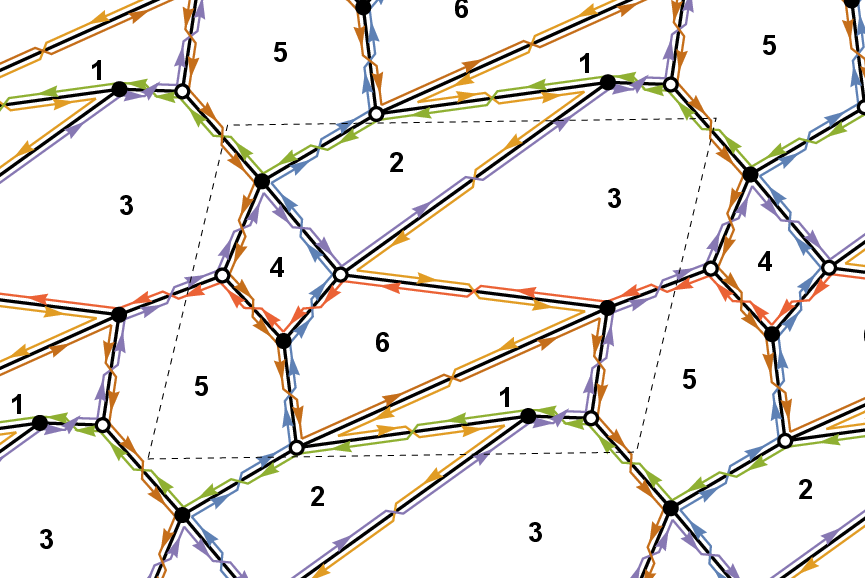}
      }
      \caption{}
      \label{fig:PdP3cB-to-PdP3bB-flow-move1}
   \end{subfigure}%
   \hspace*{\fill}%
   \begin{subfigure}[c]{0.39\columnwidth}
      \stackinset{r}{-35pt}{b}{-17pt}{%
         \includegraphics[width={0.333\textwidth}, clip]{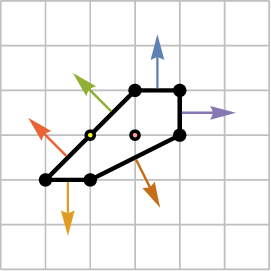}
      }{%
         \includegraphics[width={\textwidth}, trim={2.2cm 0.3cm 2.2cm 0.3cm}, clip]{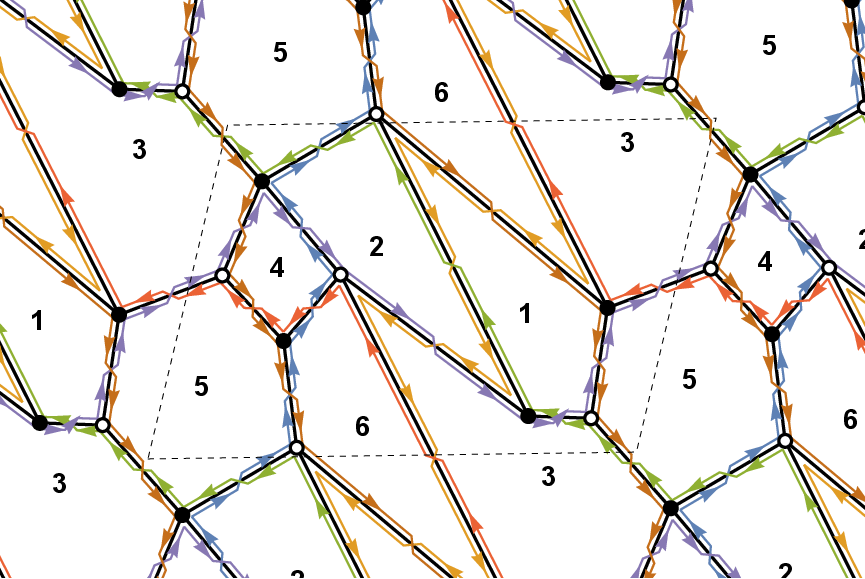}
      }
      \caption{}
      \label{fig:PdP3cB-to-PdP3bB-flow-move2}
   \end{subfigure}%
   \hspace*{35pt}%
   \caption{Tilings and toric diagrams of $\text{PdP}_{3b}$ phase B, obtained by applying to the tiling of figure \ref{fig:PdP3cB-integrated-in} the zig-zag move depicted by the purple arrows (a) and blue arrows (b), keeping the original fundamental domain fixed.}
   \label{fig:PdP3cB-to-PdP3bB-flow}
\end{figure}
There are two ways to do this, which lead to equivalent results which differ by an $SL(2,\bbZ)$ transformation, see \cref{fig:PdP3cB-integrated-in} and \cref{fig:PdP3cB-to-PdP3bB-flow}.
The zig-zag path reverses its direction (\cref{fig:PdP3cB-to-PdP3bB-flow}) because all the edges that enter the mass deformation \eqref{eq:PdP3cBdeltaW-intIn} in the brane tiling are integrated out \cite{Bianchi:2014qma}.
A key observation is the fact that the zig-zag path itself is unchanged as a cycle in the quiver, but it reverses its winding on $\bbT^2$.%
\footnote{This works for length 4 zig-zags and is a pattern throughout this work.}
Additionally, all zig-zags parallel to the one triggering the deformation also remain unaffected.

\subsubsection{Higher order non-isolated singularities}
\label{sec:higherNonIsolated}

Up to this point, we established a solid example of a toric-to-toric flow triggered by a zig-zag deformation of the form \eqref{eq:deltaWzigzag}, with mesonic operators $M_0$, $M_1$, given by non-homotopic paths on the immediate sides of a single zig-zag path in the brane tiling.
For geometries with non-isolated singularities of higher order $k$, we can generalize to
\begin{align}
   \delta W = \mu \sum_{i \in I} \,\calO_{\eta_i} \qquad\text{s.t.}\qquad \sum_{i=0}^k \calO_{\eta_i} \simeq 0 ~.
\end{align}
For $A_1$ singularities, the two possible choices for a zig-zag deformation only differ by a sign.
The deformation removes the non-isolated singularity and lifts the associated $\calN=2$ fractional brane component from the moduli.
For $A_2$ singularities, we may have three orientations to choose based on the zig-zags $\set{\eta_0, \eta_1, \eta_2}$, but all choices still amount to deforming along a single operator $\calO_{\eta_i}$.
However, the same is not true for $k>2$.

Specifically for reflexive toric geometries, there is a single example where an $A_3$ singularity is present: the orbifold $\bbC^3/(\bbZ_4 \times\bbZ_2)$ quotiented with action $(1,0,3)(0,1,1)$.
\begin{figure}[!ht]
   \centering
   \includegraphics[width=0.6\textwidth]{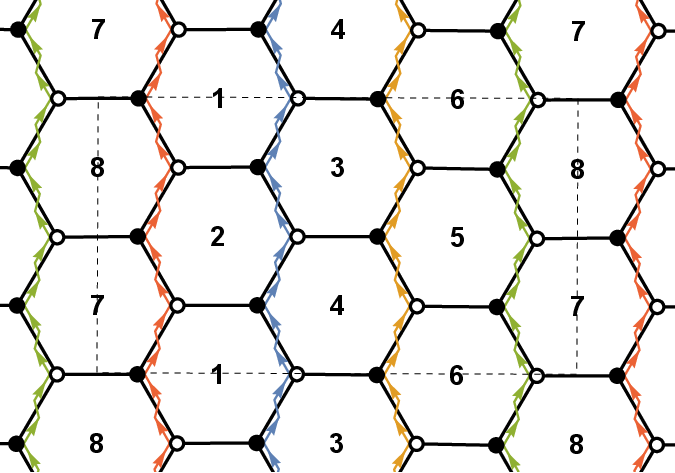}
   \caption{Tiling of $\bbC^3/(\bbZ_4 \times\bbZ_2) ~(1,0,3)(0,1,1)$, with emphasis on the zig-zah paths associated to the non-isolated $A_3$ singularity.}
   \label{fig:C3Z4Z2-til}
\end{figure}
When a regular D3-brane is on the locus of the $A_3$ singularity, it can split into four $\calN=2$ fractional branes which individually probe the non-isolated singularity.
In the worldvolume theory this fractional brane branch is parametrized by the expectation value of four mesonic operators. By the correspondence between zig-zag paths and superpotential deformations, we have the relevant zig-zag operators
\begin{align}
   \begin{aligned}
      \calO_{\eta_5} &= X_{12} X_{21} - X_{34} X_{43} \\
      \calO_{\eta_6} &= X_{34} X_{43} - X_{56} X_{65}
   \end{aligned}
   \quad\qquad
   \begin{aligned}
      \calO_{\eta_7} &= X_{56} X_{65} - X_{78} X_{87} \\
      \calO_{\eta_8} &= X_{78} X_{87} - X_{12} X_{21}
   \end{aligned}
   \quad,
   \label{eq:C3Z4Z2-ZZ-deformations}
\end{align}
with all conformal R-charges equal, $R_\text{sc}\left[\calO_\eta\right] = 4/3$.
Taking any one of these operators as the deformation $\delta W$ will trigger a flow to the toric phase A of $L^{1,3,1}/\bbZ_2 ~(0,1,1,0)$ (detailed in \cref{sec:C3Z4Z2toL131Z2}), which is expected from effects of a single zig-zag deformation on the initial tiling topology.

We can also trigger a \emph{double} zig-zag deformation of $\bbC^3/(\bbZ_4 \times\bbZ_2)$ by reversing two parallel zig-zag paths in the tiling simultaneously. This leads to the different toric phases of $\text{PdP}_{5}$ (flows \ref{sec:C3Z4Z2toPdP5}).
If we reverse two non-adjacent zig-zag paths we obtain phase A of $\text{PdP}_{5}$ \cite{Hanany:2012hi}.
There are two possible combinations that lead to the same operator, up to an overall sign,
\begin{align}
   \begin{aligned}
      \delta W &= \mu \left( \calO_{\eta_5} + \calO_{\eta_7} \right) \\
      &= \mu \left( X_{12} X_{21} - X_{34} X_{43} + X_{56} X_{65} - X_{78} X_{87} \right)   
   \end{aligned}
\end{align}
The resulting superpotential from this deformation, after integrating out massive fields, immediately obeys the toric conditions, up to a $1/\mu$ factor that can be removed by a complexified R-symmetry transformation of the fields.
From the original tiling, it is possible to perform the two zig-zag moves simultaneously as zig-zag paths (and adjacent edges) do not overlap, as described previously in \cref{fig:PdP3cB-integrated-in}.
The simultaneous reflection of the zig-zag paths of the same homology through a double deformation splits an $A_3$-type singularity splits into two of $A_1$-type with associated zig-zags of opposite homology class.
\begin{figure}[!ht]
   \centering
   \includegraphics[width=0.55\textwidth, clip]{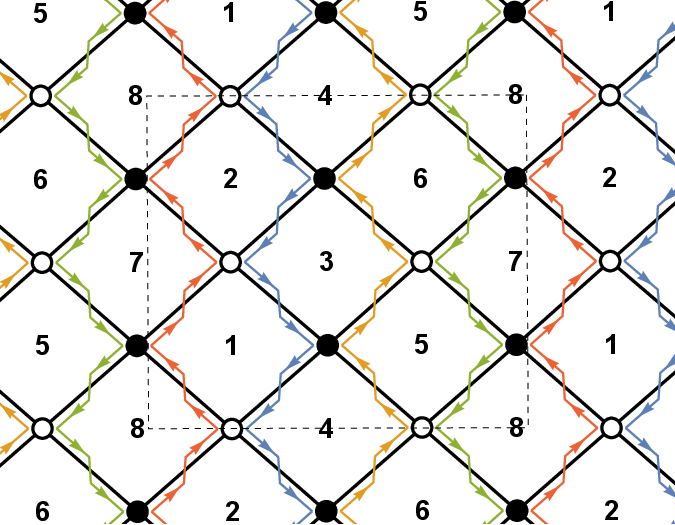}
   \caption{Brane tiling of $\text{PdP}_{5}$ phase A, showcasing the zig-zag paths of the parent model in \cref{fig:C3Z4Z2-til} and the reversed zig-zag paths after the deformation.}
   \label{fig:PdP5-td-fromC3Z4Z2}
\end{figure}

On the other hand, if we reverse two adjacent parallel zig-zag paths we obtain phase B of $\text{PdP}_{5}$.
The four possible combinations result in two possible deformations:
\begin{align}
   \delta W &= \mu \left( \calO_{\eta_5} + \calO_{\eta_6} \right) = \mu \left( X_{12} X_{21} - X_{56} X_{65} \right) 
   \label{eq:phaseB-def-1}  
   \\[0.5ex]
   \delta W &= \mu \left( \calO_{\eta_6} + \calO_{\eta_7} \right) = \mu \left( X_{34} X_{43} - X_{78} X_{87} \right)
   \label{eq:phaseB-def-2}
\end{align}
Contrasting with the previous deformation, $\text{PdP}_{5}$ phase B still contains 2-cycles in the quiver.
After integrating the massive fields in the deformed superpotential we still need to apply a field redefinition to these vector-like pairs in order to restore the $U(1)^3$ toric symmetry. 
For the first deformation in \eqref{eq:phaseB-def-1}, we apply
\begin{align}
   \begin{aligned}
       &\quad X_{34} \mapsto -\frac{1}{\mu} X_{34} + \left(\frac{1}{\mu} - \beta_1 \right) X_{31} X_{14} + \beta_1\, X_{36} X_{64} \\
       &\quad X_{43} \mapsto -\frac{1}{\mu} X_{43} + \left(\frac{1}{\mu} - \beta_1 \right) X_{45} X_{53} + \beta_1\, X_{42} X_{23} \\
       &\quad X_{78} \mapsto \frac{1}{\mu} X_{78} + \left( - \frac{1}{\mu} - \beta_2 \right) X_{72} X_{28} + \beta_2\, X_{75} X_{58} \\
       &\quad X_{87} \mapsto \frac{1}{\mu} X_{87} + \left( - \frac{1}{\mu} - \beta_2 \right) X_{86} X_{67} + \beta_2\, X_{81} X_{17}
   \end{aligned}
   \quad,
\end{align}
where the parameters $\beta_1, \beta_2 \in\bbC$ are free, as they cancel out in the superpotential.
A similar redefinition applies for the second case \eqref{eq:phaseB-def-2}.
The mesonic moduli and $\calN=2$ fractional branes for both deformations are isomorphic to the phase A case.

\begin{figure}[!ht]
   \centering
   \hspace*{\fill}
   \begin{subfigure}[c]{0.46\columnwidth}
      \includegraphics[width=\textwidth, trim={0.9cm 1.1cm 0.9cm 1.1cm}, clip]{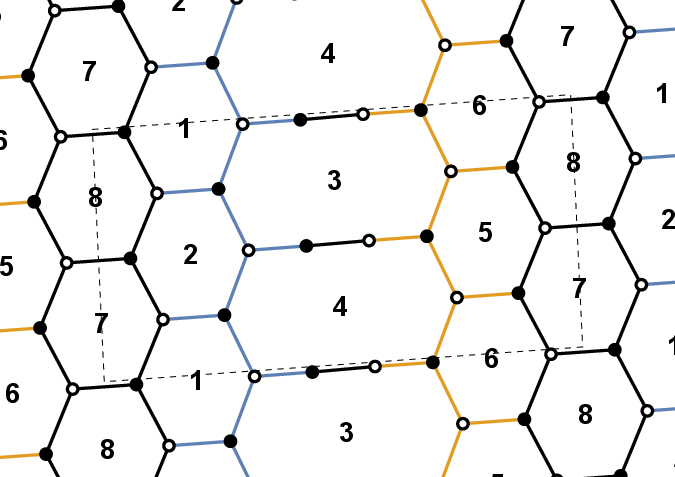}
      \caption{}
      \label{fig:C3Z4Z2-tiling-intInB}
   \end{subfigure}
   \hspace*{\fill}
   \begin{subfigure}[c]{0.46\columnwidth}
      \includegraphics[width=\textwidth, trim={0.9cm 1.1cm 0.9cm 1.1cm}, clip]{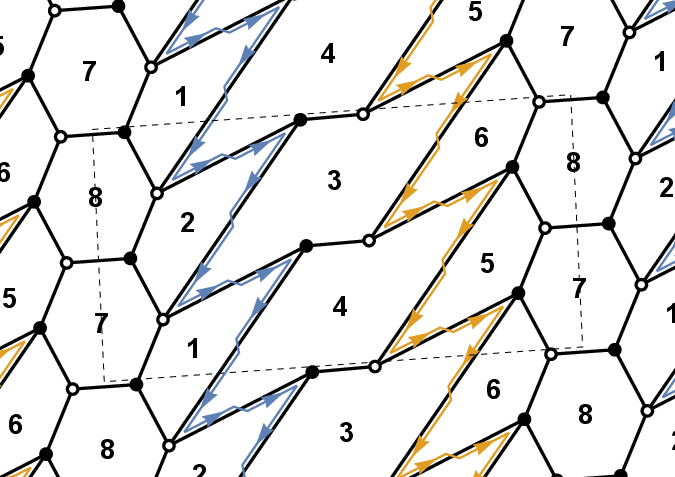}
      \caption{}
      \label{fig:PdP5B-fromC3Z4Z2-tiling}
   \end{subfigure}
   \hspace*{\fill}
   \caption{Zig-zag move to obtain PdP5 Phase B (b) from the integrated-in edges $\bbC^3/(\bbZ_4 \times\bbZ_2)$ model (a).}
   \label{fig:C3Z4Z2_to_PdP5B_flow}
\end{figure}

We note that for double zig-zag deformations that are adjacent in the tiling we need an addendum to our prescription for the brane tiling move:
besides requiring that the nodes on the zig-zag are 3-valent as before, we must also ensure that adjacent edges do not overlap for the multiple zig-zags involved in the deformation. We can do so by integrating in fields to replace an edge in the tiling by three edges connected via bivalent nodes, as exemplified in \cref{fig:C3Z4Z2-tiling-intInB}.
After resolving the overlaps the same zig-zag move as before can be performed.
In particular, the result of the deformation \eqref{eq:phaseB-def-1} and the move described above results in the tiling in \cref{fig:PdP5B-fromC3Z4Z2-tiling}.

\subsection[RG flow to a non-toric geometry: \texorpdfstring{$L^{1,3,1}/\bbZ_2$}{L131/Z2} to marginal deformation of \texorpdfstring{$\text{PdP}_{5}$}{PdP5}]
{RG flow to a non-toric geometry: from \texorpdfstring{$L^{1,3,1}/\bbZ_2$}{L131/Z2} to a marginal deformation of \texorpdfstring{$\text{PdP}_{5}$}{PdP5}}
\label{sec:deformations-marginal}

Another possibility for deformations is a flow triggered by a relevant zig-zag operator, which flows not to a toric fixed point but to an exactly marginal deformation thereof. We discuss this case since it reveals the general theory of deformations interpolating between toric quiver gauge theories.

In the previous section, the geometry of the real cone over $L^{1,3,1}/\bbZ_2 ~(0,1,1,0)$ was reached by triggering a single zig-zag deformation of $\bbC^3/(\bbZ_4 \times\bbZ_2)~(1,0,3)(0,1,1)$.
This deformation lifts one $\calN=2$ Coulomb branch modulus, flowing to a geometry with an $A_2$ singularity.
We have also seen that adding a pair of zig-zag operators to the orbifold superpotential the field theory flows to $\text{PdP}_5$, this time lifting the non-isolated $A_3$ singularity into two $A_1$ singularities.
We would then expect that by turning on a relevant zig-zag operator of $L^{1,3,1}/\bbZ_2$, we would be able to flow to $\text{PdP}_5$.

\begin{figure}[!ht]
   \centering
   \hspace*{\fill}
   \begin{subfigure}[c]{0.45\columnwidth}
      \includegraphics[width=\columnwidth, clip]{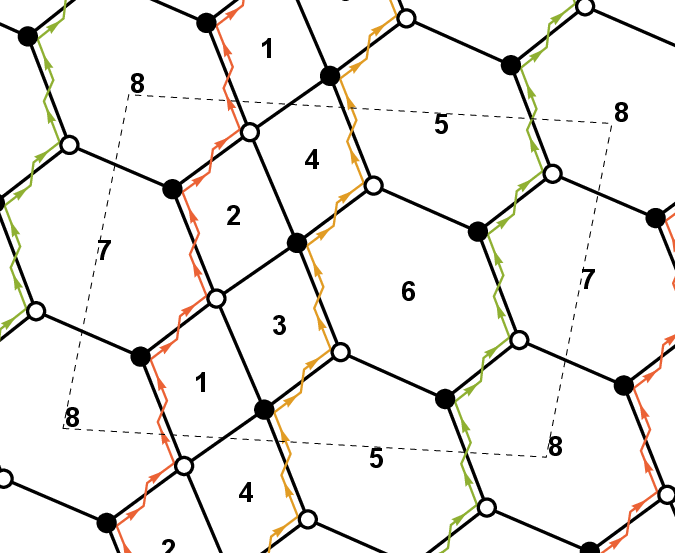}
      \caption{}
      \label{fig:L131Z2A-tiling}
   \end{subfigure}
   \hspace*{\fill}
   \hspace*{\fill}
   \begin{subfigure}[c]{0.45\columnwidth}
      \includegraphics[width=\columnwidth, clip]{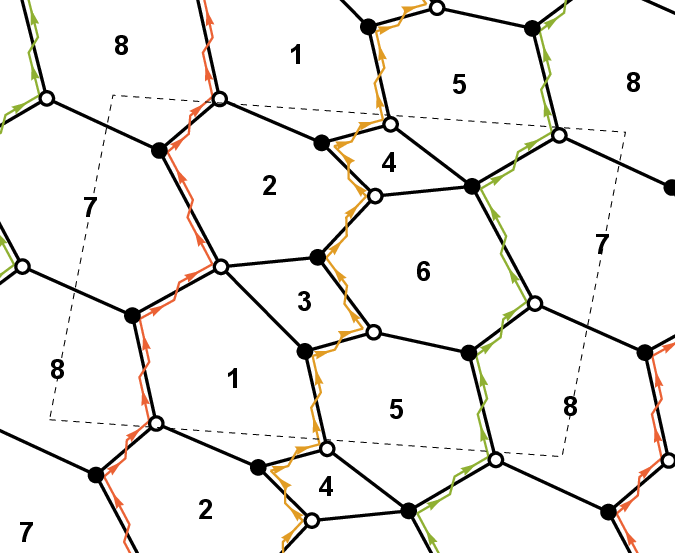}
      \caption{}
      \label{fig:L131Z2B-tiling}
   \end{subfigure}
   \hspace*{\fill}
   \caption{Brane tilings of phase A (a) and phase B (b) of $L^{1,3,1}/\bbZ_2 ~(0,1,1,0)$. The two tilings are Seiberg dual by mutating node $4$.}
\end{figure}
The $L^{1,3,1}/\bbZ_2$ 
model  has 3 relevant zig-zag operators associated to its $A_2$ non-isolated singularity, with $R_\mathrm{sc} = (10 - 2 \sqrt{7}) / 3$.
In the toric phase A in \cref{fig:L131Z2A-tiling}, they take the form
\begin{align}
   \begin{aligned}
      \calO_{\eta_6} &= X_{78} X_{87} - X_{14} X_{42} X_{23} X_{31} \\
      \calO_{\eta_7} &= X_{14} X_{42} X_{23} X_{31} - X_{56} X_{65} \\
      \calO_{\eta_8} &= X_{56} X_{65} - X_{78} X_{87}
   \end{aligned}
   \qquad\qquad
   \begin{aligned}
      \eta_6 &= X_{17} X_{72} X_{28} X_{81} \\
      \eta_7 &= X_{36} X_{64} X_{45} X_{53} \\
      \eta_8 &= X_{58} X_{86} X_{67} X_{75}
   \end{aligned}
   ~.
\end{align}
The obvious first choice is the deformation $\calO_{\eta_8}$, which according to our aforementioned zig-zag move should end up in $\text{PdP}_5$ phase A, in \cref{fig:PdP5-td-fromC3Z4Z2}.
By integrating out the massive fields $\set{X_{56}, X_{65}, X_{78}, X_{87}}$ and redefining $\set{X_{75}, X_{86}} \to \set{\mu X_{75}, \mu X_{86}}$, the resulting superpotential is
\begin{align}
   W_{\text{PdP}_5}^{(A)} + \frac{1}{\mu}\calO'_{\eta_8} ~,
\end{align}
where $\calO'_{\eta_8} = X_{17} X_{72} X_{28} X_{81} - X_{36} X_{64} X_{45} X_{53}$ is the operator generated by the reversed zig-zag $\eta_8$ (same), now obtained from the superpotential $W_{\text{PdP}_5}^{(A)}$.
We conclude that the deformation leads to an exactly marginal deformation of the $\text{PdP}_5$ phase A brane tiling, since all its zig-zag operators $\calO'_{\eta}$ have $R_\text{sc} = 2$ and are easily seen to be exactly marginal (\emph{e.g} by computing the single trace contribution to the superconformal index \cite{Eager:2012hx}). Only in the limit $\mu\to\infty$ we reach a point in the conformal manifold describing a toric SCFT.
Similarly, the relevant deformations by operators $\calO_{\eta_6}$, $\calO_{\eta_7}$ also flow to a marginal deformations $\calO'_{\eta_6}$, $\calO'_{\eta_7}$ of $\text{PdP}_5$ phase B.
Note that $\set{\eta_6, \eta_7, \eta_8}$ are the same paths before and after the deformation, no matter which zig-zag we choose to reverse. 

For $L^{1,3,1}/\bbZ_2 ~(0,1,1,0)$ phase B, the zig-zag operators $\calO_{\eta}$ in \cref{fig:L131Z2B-tiling}
\begin{align}
   \begin{aligned}
      \calO_{\eta_6} = X_{78} X_{87} - X_{12} X_{23} X_{31} \\
      \calO_{\eta_8} = X_{46} X_{65} X_{54} - X_{78} X_{87}
   \end{aligned}
\end{align}
both trigger flows to marginal deformations of $\text{PdP}_5$ phase C.
An interesting case is the deformation associated to the zig-zag $\eta_7$
\begin{align}
   \calO_{\eta_7} = X_{12} X_{23} X_{31} - X_{46} X_{65} X_{54} \qquad \eta_7 = X_{15} X_{53} X_{36} X_{62} X_{24} X_{41} ~,
   \label{eq:L131phaseB-zz-len6}
\end{align}
since we were not able to apply the previous techniques to find a way to classify the endpoint of this flow as a toric fixed point or a marginal deformation thereof.
Note that, contrary to all other deformations in $L^{1,3,1}/\bbZ_2 ~(0,1,1,0)$  phase A (and all other shown above), this zig-zag path $\eta_7$ has length 6 instead of 4 due to the toric/Seiberg duality. We will see this difference play a crucial role in the classification of zig-zag deformations.

\subsection{Zig-zag deformation under Mirror Symmetry}
\label{sec:main-argument}
 
Now that we have considered what happens for all possible zig-zag deformations of reflexive models, we will make a general statement to summarize the results, and we will also present how these are connected to the mirror geometries, more specifically, to the tilings of the mirror curve $\Sigma$ obtain by specular duality \cite{Hanany:2012vc}, making connection with the work of \cite{Franco:2023flw}.

In the previous section and in appendix \ref{sec:RGflows}, we studied all possible deformations by zig-zag operators $\calO_\eta$ of toric quiver gauge theories associated to reflexive toric diagrams, with emphasis on relevant and exactly marginal deformations.
The common threads we found can be summarized as follows:
\begin{itemize}
   \item Whenever the zig-zag $\eta$ generating a relevant $\calO_\eta$ had length 4, we could relate the deformation of the given toric quiver gauge theory $(\mathcal{Q},W)$ with quiver $\mathcal{Q}$ and superpotential $W$, to another toric model $(\mathcal{Q}',W')$ in this class (or an exactly marginal deformation thereof).%
   \footnote{Recall that for the deformation \eqref{eq:L131phaseB-zz-len6} of $L^{1,3,1}/\bbZ_2 ~(0,1,1,0)$ phase B, associated to a zig-zag path of length 6, we could not match the result to a zig-zag marginal deformation of a toric phase for $\text{PdP}_5$, as expected from phase A results, though we were able to match the deformed geometries on the two sides.}
   \item The zig-zag $\eta$ is reversed in the endpoint model $(\mathcal{Q}',W')$, and is described by the same closed path in the quiver. However, the latter is a coincidence of length 4 zig-zags.
   If we consider $\text{PdP}_{3c}/\bbZ_2$ by extending the tiling in \cref{fig:PdP3cB-integrated-in} along the zig-zag, the brane tiling move still reverses the zig-zag, but leads to a different cycle $\eta'$ in $\mathcal{Q}'$.
   \item Deforming a UV toric model $(\mathcal{Q},W)$ by a relevant zig-zag operator $\calO_\eta$ triggers an RG flow that  approaches another toric model $(\mathcal{Q}',W')$ in the IR from an irrelevant/marginal direction $\calO_{\eta'}'$. For example, in the previously discussed deformation of $\text{PdP}_{3c}$ phase B for the choice of zig-zag path $\eta_5 = X_{12} X_{23} X_{36} X_{61} = \eta_5'$, we can find field redefinitions such that
   \begin{align}
      \begin{aligned}
         W_{\text{PdP}_{3c}^{(B)}} + \mu\calO_{\eta_5}
         \xrightarrow[\substack{\quad X_{31} \mapsto -\frac{1}{\mu} X_{31} + X_{36} X_{61} \quad\\%
         \quad X_{23} \mapsto \mu X_{23} \,,~ X_{61} \mapsto \mu X_{61} \quad}]{}
         W_{\text{PdP}_{3b}^{(B)}} + \frac{1}{\mu}\calO'_{\eta_5'}
      \end{aligned}
      \label{eq:PdP3cdef-redef-to}
   \end{align}
   We were able to find additional field redefinitions resulting in the toric fixed point in \eqref{eq:PdP3c-def-field-refef} because the zig-zag operator $\calO'_{\eta_5'} = X_{15} X^{2}_{53} X_{31} - X_{25} X_{56} X_{64} X_{42} \simeq 0$ under the IR F-terms. In contrast, if the reversed zig-zag path is parallel to another zig-zag path, signalling a non-isolated singularity of the IR geometry, as in section \ref{sec:deformations-marginal}, the deformation $\calO'_{\eta'}$ is non-trivial in the chiral ring and cannot be absorbed by a field redefinition. 
\end{itemize}

\begin{figure}[!ht]
   \centering
   \begin{tikzpicture}[thick, scale=1.1, every node/.style={scale=1.1}]
      \tikzstyle{rect}=[rectangle, minimum height=2em, inner sep=3pt, rounded corners=.5em, fill=gray!20]
      \tikzstyle{bigrect}=[rectangle, minimum height=2.8em, inner sep=3pt, rounded corners=.5em, fill=gray!20]
      \tikzstyle{arrow}=[->, shorten >=0.5em, shorten <=0.5em]
      \tikzstyle{twosidearrow}=[<->, shorten >=0.5em, shorten <=0.5em]
      \tikzstyle{arrowlbl}=[midway, font=\scriptsize, fill=white, inner sep=2pt]
      
      \node[rect] (WI) at (0em, 0em) {$(\calQ, W)$};
      \node[bigrect, above right=4.5em and -2.5em of WI] (WIdef) {$\left(\calQ, W + \mu\sum_{i\in I} \mkern-1mu \calO_{\eta_{i}}\right)$};
      \node[rect, below right=4.5em and 1em of WI] (WIspec) {$(\hat{\calQ}, \hat{W})$};

      \node[rect] (WF) at (30em, 0em) {$(\calQ', W')$};
      \node[bigrect, above left=4.5em and -2.5em of WF] (WFdef) {$\left(\calQ', W' + \frac{1}{\mu}\sum_{i\in I} \mkern-1mu \calO_{\eta_{i}'}'\right)$};
      \node[rect, below left=4.5em and 1em of WF] (WFspec) {$(\hat{\calQ'}, \hat{W'})$};

      \draw[arrow] (WI) -- node[arrowlbl] {Zig-zag def.} (WIdef);
      \draw[twosidearrow] (WIdef) -- node[arrowlbl, below] {Field redefinitions} (WFdef);
      \draw[arrow] (WF) -- node[arrowlbl] {Zig-zag def.} (WFdef);
      \draw[twosidearrow] (WI) -- node[arrowlbl] {Specular} (WIspec);
      \draw[twosidearrow] (WIspec) -- node[arrowlbl] {Toric-Seiberg on $U(N)_{i\in I}$} (WFspec);
      \draw[twosidearrow] (WF) -- node[arrowlbl] {Specular} (WFspec) ;
   \end{tikzpicture}
   \caption{Diagram representing the connection between length 4 zig-zag deformations and Seiberg duality on the specular dual models.}
   \label{fig:mirror-argument}
\end{figure}
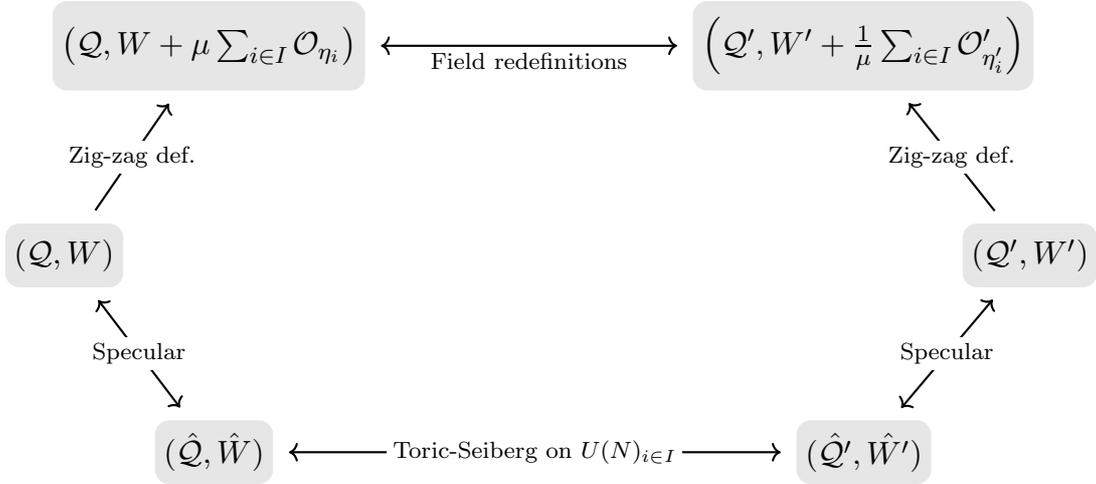

In order to relate to the mirror (or specular dual) geometry, the first key insight is that the zig-zag path $\eta$ that triggers the flow must be of length 4.
Under specular duality, a zig-zag path $\eta_i$ becomes the boundary of a face in the tiling representing the gauge group $U(N)_i$, and vice-versa. 
Thus, from the perspective of the specular dual, we need to consider operations that ``reverse'' the cycle associated to square faces.
This operation is exactly the toric-Seiberg duality on the node $U(N)_i$. Therefore, it is natural to expect that the specular dual of the UV toric fixed point is toric-Seiberg dual to the specular dual to the IR toric fixed point of the zig-zag flow. Indeed this was first proposed in \cite{Franco:2023flw},%
\footnote{This point was also realized independently by JS.} where the quiver obtained by specular duality was dubbed \emph{twin quiver}, and mutations of (generalized) toric polytopes were related to mutations of twin quivers.%
\footnote{%
From the perspective of \cite{Franco:2023flw}, reversing a zig-zag path of length $\ell > 4$ corresponds to a Seiberg duality of $SU(N)$ gauge group with $N_{f} = N \ell/2$ flavours, resulting in a dual gauge group $SU(N(\ell/2-1))$ of higher rank. Finding the twin model of this Seiberg dualized twin, which presumably describes the $\mu\to\infty$ limit of our zig-zag deformation, is an important open problem.}
This is depicted in the bottom half of \cref{fig:mirror-argument}. The main contribution of this paper is to complete \cref{fig:mirror-argument} by adding the top half: in the special case where the two geometries are toric (not generalized toric) and the reversed zig-zag path has length 4, we give a systematic prescription for finding the deformations of the superpotential that relate the two toric models $(\mathcal{Q},W)$ and $(\mathcal{Q}',W')$. We have checked that this expectation is correct for the brane tilings associated to reflexive polygons (and more, see \cite{Sa2023}). We also note that the zig-zag deformation operators $\calO_\eta$ which play a crucial role in our story do not map to gauge invariant mesonic (single trace) operators in the twin (or specular dual) models. This is perhaps unsurprising, since specular duality is not a quantum field theory duality, but rather a duality of graphs, which swaps mesonic symmetries with (hidden or anomalous) baryonic symmetries of twin models \cite{Hanany:2012vc}.%
\footnote{At best, one could relate the mesonic deformation operator $\calO_{\eta}$ to a baryonic operator in the twin model with $SU$ gauge groups (this operator would not be gauge invariant with $U$ gauge groups). However, the two operators in the twin models would have very different dimensions at large $N$.}

The deformation parameter $\mu$ can be viewed as the inhomogeneous coordinate of a base $\bbP^1$, over which we fibre a quiver with deformed superpotential. The two toric models correspond to the two poles. Each zig-zag deformation of a toric model describes how the fibre varies over a patch of $\bbP^1$, which excludes the other pole. On the overlap of the two patches, we can find field redefinitions relating the two deformed quivers and superpotentials: these are the transition functions for the fibre. 

In this paper we considered relevant zig-zag deformations of UV toric models and matched them to trivial/irrelevant/marginal  zig-zag deformations of IR toric models, so there is a clear RG-flow direction. It turns out however that  \cref{fig:mirror-argument} describes more generally 1-parameter families of deformations relating a pair of toric models, with no reference to an RG flow direction. This structure and the underlying geometry will be explored in a companion paper \cite{CCS2023}.

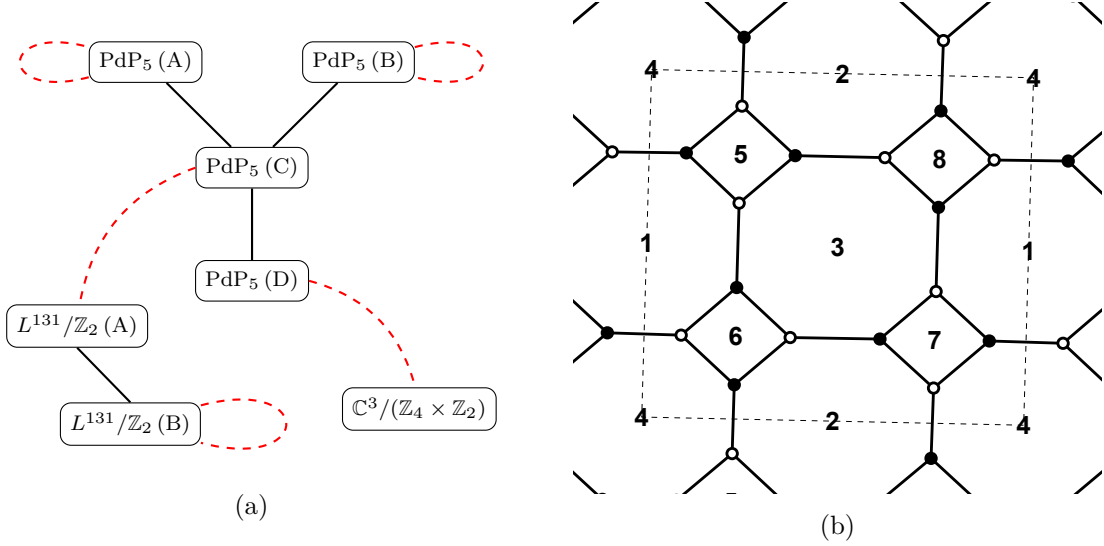
\begin{figure}[!ht]
   \centering
   \begin{subfigure}[c]{0.52\columnwidth}
      \centering
      \begin{tikzpicture}
         \begin{scope}[every node/.style={rectangle,rounded corners=4,draw, font={\scriptsize#1}}]
            \node (PdP5A) at (0.2,0.3) {$\text{PdP}_{5}\,\text{(A)}$};
            \node (PdP5B) at (3.0,0.3) {$\text{PdP}_{5}\,\text{(B)}$} ;
            \node (PdP5C) at (1.6,-1.1) {$\text{PdP}_{5}\,\text{(C)}$};
            \node (PdP5D) at (1.6,-2.6) {$\text{PdP}_{5}\,\text{(D)}$};
            \node (L131Z2A) at (-0.7,-3.2) {$L^{131}/\bbZ_2\,\text{(A)}$};
            \node (L131Z2B) at (0,-4.5) {$L^{131}/\bbZ_2\,\text{(B)}$};
            \node (C3Z4Z2) at (3.8,-4.3) {$\bbC^3/(\bbZ_4\times\bbZ_2)$};
        \end{scope}
        \begin{scope}[
         every node/.style={fill=white,font={\scriptsize#1}},
         every path/.style={draw=black, thick}]
            \path[-] (PdP5A) to (PdP5C);
            \path[-] (PdP5B) to (PdP5C);
            \path[-] (PdP5D) to (PdP5C);
            \path[-] (L131Z2A.south) to (L131Z2B.north);
         \end{scope}
         \begin{scope}[
            every node/.style={fill=white,font={\scriptsize#1}},
            every path/.style={draw=red, thick, dashed}]
               \path[-] (PdP5C.west) to[bend right] (L131Z2A);
               \path[-] (PdP5D.east) to[bend left] (C3Z4Z2);
               \path[-] (PdP5A) to[loop left] (PdP5A);
               \path[-] (PdP5B) to[loop right] (PdP5B);
               \path[-] (L131Z2B) to[loop right] (L131Z2B);
            \end{scope}
      \end{tikzpicture}
      \caption{}
      \label{fig:G8-specular-Seiberg-graph}
   \end{subfigure}
   \hspace*{\fill}
   \begin{subfigure}[c]{0.46\columnwidth}
      \includegraphics[width=\columnwidth, trim=29 29 29 29, clip]{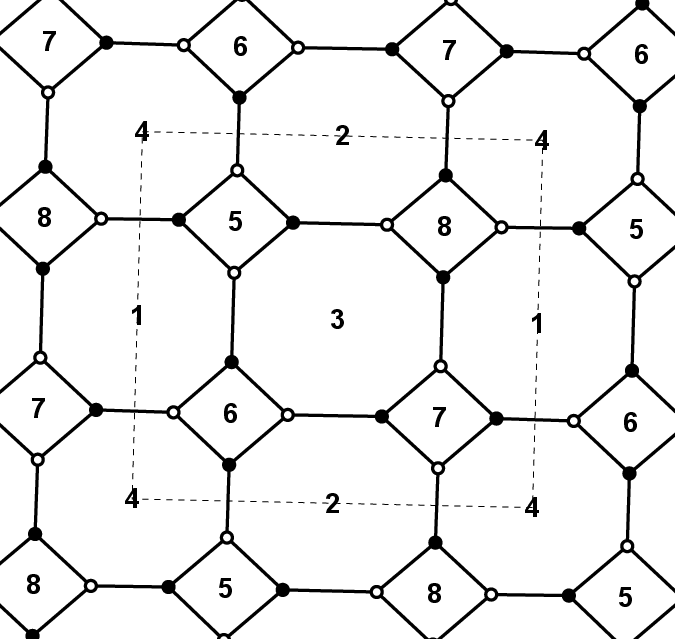}
      \caption{}
      \label{fig:PdP5D-tiling}
   \end{subfigure}
   \caption{(a) Graph with toric-Seiberg dualities (solid black edges ) and specular dualites (dashed red edges) of $G=8$ reflexive models. (b) Brane tiling of $\text{PdP}_{5}$ phase D, specular dual to the model $\bbC^3/(\bbZ_4 \times \bbZ_2) \,(1,0,3)(0,1,1)$.}
\end{figure}

The relation to specular duality and (toric) Seiberg duality which was appreciated in \cite{Franco:2023flw} also explains why triggering different or multiple zig-zag deformations associated the same non-isolated singularity may lead to different models.
Take for example the reflexive model $\bbC^3/(\bbZ_4 \times \bbZ_2) \,(1,0,3)(0,1,1)$, which is specular dual to model $\text{PdP}_{5}$ phase D.
This geometry has a non-isolated $A_3$ singularity, as manifested by the 4 zig-zag paths with the same winding in its brane tiling.
The specular dual of these zig-zags are represented by 4 square polygons symmetrically placed around an octagon in $\text{PdP}_{5}$ phase D, also shown in \cref{fig:PdP5D-tiling}.
The symmetry of $\text{PdP}_{5}$ (D) tiling means that no matter the $\bbC^3/(\bbZ_4\times\bbZ_2)$ zig-zag deformation chosen, the model flows to $L^{131}/\bbZ_2$ phase A, since $\text{PdP}_{5}$ phase D is connected to phase C by a single toric-Seiberg duality, which in turn is specular dual to $L^{131}/\bbZ_2$ phase A (\cref{fig:G8-specular-Seiberg-graph}).
On the other hand, double deformations of $\bbC^3/(\bbZ_4\times\bbZ_2)$ flow to either of the 2 specular self-dual phases of $\text{PdP}_{5}$, depending on the pair of zig-zag operators that trigger the flow: the pairs $\set{\eta_5, \eta_7}$, $\set{\eta_6, \eta_8}$ flow to phase A, while $\set{\eta_5, \eta_6}$, $\set{\eta_6, \eta_7}$, $\set{\eta_7, \eta_8}$, $\set{\eta_8, \eta_5}$ flow to phase B.
By the argument above, the split among the possible pairs occurs because applying toric-Seiberg dualities on opposite squares faces in the octagon in \cref{fig:G8-specular-Seiberg-graph} leads to a different tiling from the brane tiling resulting from dualizing adjacent square faces.

Additionally, the flows from $L^{1,3,1}/\bbZ_2\,(0,1,1,0)$ to marginal deformations of $\text{PdP}_5$ can also be understood using \cref{fig:G8-specular-Seiberg-graph}.
Listing the possible paths from these two geometries
\begin{equation}
   \begin{array}{ccccccc}
      L^{1,3,1}/\bbZ_2\,\text{(A)} & \xrightarrow{~~\text{spec.}\,~} & \text{PdP}_5\,\text{(C)} & \xrightarrow{~~\text{Seib.}\,~} & \text{PdP}_5\,\text{(A)} 
      & \xrightarrow{~~\text{spec.}\,~} & \text{PdP}_5\,\text{(A)} \\
      L^{1,3,1}/\bbZ_2\,\text{(A)} & \xrightarrow{~~\text{spec.}\,~} & \text{PdP}_5\,\text{(C)} & \xrightarrow{~~\text{Seib.}\,~} & \text{PdP}_5\,\text{(B)} 
      & \xrightarrow{~~\text{spec.}\,~} & \text{PdP}_5\,\text{(B)} \\
      L^{1,3,1}/\bbZ_2\,\text{(B)} & \xrightarrow{~~\text{spec.}\,~} & L^{1,3,1}/\bbZ_2\,\text{(B)} & \xrightarrow{~~\text{Seib.}\,~} & L^{1,3,1}/\bbZ_2\,\text{(A)} 
      & \xrightarrow{~~\text{spec.}\,~} & \text{PdP}_5\,\text{(C)} \\
   \end{array}
\end{equation}
we see these correspond exactly to all the possible flows in the previous section (and \cref{sec:L131Z2toPdP5}).

%% file: Sections/Resolutions.tex

\section{Zig-zag deformation and resolutions}
\label{sec:Resolutions}

In this section we study the interplay between the zig-zag deformations discussed previously and crepant resolutions $\tilde{\calY}$ of singular Calabi-Yau cones $\calY$.

\subsection{Minimal GLSM}
\label{sec:KC-minGLSM}

A four-dimensional $\calN=1$ quiver gauge theory with abelian gauge group $\calG$, quiver $\calQ$ and superpotential $W$  is the low energy worldvolume description of a regular D3-brane probing the cone $\calY$.
Resolutions of the singular cone can be viewed as fibres $\tilde{\calY}$ in the master space
\begin{align}
    \calF = \setst{(X_e)_e}{\partial W = 0} ~.
\end{align}
Algebraically, $\calY$ is fully reproduced by the moduli space of the abelian gauge theory with $\calG = U(1)^G$.
Similarly, resolving the cone $\tilde{\calY}$ corresponds to turning on Fayet-Iliopoulos terms in the action.
The FI parameters affect D-term equations, leading to non-zero levels for the moment map $\mu : \calF \to \mathfrak{g}^{*} \cong (\bbC^{\times})^G$  
\begin{align}
    \mu_i(X) = \sum_e d_{ie} \abs{ X_e }^{2} ~,
\end{align}
where $d_{ie} = \delta_{i,s(e)} - \delta_{i,t(e)}$ is the incidence matrix of the quiver $\calQ$.
The K\"{a}hler quotient description of the moduli space $\calM(\calQ, W; \xi)_\text{K}$ is thus given by%
\footnote{For $\xi$ a regular value of $\mu$. More generally, if $\calG$ is non-abelian, one has to quotient by the co-adjoint stabilizer at the level $\xi$, given by $\calG_\xi = \setst{g\in\calG}{\mathrm{Ad}_{g}^{*}(\xi) = \xi}$.}
\begin{align}
    \calF \mkern1mu\big/\mkern-6mu\big/_{\mkern-4mu\xi}\, \calG 
    \equiv \coset{\mu^{-1}(\xi)}{\calG}~.
    \label{eq:KahlerReduction}
\end{align}

For toric quiver gauge theories, we can exploit dimer model technology and perfect matchings. 
We can introduce a GLSM \eqref{eq:bifund_from_pm} with no superpotential, that trivializes the F-term equations $\partial W = 0$.
Since this description is redundant it comes at a cost of extra D-term equations of a spurious $U(1)^{c-G-2}$ gauge symmetry, with charges $Q_F$ defined in \eqref{eq:Q_F}.
These additional gauge symmetries do not have FI parameters as they only serve as connection between the GLSM and the toric variety.
Similarly, in the basis of perfect matchings $\set{p_\alpha}$, we can obtain the charges $Q_D$ from the incidence matrix using \eqref{eq:Q_D}.
We can group all the D-terms as in \cref{tb:fi}.
\begin{table}[!ht]
    \centering
    \tabulinesep=1mm
    \begin{tabu}{@{\hskip 0.1cm}l|ccc|c@{\hskip 0.1cm}}
        & $p_1$ & \dots & $p_c$ & FI \\
        \hline\hline
        $U(1)_{i}^F$ & $(Q_F)_i^1$ & \dots & $(Q_F)_i^c$ & $0$ \\
        $U(1)_{j}^D$ & $(Q_D)_j^1$ & \dots & $(Q_D)_j^c$ & $\xi_j$
    \end{tabu}
    \caption{Description of the D-terms for the GLSM associated to a toric model.}
    \label{tb:fi}
\end{table}
Each perfect matching $p_\alpha$ corresponds to a point in the toric diagram $\Delta$ of the singular $\calY$.
Multiple perfect matchings are associated to non-extremal points in $\Delta$.
It is possible to eliminate some of the $p_\alpha$ in the D-term equations such that exactly one variable remains per point in $\Delta$.
Because all perfect matchings appear in the D-terms as linear combinations of $\abs{p_\alpha}^2$, each choice of a single p.m. per point in the toric diagram determines an \emph{open string K\"{a}hler chamber} in FI parameter space \cite{Closset:2012ep}.
Conversely, a generic choice of FI parameters $\xi$ falls in the interior of a K\"{a}hler chamber, which determines a p.m. variable for each point in the toric diagram.
\begin{table}[H]
    \centering
    \tabulinesep=1mm
    \begin{tabu}{@{\hskip 0.1cm}l|ccc|c@{\hskip 0.1cm}}
        & $p_{x_1}$ & \dots & $p_{x_\ell}$ & FI \\
        \hline\hline
        $U(1)_{a}$ & $Q_a^1$ & \dots & $Q_a^\ell$ & $\zeta_a(\xi)$
    \end{tabu}
    \caption{Minimal GLSM, with one p.m. variable per lattice point in $\Delta$. The resolution parameter $\zeta_a(\xi)$ control the K\"{a}hler volumes of a basis of holomorphic 2-cycles, which depend linearly on the FI parameters in each open string K\"{a}hler chamber.}
    \label{tb:fi_minimal}
\end{table}

In order to visualize the K\"{a}hler chambers of a toric model let us look at the example of the \emph{pseudo del Pezzo 1} ($\calY = \coset{\bbC^3}{\bbZ_4} \,(1,1,2)$) model, with superpotential
\begin{align}
    \begin{aligned}
        W &= X_{13} X_{34}^1 X_{41}^2 + X_{24} X_{41}^1 X_{12}^2 + X_{31} X_{12}^1 X_{23}^2 + X_{42} X_{23}^1 X_{34}^2 \\
        & \qquad - X_{13} X_{34}^2 X_{41}^1 - X_{24} X_{41}^2 X_{12}^1 - X_{31} X_{12}^2 X_{23}^1 - X_{42} X_{23}^2 X_{34}^1~.
    \end{aligned}
    \label{eq:PdP1-potential}
\end{align}
The perfect matchings are 
\begin{align}
    \begin{aligned}
        p_1 &= \set{X_{12}^1,X_{23}^1,X_{34}^1,X_{41}^1} \\
        p_2 &= \set{X_{13},X_{24},X_{31},X_{42}} \\
        p_3 &= \set{X_{12}^2,X_{23}^2,X_{34}^2,X_{41}^2} \\
        f_1 &= \set{X_{12}^1,X_{12}^2,X_{34}^1,X_{34}^2} \\
        f_2 &= \set{X_{23}^1,X_{23}^2,X_{41}^1,X_{41}^2} \\
    \end{aligned}
    \hspace{1em}
    \begin{aligned}
        s_1 &= \set{X_{12}^1,X_{12}^2,X_{13},X_{42}} \\
        s_2 &= \set{X_{13},X_{23}^1,X_{23}^2,X_{24}} \\
        s_3 &= \set{X_{24},X_{31},X_{34}^1,X_{34}^2} \\
        s_4 &= \set{X_{31},X_{41}^1,X_{41}^2,X_{42}} \\
    \end{aligned}
\end{align}
and their associated perfect matching variables are shown in the toric diagram in \cref{tb:PdP1-glsm-td}.
From this we can extract the perfect matching matrix $P$ and rewrite the bifundamentals in terms of the GLSM fields $\set{p_\alpha}$, so that all F-term equations hold. 
From the perfect matching matrix and D-terms of the toric model we obtain (recall \cref{eq:Q_F,eq:Q_D}) a GLSM with charges and FI parameters given in \cref{tb:PdP1-glsm-td}.
\begin{table}[H]
    \centering
    \begin{minipage}[c]{0.67\textwidth}
        \tabulinesep=0.75mm
        \begin{tabu}{@{\hskip 0.1cm}l|ccccccccc|c@{\hskip 0.1cm}}
            & $p_1$ & $p_2$ & $p_3$ & $f_1$ & $f_2$ & $s_1$ & $s_2$ & $s_3$ & $s_4$ & FI \\
            \hline\hline
            $U(1)^F_1$ & $-1$ & $-1$ & $-1$ & $1$ & $0$ & $0$ & $1$ & $0$ & $1$ & $0$ \\
            $U(1)^F_2$ & $0$ & $-1$ & $0$ & $-1$ & $0$ & $1$ & $0$ & $1$ & $0$ & $0$\\
            $U(1)^F_3$ & $-1$ & $0$ & $-1$ & $1$ & $1$ & $0$ & $0$ & $0$ & $0$ & $0$\\
            $U(1)^D_1$ & $-1$ & $-1$ & $-1$ & $1$ & $0$ & $1$ & $1$ & $0$ & $0$ & $\xi_1$ \\
            $U(1)^D_2$ & $0$ & $0$ & $0$ & $0$ & $0$ & $-1$ & $1$ & $0$ & $0$ & $\xi_2$\\
            $U(1)^D_3$ & $0$ & $1$ & $0$ & $1$ & $0$ & $-1$ & $-1$ & $0$ & $0$ & $\xi_3$\\
            $U(1)^D_4$ & $1$ & $0$ & $1$ & $-2$ & $0$ & $1$ & $-1$ & $0$ & $0$ & $\xi_4$
        \end{tabu}
    \end{minipage}
    \begin{minipage}[c]{0.22\textwidth}
        \includegraphics[width=\columnwidth, clip]{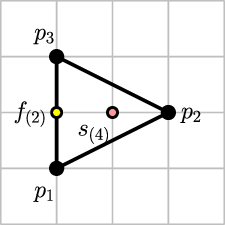}
    \end{minipage}
    \caption{Table encoding D-term equations of the GLSM and toric diagram of $\text{PdP}_1$.}
    \label{tb:PdP1-glsm-td}
\end{table}

\begin{figure}[!ht]
    \centering
    \includegraphics[width=0.78\textwidth, clip]{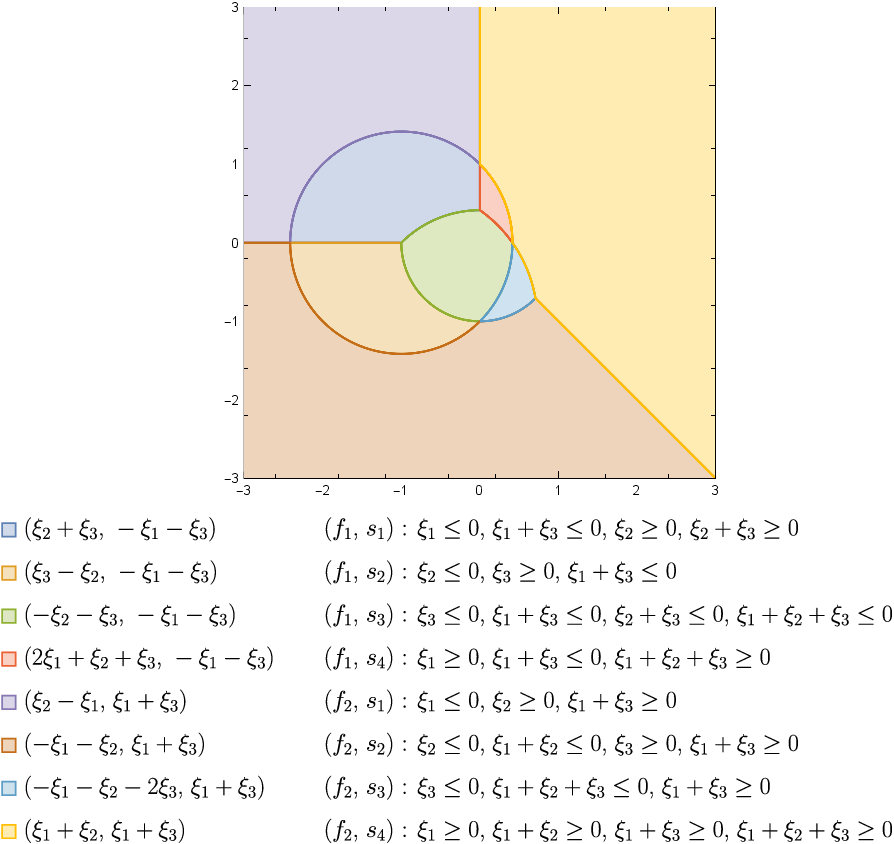}
    \caption{Region plot of open string K\"{a}hler chambers for $\text{PdP}_1$, using stereographic projection of $(\xi_1, \xi_2, \xi_3) $.
    Swatch legend shows the resolution parameters $(\zeta_1, \zeta_2)$ as piecewise linear functions of the FI parameters for each of the choices of $(p_\text{(-1,0)}, p_\text{(0,0)})$.}
    \label{fig:PdP1-KC-stereo}
\end{figure}

The GLSM fields associated to extremal points on the toric diagram are unique by consistency.
We can eliminate those and obtain D-terms with the remaining fields of higher multiplicity.
In the $\text{PdP}_1$ case, we have
\begin{gather}
    \begin{gathered}
        \abs{s_2}^2 - \abs{s_1}^2 = \xi_2 \qquad \abs{s_3}^2 - \abs{s_2}^2 = \xi_3 \qquad \abs{s_4}^2 - \abs{s_3}^2 = \xi_4
    \end{gathered}
    \label{eq:PdP1_s00_pm} \\
    \abs{f_2}^2 - \abs{f_1}^2 = \xi_4 + \xi_2
    \label{eq:PdP1_f-10_pm}
\end{gather}
subject to the condition $\xi_1 + \xi_2 + \xi_3 +\xi_4 = 0$, coming from the decoupled center-of-mass $U(1)$.
From these D-terms, we can choose p.m. variables $(p_\text{(-1,0)}, p_\text{(0,0)})$ for the points $(-1,0), (0,0)$ to obtain the conditions that define the corresponding open string K\"{a}hler chamber.
For example, the choice $p_\text{(-1,0)}= f_1$ requires that $\xi_2 + \xi_4 \ge 0$, because in that case $f_2$ can be solved for in terms of $f_1$, 
while $p_\text{(0,0)}= s_1$ requires $\xi_2 \ge 0 \wedge \xi_2 + \xi_3 \ge 0 \wedge \xi_2 + \xi_3 + \xi_4 \ge 0$.
By repeating this process for all choices of perfect matching variables, we obtain conditions that divide the FI parameter space $\bbR^3$ into 8 polyhedral cones that intersect at the origin $\xi=0$.
This particular example allows us to visualize open string K\"{a}hler chambers using the stereographic projection (\cref{fig:PdP1-KC-stereo}).

Eliminating the redundant D-term equations and GLSM fields using \cref{eq:PdP1_s00_pm,eq:PdP1_f-10_pm} onto the original D-terms in \cref{tb:PdP1-glsm-td} we obtain the minimal GLSM in \cref{tb:PdP1_minimal_GLSM}, by fixing the same $U(1)$ charges for all K\"{a}hler chambers.
\begin{table}[!ht]
    \centering
    \tabulinesep=0.75mm
    \begin{tabu}{@{\hskip 0.1cm}l|ccccc|c@{\hskip 0.1cm}}
        & $p_\text{(-1,-1)}$ & $p_\text{(1,0)}$ & $p_\text{(-1,1)}$ & $p_\text{(-1,0)}$ & $p_\text{(0,0)}$ & FI \\
        \hline\hline
        $U(1)_1$ & $0$ & $1$ & $0$ & $1$ & $-2$ & $\zeta_1(\xi)$ \\
        $U(1)_2$ & $1$ & $0$ & $1$ & $-2$ & $0$ & $\zeta_2(\xi)$ \\
    \end{tabu}
    \caption{Minimal GLSM for $\text{PdP}_1$, with $\zeta_a(\xi)$ given in \cref{fig:PdP1-KC-stereo}.}
    \label{tb:PdP1_minimal_GLSM}
\end{table}

The singularity $\coset{\bbC^3}{\bbZ_4} \,(1,1,2)$ is a fairly simple example that allows visualization of the K\"{a}hler chambers, but it misses the complication of compatibility of the K\"{a}hler chambers with different triangulations of the toric diagram.
We will describe an elegant and more systematic way of obtaining these wedge regions in the FI parameter space, using the fact that K\"{a}hler chambers stem from the quiver $\calQ$ and modules of a path subalgebra of $\bbC\calQ$.

\subsection{K\"{a}hler chambers from $\theta$-stability}

The stability condition for the orbits of the complexified gauge group $\calG_\bbC$ needed to construct a given resolution divides the moduli space of resolutions into chambers as before.
We now construct the possible open string K\"{a}hler chambers from the perspective of quiver representations of a quiver $\calQ$, with path relations encoded in the superpotential $W$.

Through the Kempf–Ness theorem, the K\"{a}hler quotient \eqref{eq:KahlerReduction} is related to the moduli space of quiver representations computed via the GIT quotient,
\begin{align}
    \calM(\calQ,W; \xi)_\text{K} = \calM(\calQ,W; \theta)_\text{GIT} ~, \quad\text{for}\quad \xi = \theta \in \bbZ^{G} ~.
\end{align}
Following results of \cref{sec:quiver-rep-moduli-stability}, closed points of $\calM(\calQ,W; \theta)_\text{GIT}$ are in correspondence with semistable representations of the quiver $\calQ$.
We are interested in the representation space $\Rep(\calQ, \alpha)$ with dimension vector $\alpha = \mathbbl{1}_G=(1,\dots,1)$.
For a given $\theta \in \bbR^G$, a representation $V$ of $\calQ$ with nonzero dimension vector $\alpha$ is called \emph{$\theta$-semistable} if $\theta\cdot\alpha = 0$ and for any proper subrepresentation $W \subset V$, with dimension vector $\beta = \dim W$, we have $\theta\cdot\beta \le 0$.
We say that $V$ is \emph{$\theta$-stable} if under the previous assumptions $\theta\cdot\beta < 0$ for any nontrivial proper subrepresentation $W \subset V$.

A choice of K\"{a}hler chamber $K$ corresponds to a choice of a perfect matching for each point $x$ in the toric diagram $\Delta$, denoted by $K_x$. We can restrict to the exceptional divisor for this resolution by vanishing the corresponding field in the GLSM, thus setting $X_e = 0$ for all $X_e \in K_x$.
We define the subquiver $\calQ_{K_x}$ as the quiver $\calQ$ with edges not in the perfect matching $K_x$, $\setst{e \in \calQ_1}{X_e \notin K_x}$.
A representation $V$ of $\calQ_{K_x}$ with dimension $\alpha$ is also a subrepresentation of the quiver $\calQ$, with
\begin{align}
    X_e = 0 ~\quad \forall\, X_e \in K_x ~.
\end{align}
The intersection of the $\theta$-semistability%
\footnote{Note that saturating the $\theta$-semistability lands us on the boundary between multiple open string K\"{a}hler chambers. Thus, full resolutions are obtain by considering $\theta$-stability.}
conditions for a general module in all the subquiver representation spaces $\Rep(\calQ_{p},\alpha)$, $p \in K$, defines the region of compatibility in the resolutions space, $\xi = \theta$, for the chamber $K$.
More concretely, we can write this as
\begin{align}
    \calR(K) = \bigcap_{x \in \Delta} \calR( \calQ_{K_x} )
    \label{eq:oskcConditions-1}
\end{align}
with
\begin{align}
    \calR(\calQ) = \setst*{ \xi \in\bbR^G }{
    \xi \cdot \dim V \le 0 ,~ \xi\cdot\alpha = 0, ~\forall\, V \in \Rep(\calQ,\alpha) } ~.
    \label{eq:oskcConditions-2}
\end{align}

\begin{figure}[!ht]
    \centering
    \hspace*{\fill}
    \begin{subfigure}[c]{0.24\columnwidth}
       \includegraphics[width=\textwidth, clip]{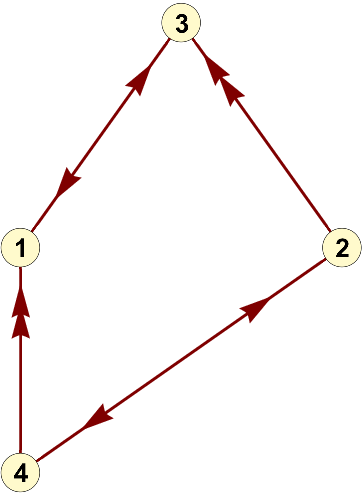}
       \caption{$\calQ_{\set{f_1}}$}
       \label{fig:Q-PdP1-sub-f1}
    \end{subfigure}
    \hspace*{\fill}
    \begin{subfigure}[c]{0.24\columnwidth}
       \includegraphics[width=\textwidth, clip]{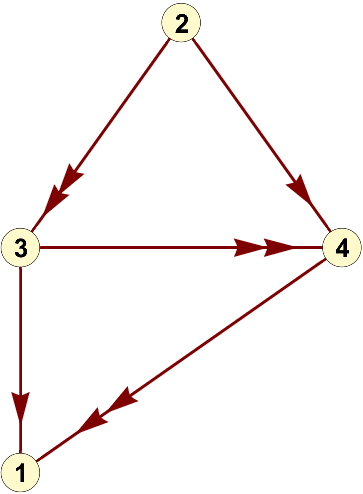}
       \caption{$\calQ_{\set{s_1}}$}
       \label{fig:QPdP1-sub-s1}
    \end{subfigure}
    \hspace*{\fill}
    \caption{Subquivers of $\text{PdP}_1$ obtained by deleting edges from the perfect matching of the chamber $(f_\text{(-1,0)}, s_\text{(0,0)}) = (f_1,s_1)$. }
    \label{fig:QPdP1-sub}
\end{figure}

Take the example of $\text{PdP}_1$ in \cref{sec:KC-minGLSM}. 
For the the choice $(f_\text{(-1,0)}, s_\text{(0,0)}) = (f_1,s_1)$, we obtain the subquivers in \cref{fig:QPdP1-sub}.
Quiver subrepresentations correspond to quiver subdiagrams which are invariant under outward flow. 
For example, the subquiver $\calQ_{f_1}$ has a single proper subrepresentation with dimension $\beta = (1,0,1,0)$.
$\calQ_{s_1}$ has proper subrepresentations with dimensions $\beta \in \set{(1,0,1,1), (1,0,0,1), (1,0,0,0)}$.
Together with $0 = \alpha\cdot\xi = \xi_1 + \xi_2 + \xi_3 + \xi_4$, the semistability conditions become $\xi_2 + \xi_4 \ge 0$ and $\xi_2 \ge 0 \wedge \xi_2 + \xi_3 \ge 0 \wedge \xi_2 + \xi_3 + \xi_4 \ge 0$ respectively, matching the result obtained above.

\subsection{\texorpdfstring{$(p,q)$}{(p,q)}-webs, 2-cycles and zig-zags}

A crepant resolution of a toric Calabi-Yau singularity $\calY$ to $\tilde{\calY}$ consists of a ``blow-up'' of multiple $\bbP^1$.
There are only a few ways to resolve the singularity consistently with the toric structure. Each toric crepant (partial) resolution is encoded in a $(p,q)$-web diagram.

$(p, q)$-webs are a geometric representation of 5-brane configurations in type IIB string theory, which engineer 5-dimensional field theories.
In this context, we are allowed to create bound states of D5-branes and NS5-branes, which we assign charges $(1,0)$ and $(0,1)$ respectively.
If the 5-branes worldvolumes share $4+1$ dimensions, in order to preserve eight supercharges the remaining worldvolume directions form segments in a 2d plane $(x,y)$ oriented according to the brane charges, \emph{i.e.} $\Delta x + i \Delta y \,\lVert\, p + \tau q$ for the axio-dilaton $\tau=C_0 + i e^{-\phi}$ of type IIB string theory. As is customary, we depict all web diagrams at $\tau=i$, so that D5/NS5 pure states align with the horizontal/vertical axis: the effect of changing $\tau$ is to perform a general linear transformation in the $(x,y)$ plane.
The balance of forces require RR-NSNS charge conservation at each vertex: if all fivebranes are incoming, 
\begin{align}
    \sum_{i} (p_i,q_i) = (0, 0) ~.
    \label{eq:pq-vertex-condition}
\end{align}

The toric Calabi-Yau 3-folds studied in this paper are affine toric varieties given by a complex cone over a compact toric surface.
The mesonic toric action $U(1)^2 \subset U(1)^3$ acts naturally on a torus fibre $\bbT^2$ over the complex plane.
The $(p,q)$-web associated to this geometry is a given by a set of segments and vertices, in which one or both $S^1$ in the fibration shrink to a point, respectively.
Resolving a toric singularity corresponds to blowing up a point and replacing it by a $\bbP^1$, \emph{i.e.} replacing a vertex in the $(p,q)$-web by a segment (with $S^1$ fibre).

Given a complete triangulation of the toric diagram of $\calY$, we can construct a consistent $(p,q)$-web of a fully resolved toric singularity.
Every unit triangle in the triangulation of the toric diagram is dual to a 3-valent vertex in the $(p,q)$-web, in such a way that every connected line is perpendicular to the triangle edges, which automatically satisfy \cref{eq:pq-vertex-condition}. 
Edges of the boundary of the toric diagram are dual to the semi-infinite external legs of the web, and represent non-compact holomorphic 2-cycles.
Internal edges are dual to finite segments in the web, which represent holomorphic 2-cycles with a finite volume.

The volume of holomorphic 2-cycles can be quickly computed from the GLSM.
Recall that a toric Calabi-Yau 3-fold has a collection of toric divisors $\calD_{p_\alpha}$, defined by setting
\begin{align}
    p_\alpha = 0 ~
\end{align}
in the GLSM in \cref{tb:fi}.
Toric curves, which are $\bbP^1$, arise as transverse intersection of two toric divisors.
The volume of a holomorphic 2-cycle $\calC = \calD_{p_\alpha} \cdot \calD_{p_\beta} \in H_2(\calY,\bbZ)$ is 
\begin{align}
    \mathrm{vol}(\calC)
    = \int_{\calC} \omega 
    = \int_{\tilde{\calY}} \mathrm{PD}(\calD_{p_\alpha}) \wedge \mathrm{PD}(\calD_{p_\beta}) \wedge \omega ~,
\end{align}
where $\mathrm{PD}(\calD_{p_\alpha})$ is the Poincar\'e dual in $H^2
(\tilde{\calY}, \bbZ)$ to the divisor $\calD_{p_\alpha}$ and $\omega$ is the K\"{a}hler form.
The result is a positive linear combination of the resolution parameters.
By setting to zero the variables in the GLSM associated to the intersecting divisors, $p_\alpha = p_\beta = 0$, from a linear combination of the D-term equations of the GLSM we can obtain an equation
\begin{align}
    \abs{p_N}^2 + \abs{p_S}^2 = \mathrm{vol}(\calC) ~,
\end{align}
where the GLSM fields $p_N$, $p_S$ vanish on the toric divisors that intersect $\calC$ at the poles.

For any triangulation $T_{\Delta}$ of the toric diagram, we can find the tuples $(p_\alpha, p_\beta, p_N, p_S)$ associated to each internal edge.
Furthermore, the positivity condition
\begin{align}
    \int_\calC \omega \ge 0 ~~,
    \label{eq:effective-curve-positivity}
\end{align}
for all the holomorphic 2-cycles for a given fully resolved singularity $\tilde{\calY}$, gives us the compatibility conditions for the associated triangulated toric diagram $T_{\Delta}$, which are a set of inequalities for the FI parameters in the GLSM.  
The quiver representation machinery also allows us to quickly obtain these regions in FI parameter space.
A triangulation $T_\Delta$ is \emph{compatible} with an open string K\"{a}hler chamber $K$ if the semistability condition for all subquivers $\calQ_{K_x, K_y}$ and $\calQ_{K_x}$, defined by
\begin{align}
    \calR(T_{\Delta}, K) = \bigcap_{(x,y) \in T_\Delta} \calR(\calQ_{K_x, K_y}) \cap \calR(K)
    \label{eq:oskcConditions-3}
\end{align}
results in a region of codimension 0, where $\calR(K)$ and $\calR(\calQ)$ are defined in \cref{eq:oskcConditions-1} and \eqref{eq:oskcConditions-2}.
The union of all $\calR(T_{\Delta}, K)$ for all open string K\"ahler chambers $K$ is equivalent to the positivity conditions \eqref{eq:effective-curve-positivity} for the effective curves.

\begin{figure}[!ht]
    \centering
    \hspace*{\fill}
    \begin{subfigure}[c]{0.35\columnwidth}
        \stackinset{l}{-18pt}{b}{-5pt}{%
            \includegraphics[width={0.28\textwidth}, clip, fbox]{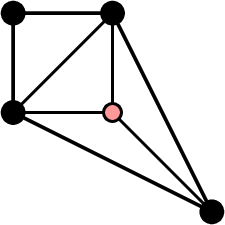}
        }{%
            \includegraphics[width=\columnwidth, clip]{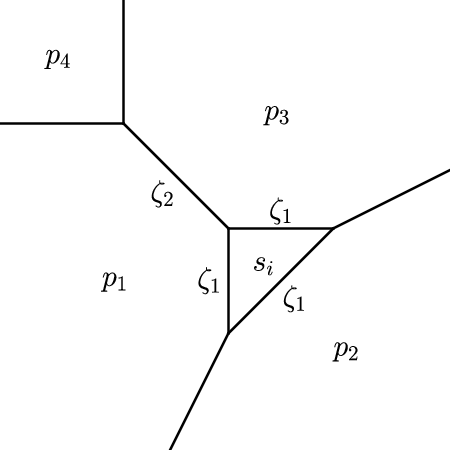}
        }
        \caption{$T^{(1)}_{\Delta}$}
        \label{fig:dP1-pq-web-1}
    \end{subfigure}
    \hspace*{\fill}
    \begin{subfigure}[c]{0.35\columnwidth}
        \stackinset{r}{-10pt}{b}{-5pt}{%
            \includegraphics[width={0.28\textwidth}, clip, fbox]{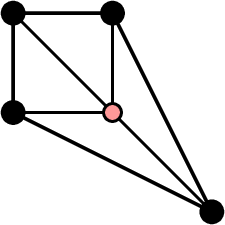}
        }{%
            \includegraphics[width=\columnwidth, clip]{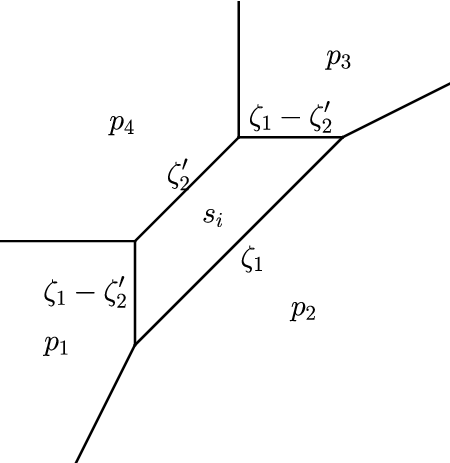}
        }
        \caption{$T^{(2)}_{\Delta}$}
        \label{fig:dP1-pq-web-2}
    \end{subfigure}
    \hspace*{\fill}
    \caption{$(p,q)$-webs associated to the two resolutions of the complex cone over $\text{dP}_1$, related by a flop transition (at $\zeta_2' = -\zeta_2 = 0$). For the choice of $p_{(0,0)} = s_3$, we have $\zeta_1 = \xi_1 - \xi_2$ and $\zeta_2 = - \zeta_2'= \xi_2 + \xi_3$.}
    \label{fig:dP1-pq-webs}
\end{figure}

Take, for example, the complex cone over the \emph{del Pezzo 1} surface, with toric diagram $\Delta = \mathrm{conv}\set{(-1,0),(1,-1),(0,1),(-1,1)}$.
\begin{figure}[!ht]
    \centering
    \hspace*{\fill}
    \begin{subfigure}[c]{0.6\columnwidth}
        \includegraphics[width=\columnwidth]{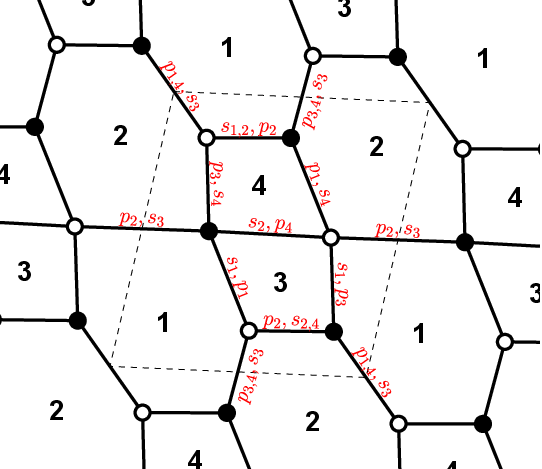}
        \caption{}
        \label{fig:dP1-tiling}
    \end{subfigure}
    \hspace*{\fill}
    \begin{subfigure}[c]{0.28\columnwidth}
        \includegraphics[width=\columnwidth]{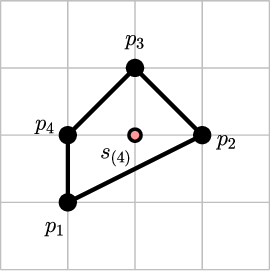}
        \caption{}
        \label{fig:dP1-td}
    \end{subfigure}
    \hspace*{\fill}
    \caption{(a) Brane tiling of $\text{dP}_1$, with each edge labelled with perfect matchings it belongs to. (b) Toric diagram of $\text{dP}_1$.}
    \label{fig:dP1-model}
\end{figure}
Using p.m. variables from the brane tiling in \cref{fig:dP1-tiling}, we find a GLSM with $D$-term equations 
\begin{align}
\begin{aligned}
    \abs{p_1}^2 + \abs{p_2}^2 + \abs{p_3}^2 - \abs{s_1}^2 - \abs{s_3}^2 - \abs{s_4}^2 &= 0 \\
    \abs{p_2}^2 + \abs{p_4}^2 - \abs{s_2}^2 - \abs{s_3}^2 &= 0 \\
    \abs{s_1}^2 - \abs{s_3}^2 &= \xi_1 \\
    \abs{s_3}^2 - \abs{s_4}^2 &= \xi_2 \\
    \abs{s_2}^2 - \abs{s_1}^2 &= \xi_3 \\
    \abs{s_4}^2 - \abs{s_2}^2 &= \xi_4
\end{aligned}
\end{align}
This model has four open string K\"{a}hler chambers, corresponding to the choice of perfect matching variable for the internal point in the toric diagram, $K_i = \set{s_i}$, $1 \le i \le 4$.
Triangulations of the toric diagram are associated to two complete resolutions, represented along with their dual $(p,q)$-webs in \cref{fig:dP1-pq-webs}.
The triangulations can be defined by the single internal segment where a flop transition is possible, thus we have $T^{(1)}_{\Delta} = \set{((-1,0), (0,1))}$ and $T^{(2)}_{\Delta} = \set{((0,0), (-1,1))}$.
For definiteness, let's choose the K\"ahler chamber such that $p_{(0,0)} = s_3$, for which we obtain
\begin{align}
    \calR(\set{s_3}) = \set{\xi_1 \ge 0, \xi_2 \le 0, \xi_1 + \xi_3 \ge 0} ~.
\end{align}
After eliminating the redundant p.m. variables, we obtain the minimal GLSM
\begin{align}
\begin{aligned}
    \abs{p_4}^2 + \abs{s_3}^2 - \abs{p_1}^2 - \abs{p_3}^2 &= \xi_2 + \xi_3 ~,\\
    \abs{p_2}^2 + \abs{p_4}^2 - 2 \abs{s_3}^2 &= \xi_1 + \xi_3 ~
\end{aligned}
\end{align}
that describes the geometry. In this FI parameter region, we can obtain the holomorphic 2-cycles $\calD_{p_\alpha} \cdot \calD_{p_\beta}$, where $\calD_p$ is the toric divisor associated to the GLSM fields $p$,
by setting $p_\alpha=p_\beta= 0$ in the $D$-term equations of the GLSM.
For the resolution associated with $T^{(1)}_{\Delta}$, which describes a finite size $\bbP^2$ intersecting $\bbP^1$, the holomorphic 2-cycles have volumes
\begin{align}
\begin{aligned}
    \mathrm{vol}(\calD_{p_1} \cdot \calD_{s_3}) = \mathrm{vol}(\calD_{p_3} \cdot \calD_{s_3}) = \mathrm{vol}(\calD_{p_2} \cdot \calD_{s_3}) &= \xi_1 - \xi_2 ~,\\
    \mathrm{vol}(\calD_{p_1} \cdot \calD_{p_3}) &= \xi_2 + \xi_3 ~.
\end{aligned}
\end{align}
On the other hand, for the resolution associated with $T^{(2)}_{\Delta}$, which has a finite size $\text{dP}_1$, 
\begin{align}
\begin{aligned}
    \mathrm{vol}(\calD_{p_2} \cdot \calD_{s_3}) &= \xi_1 - \xi_2  ~, \\
    \mathrm{vol}(\calD_{p_4} \cdot \calD_{s_3}) &= -\xi_2 - \xi_3  ~, \\
    \mathrm{vol}(\calD_{p_1} \cdot \calD_{s_3}) = \mathrm{vol}(\calD_{p_3} \cdot \calD_{s_3}) &= \xi_1 + \xi_3 ~.\\
\end{aligned}
\end{align}

In the chamber $\calR(\set{s_3})$, the parameter $\xi_2 + \xi_3$ can be positive/negative and its sign determines two resolutions in \cref{fig:dP1-pq-webs}, related by the flop transition
\begin{align}
    \calD_{p_1} \cdot \calD_{p_3} \leftrightarrow \calD_{p_4} \cdot \calD_{s_3} ~.
\end{align}
If $\xi_2 + \xi_3 > 0$, we lie on the resolution $T^{(1)}_{\Delta}$, otherwise for $\xi_2 + \xi_3 < 0$ we land in $T^{(2)}_{\Delta}$, consistently with the positivity conditions of the volumes of $\calD_{p_1} \cdot \calD_{p_3}$ and $\calD_{p_4} \cdot \calD_{s_3}$ respectively.
If $\xi_2 + \xi_3 = 0$ the singularity is not fully resolved.
Note that for the curve $\calD_{p_1} \cdot \calD_{s_3}$, either $(p_N, p_S) = (p_2, p_3)$ or $(p_2, p_4)$, since $\abs{p_4}^2 - \abs{p_3}^2 = \xi_2 + \xi_3$ for $p_1 = s_3 = 0$.
A similar result holds for $\calD_{p_3} \cdot \calD_{s_3}$.
Compatibility conditions of the resolution $T^{(1)}_{\Delta}$ with the K\"{a}hler chamber $K = \set{s_3}$ can be directly obtained from the $\theta$-stability of representation of $\calQ_{p_1, p_3}$, from which we have the proper subrepresentation with dimension vector $(0,1,1,0)$.
For the subquiver $\calQ_{p_4, s_3}$, we have the complementary dimension vector $(1,0,0,1)$.
These determine
\begin{align}
    \begin{aligned}
        \calR(T^{(1)}_{\Delta}, \set{s_3}) &= \set{\xi_2 + \xi_3 \ge 0} \cap \calR(\set{s_3}) ~,\\
        \calR(T^{(2)}_{\Delta}, \set{s_3}) &= \set{\xi_2 + \xi_3 \le 0} \cap \calR(\set{s_3}) ~.
    \end{aligned}
\end{align}

Semi-infinite legs of fivebrane webs are related to zig-zag paths in the tiling, by matching $(p,q)$ charges with homology classes (up to $SL(2,\bbZ)$).
Moreover, by taking successive pairwise counterclockwise differences of perfect matchings $p_i - p_{i+1}$ along the boundary of $\Delta$, we can reconstruct the zig-zag using edges $X_e$ alternating from the matchings $p_i$ and $p_{i+1}$ with opposite orientation.%
\footnote{Edges present in the both $p_i$ and $p_{i+1}$ will cancel, so only $X_e$ in the symmetric difference of the matching sets will form the path.}
For isolated toric singularities, these ordered pairs of external matchings are in one-to-one correspondence to the zig-zag paths \cite{Butti:2006nk, Kennaway:2007tq, Gulotta:2008ef}.
The case of non-isolated singularities is more interesting, since the corresponding toric diagrams have external perfect matchings of higher multiplicity.

\begin{figure}[!ht]
    \centering
    \hspace*{\fill}
    \begin{subfigure}[c]{0.58\columnwidth}
       \includegraphics[width=\columnwidth, clip]{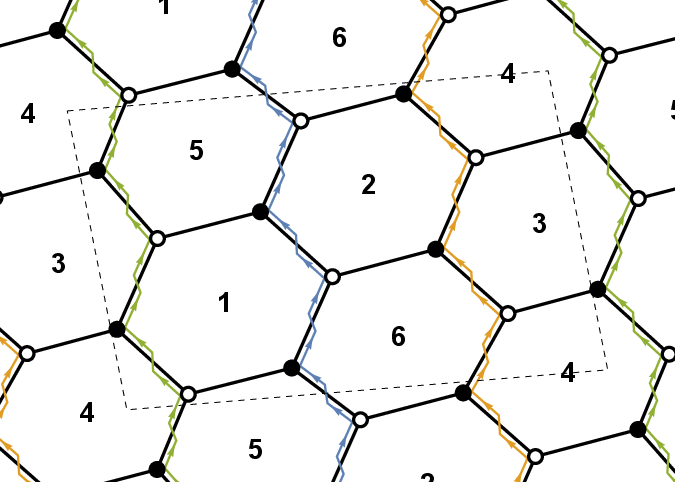}
       \caption{}
       \label{}
    \end{subfigure}
    \hspace*{\fill}
    \hspace*{\fill}
    \begin{subfigure}[c]{0.21\columnwidth}
       \includegraphics[width=\columnwidth, clip]{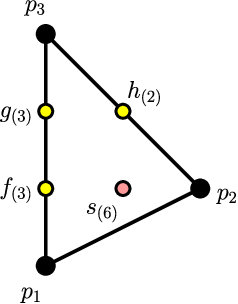}
       \caption{}
       \label{fig:PdP3a-td-pms}
    \end{subfigure}
    \hspace*{\fill}
    \caption{Brane tiling and toric diagram for $\coset{\bbC^3}{\bbZ_6} \,(1,2,3)$, highlighting the three parallel zig-zag paths associated to the $A_2$ singularity.}
    \label{fig:PdP3a-til-td}
\end{figure}
\begin{figure}[H]
    \centering
    \hspace*{\fill}
    \includegraphics[height=0.33\columnwidth, clip]{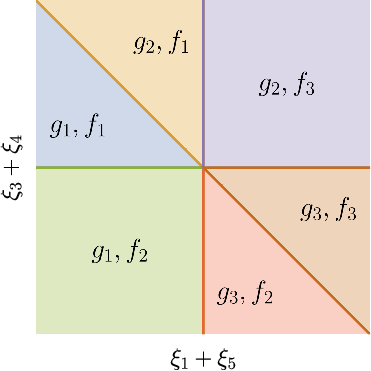}
    \hspace*{\fill}
    \caption{Regions formed by the stability of reps. of $Q_{g_i}, Q_{f_j}, Q_{g_i, f_j}$ in the $\text{PdP}_{3a}$ quiver.}
    \label{fig:PdP3a-A2-KCcomp}
\end{figure}

For instance, the complex cone over \emph{pseudo del Pezzo 3a} ($\calY = \coset{\bbC^3}{\bbZ_6} \,(1,2,3)$) has an $A_2$ singularity, as signalled by the length $3$ side in its toric diagram, see \cref{fig:PdP3a-til-td}.
The $\text{PdP}_{3a}$ theory flows to another toric quiver gauge theory by the deformation associated to any of the parallel zig-zag paths normal to the side:
\begin{align}
    \begin{aligned}    
        \eta_4 &= X_{12} X_{25} X_{56} X_{61} \quad:\qquad\set{p_3 - g_1,\, g_2 - f_1,\, g_3 - f_2,\, f_3 - p_1} \\
        \eta_5 &= X_{24} X_{46} X_{63} X_{32} \quad:\qquad\set{p_3 - g_2,\, g_1 - f_1,\, g_3 - f_3,\, f_2 - p_1} \\
        \eta_6 &= X_{13} X_{35} X_{54} X_{41} \quad:\qquad\set{p_3 - g_3,\, g_1 - f_2,\, g_2 - f_3,\, f_1 - p_1} \\
    \end{aligned}
    \label{eq:PdP3a-zz-pms}
\end{align}
We can infer the p.m. differences in \eqref{eq:PdP3a-zz-pms} from the perfect matching submatrix \eqref{eq:PdP3a-pm-submatrix} for the relevant matchings and bifundamentals:
\begin{align}
    \left(
    \scalebox{0.85}{$\begin{array}{c|cccccccccccc}
        \text{} & X_{12} & X_{13} & X_{24} & X_{25} & X_{32} & X_{35} & X_{41} & X_{46} & X_{54} & X_{56} &
        X_{61} & X_{63} \\
        \hline
        p_3 & 0 & 1 & 0 & 1 & 1 & 0 & 0 & 1 & 1 & 0 & 1 & 0 \\
        \hline
        g_1 & 1 & 1 & 0 & 0 & 1 & 0 & 0 & 1 & 1 & 1 & 0 & 0 \\
        g_2 & 0 & 1 & 1 & 1 & 0 & 0 & 0 & 0 & 1 & 0 & 1 & 1 \\
        g_3 & 0 & 0 & 0 & 1 & 1 & 1 & 1 & 1 & 0 & 0 & 1 & 0 \\
        \hline
        f_1 & 1 & 1 & 1 & 0 & 0 & 0 & 0 & 0 & 1 & 1 & 0 & 1 \\
        f_2 & 1 & 0 & 0 & 0 & 1 & 1 & 1 & 1 & 0 & 1 & 0 & 0 \\
        f_3 & 0 & 0 & 1 & 1 & 0 & 1 & 1 & 0 & 0 & 0 & 1 & 1 \\
        \hline
        p_1 & 1 & 0 & 1 & 0 & 0 & 1 & 1 & 0 & 0 & 1 & 0 & 1 \\
    \end{array}$}
    \right)
    \label{eq:PdP3a-pm-submatrix}
\end{align}
For non-isolated singularities, the multiplicity of external non-extremal perfect matchings leads to multiple successive boundary pairs that construct the same zig-zag path.
A choice of K\"{a}hler chamber fixes the perfect matching variables for each lattice point in $\Delta$ and selects unique boundary pairs.
However, some chambers do not have compatible resolutions.
For incompatible chambers, we may have multiple differences $p_i - p_{i+1}$ which give rise to the same path or one which describes a disconnected set of paths in the brane tiling.
In the example above, zig-zag paths cannot be constructed by differences of boundary matchings when any of the pairs $(g_3, f_1)$, $(g_2, f_2)$, $(g_1, f_3)$ are chosen for the points $(-1,-1), (-1,0) \in\Delta$.
This is consistent with the fact that the open string K\"{a}hler chambers
\begin{align}
    \set{g_3, f_1, h_i, s_j} ~,\quad \set{g_2, f_2, h_i, s_j} ~,\quad \set{g_1, f_3, h_i, s_j}~,
\end{align}
for $1 \le i \le 2$ and $1 \le j \le 6$, are incompatible with all five possible triangulations of \cref{fig:PdP3a-td-pms}.
We can visualize the region of resolution parameters in which the p.m.s of an $A_k$ singularity side are compatible as the total space of a fibration, with the fibre given by the perfect matchings.
In particular, for $\text{PdP}_{3a}$ we can represent the regions and boundaries in a plane (see \cref{fig:PdP3a-A2-KCcomp}).

Using the data from the GLSM and subquiver representation stability, we obtain the volumes of 2-cycles in terms of FI parameters and associate external legs of the fivebrane web to zig-zag paths in the brane tiling, and hence chiral operators in the gauge theory. Geometrically, we can apply the zig-zag deformation to the quiver gauge theory with FI parameters, therefore it should be possible to map K\"{a}hler chambers between UV and IR toric models. We analyze this problem next.

\subsection{Zig-zag deformation as a Hanany-Witten move}

We have seen that for a zig-zag deformation $\calO_\eta$ associated to a zig-zag path $\eta$ of length 4, the deformed UV toric theory flows to another toric gauge theory in the IR, which has $\eta$ reversed relative to the original brane tiling.
The UV geometry has at least one non-isolated $A_k$ singularity, $k\ge1$, meaning that the tiling has $k+1$ parallel zig-zags of the same homology $(p,q)$. At the IR endpoint of the deformation, each reversed zig-zag has homology $(-p,-q)$.
The other parallel zig-zag paths are unaffected in the tiling, while some of the other zig-zag paths are rearranged and have different homology, as seen by comparing \cref{fig:PdP3cB-til-quiver-td} and \cref{fig:PdP3cB-to-PdP3bB-flow}.

Translating the zig-zag move in the brane tiling to the $(p,q)$-web suggests that the deformation is described by a Hanany-Witten move for a 7-brane on which the 5-brane associated to the deformation zig-zag ends.%
\footnote{We thank Michele Del Zotto for this suggestion.}
Indeed, an external $(p,q)$ fivebrane can terminate on a $[p,q]7$-brane, which is a point in the plane of the fivebrane web \cite{DeWolfe:1999hj}. 
By using $SL(2,\bbZ)$, we can assume without loss of generality that the external legs associated to the relevant parallel zig-zags are D5-branes with charge $(-1,0)$. We can now end one of the aforementioned D5-branes on a D7-brane, which we then move along the line of the D5-brane, until it ends on the opposite side of the $(p,q)$ web diagram. This corresponds to reversing the orientation of the zig-zag path. 

Every time the D7 (or $[1,0]7$) brane crosses an $(r,s)5$-brane, by the Hanany–Witten effect a number $|s|$ of  $(1,0)5$-branes are created, which are suspended between the crossed 5-brane and the D7-brane. For the reflexive geometries studied in this paper, reversing the D7-brane horizontally to the opposite side has it cross exactly two fivebranes with NSNS charge $|s|=1$: the first HW transition annihilates the original $(-1,0)5$-brane, while the second HW transition creates a new $(1,0)5$-brane. 
We then can take the limit of the 7-brane going to infinity in the opposite direction to recover a new toric geometry.%
\footnote{It is natural to interpret the position of the $7$-brane along the line of the 5-branes as a monotonic function of $|\mu|$ which tends to $\mp\infty$ at the two ends of the line (for example, $\log|\mu|$). We will derive this fact and elaborate on the precise relation in \cite{CCS2023}.} We will provide precise evidence that the resulting fivebrane web describes the fully resolved toric geometry obtained at the endpoint of the Klebanov-Witten deformation. 

The 7-brane sources an $SL(2,\bbZ)$ monodromy which affects the rest of the web. We can take the monodromy cut to extend from the 7-brane to infinity along the line of the 5-brane we attached it to. As we slide the 7-brane to the opposite end of the line, we also need to rotate the monodromy cut by $\pm 180$ degrees to the opposite side, so that the cut disappears when the 7-brane goes to infinity in the opposite direction. Every time the monodromy cut of a $[p,q]7$-brane crosses a 5-brane, the charge of the 5-brane is acted upon by the $SL(2,\bbZ)$ monodromy matrix
\begin{align}
    M_{p,q} =
    \begin{pmatrix}
        1 - p q & p^2      \\
        -q^2    & 1 + p q
    \end{pmatrix}
    ~.
\end{align}
If the branch cut is moved counterclockwise then we apply $M_{p,q}$ to the affected segments, otherwise we act with $M_{p,q}^{-1}$ moving the cut clockwise. Rotating the monodromy cut by 180 degrees clockwise or counterclockwise and sending the 7-brane to infinity, the two resulting $(p,q)$-webs are equivalent up to an $SL(2,\bbZ)$ transformation.
This matches exactly the choice of tiling move in \cref{fig:PdP3cB-integrated-in} for the zig-zag of homology $(0,1)$, which results either in \cref{fig:PdP3cB-to-PdP3bB-flow-move1} (clockwise) or \cref{fig:PdP3cB-to-PdP3bB-flow-move2} (anticlockwise).
Similarly to the zig-zag deformation, parallel D5-branes of the same charge as the $(p,q)$ 7-brane are unaffected by the monodromy.
\begin{figure}[!ht]
    \begin{tikzpicture}
        \node (fig) at (0,0) {\includegraphics[width=0.98\columnwidth,clip]{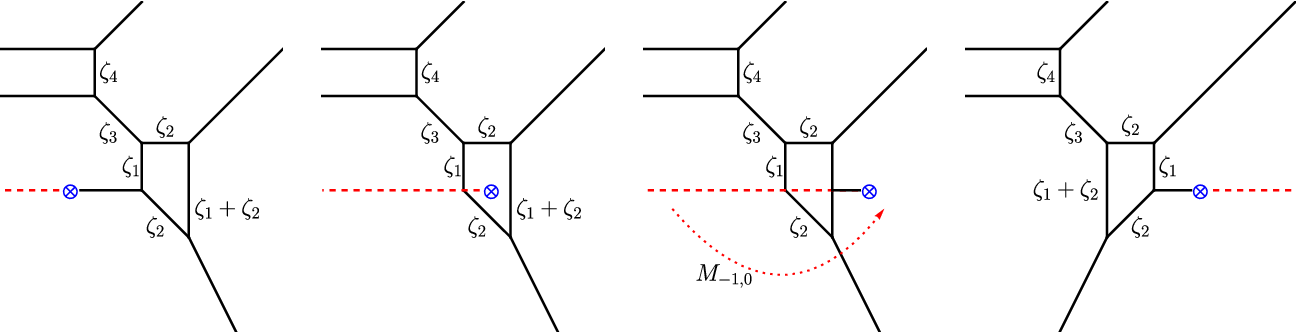}};
        \node[fill=white] (tdI) at (-6.5,-2.4) {\includegraphics[width={0.09\columnwidth}, clip, fbox]{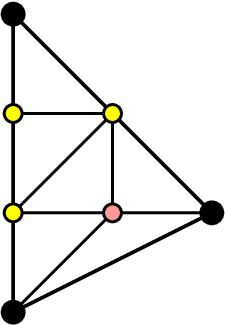}};
        \node[fill=white] (tdF) at (+6.5,-2.4) {\includegraphics[width={0.09\columnwidth}, clip, fbox]{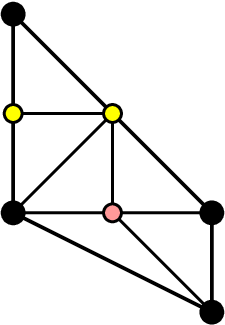}};
    \end{tikzpicture}
    \caption{Hanany-Witten move mapping resolution $T_\Delta$ of $\text{PdP}_{3a}$ to $T_\Delta'^{(2)}$ of $\text{PdP}_{3c}$. $5$-branes are in black, $7$-branes are in blue, and monodromy cuts in dashed red. Volumes of holomorphic cycles $\zeta_a$ are superimposed.}
    \label{fig:PdP3a-pq-move2}
\end{figure}

It is straightforward to see that zig-zag deformations which relate fully resolvable toric geometries with a single exceptional divisor can be generalized to non-reflexive toric geometries if and only if the lattice width $w$ of the toric diagram normal to the $(p,q)$ 5-brane being reversed is exactly 2.%
\footnote{If $w>2$, at the endpoint of the Hanany-Witten move the D7 brane would become attached to $w-1>1$ coincident D5 branes, leading to a toric geometry with frozen resolutions, which is described by a \emph{generalized toric diagram} (or \emph{polytope}) \cite{Benini:2009gi, vanBeest:2020kou, Bourget:2023wlb}. We will study those situations in a companion paper \cite{CCS2023}.}
Infinite families of these deformations have already been verified, for example from $(\bbC^2/\bbZ_n \times \bbC)/\bbZ_2$ to $L_{1,n-1,1}/\bbZ_2$, for all $n \ge 2$ \cite{Bianchi:2014qma}, and more examples will be provided in \cite{Sa2023}.
\begin{figure}[!ht]
    \centering
    \begin{tikzpicture}
        \node (fig) at (0,0) {\includegraphics[width=0.98\columnwidth,clip]{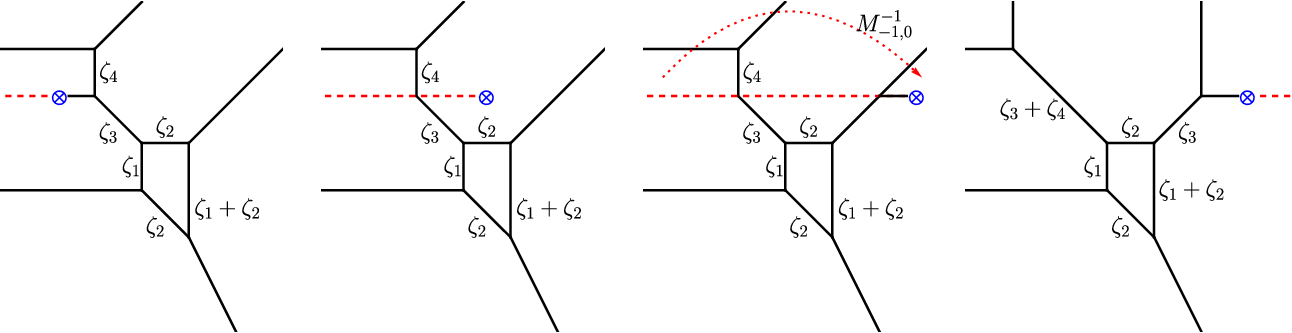}};
        \node[fill=white] (tdI) at (-6.5,-2.4) {\includegraphics[width={0.09\columnwidth}, clip, fbox]{PdP3a-td-trig1}};
        \node[fill=white] (tdF) at (+6.5,-2.75) {\includegraphics[width={0.09\columnwidth}, clip, fbox]{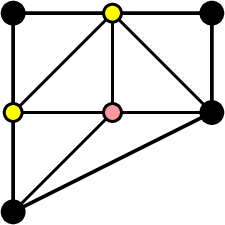}};
    \end{tikzpicture}
    \caption{Hanany-Witten move mapping resolution $T_\Delta$ of $\text{PdP}_{3a}$ to $T_\Delta'^{(1)}$ of $\text{PdP}_{3c}$.
Volumes of holomorphic cycles $\zeta_a$ are superimposed.}
    \label{fig:PdP3a-pq-move1}
\end{figure}

We can show that the Hanany-Witten move is consistent with the quiver gauge theory analysis for the regions $\calR(T_\Delta, K)$, defined in \cref{eq:oskcConditions-3}, of the field theories involved in the zig-zag deformation.
In \cref{fig:PdP3a-pq-move1,fig:PdP3a-pq-move2}, we have $(p,q)$-webs associated to the triangulated toric diagram $T_{\Delta}$ in which and we perform a Hanany-Witten move for two different parallel 5-branes.
The resulting $(p,q)$-webs are dual to the triangulated toric diagrams $T_{\Delta'}^{(1)}$ or $T_{\Delta'}^{(2)}$ of $\text{PdP}_{3c}$, as expected.

For any open string K\"ahler chamber, we know exactly which zig-zag path corresponds to which semi-infinite 5-brane. We can therefore identify which 5-brane we need to reverse to match the zig-zag deformation.
The result of the Hanany-Witten move can then be compared with resolved geometry obtained from the quiver gauge theory analysis.

In the first move represented in \cref{fig:PdP3a-pq-move1}, there exist multiple pairs of open string K\"{a}hler chambers $K^{(1)}_1, K^{(1)}_2$ such that the chosen zig-zag $\eta$ corresponds to either of the top two parallel 5-branes.
As a result,
\begin{align}
    \calR\left(
    \begin{minipage}[h]{0.14\columnwidth}
        \vspace{0pt}\includegraphics[width=\columnwidth]{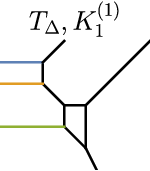}
    \end{minipage}
    \right)
    ~\bigcup~
    \calR\left(
    \begin{minipage}[h]{0.14\columnwidth}
        \vspace{0pt}\includegraphics[width=\columnwidth]{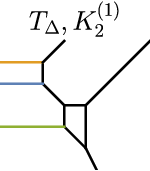}
    \end{minipage}
    \right)
    =
    \calR\left(
    \begin{minipage}[h]{0.14\columnwidth}
        \vspace{0pt}\includegraphics[width=\columnwidth]{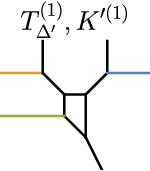}
    \end{minipage}
    \right) ~.
\end{align}
For example, for the reversal of $\eta_4 = X_{12} X_{25} X_{56} X_{61}$ in $\text{PdP}_{3a}$ ($\mu\to0$), see \ref{eq:PdP3atoPdP3c-eta4}, the two regions
\begin{align}
\begin{aligned}
    \calR\left(T_{\Delta}, \set{g_1, f_1, h_2, s_2}\right) &= \set{\xi_3\leq 0, \xi_1\leq 0, \xi_2+\xi_5\geq 0, \xi_1+\xi_3+\xi_4\geq 0, \xi_1+\xi_3+\xi_4+\xi_5\leq 0} \\
    \calR\left(T_{\Delta}, \set{g_2, f_1, h_2, s_2}\right) &= \set{\xi_3\leq 0, \xi_1\leq 0, \xi_2+\xi_5\geq 0, \xi_5\leq 0, \xi_1+\xi_3+\xi_4+\xi_5\geq 0}
\end{aligned}
\label{eq:oscsKC-pair-PdP3a-eg}
\end{align}
join into the single region
\begin{align}
    \calR\left(T_{\Delta'}^{(1)}, \set{f'_2, g'_1, s'_2}\right) &= \set{\xi_5\leq 0, \xi_3\leq 0, \xi_1\leq 0, \xi_2+\xi_5\geq 0, \xi_1+\xi_3+\xi_4\geq 0} ~,
\end{align}
for phase A of $\text{PdP}_{3c}$ ($\mu\to\infty$).
In these subregions, the volumes of 2-cycles (see \cref{fig:PdP3a-pq-move2,fig:PdP3a-pq-move1}) can be written as
\begin{align}
\begin{aligned}
    \zeta_1 &= -\xi_1 \\
    \zeta_2 &= \xi_2 - \xi_3 + \xi_5
\end{aligned}
\qquad
\begin{aligned}
    \zeta_3 &= -\xi_5 + \min\left(0, \xi_1+\xi_3+\xi_4+\xi_5\right) \\
    \zeta_4 &= \abs{ \xi_1+\xi_3+\xi_4+\xi_5 }
\end{aligned}
\end{align}
The intersection between the two regions in \cref{eq:oscsKC-pair-PdP3a-eg} corresponds to the $\theta$-stability conditions for subquiver $\calQ_{g_1, g_2}$, which is when the volume $\zeta_4(\xi)$ vanishes, or equivalently when the two parallel 5-branes coincide.

In general, each resolution partitions the $k$ parallel external legs of the nonisolated $A_{k-1}$ singularity into groups of $k_1, \dots, k_m$ parallel 5-branes.
Moving a parallel 5-brane from a given partition to another requires a flop transition, thus moving us to another triangulation.
For a given resolution $T_{\Delta}$, we have
\begin{align}
    \bigcup_{i=1}^{k_a} \calR\left( T_{\Delta}, K_{i}^{(a)} \right) =
    \calR\left( T_{\Delta'}^{(a)}, K'^{(a)} \right)
    \quad\text{for}\quad a\in\set{1, \dots, m} ~,
\end{align}
for $T_{\Delta'}^{(a)}$ and $K'^{(a)}$ triangulations and open string K\"{a}hler chambers of the $\mu\to\infty$ geometry.
The move represented in \cref{fig:PdP3a-pq-move2} is when the zig-zag corresponds to an external leg in a $k_a = 1$ partition, in which case
\begin{align}
    \calR\left(
    \begin{minipage}[h]{0.14\columnwidth}
        \vspace{0pt}\includegraphics[width=\columnwidth]{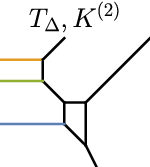}
    \end{minipage}
    \right)
    =
    \calR\left(
    \begin{minipage}[h]{0.14\columnwidth}
        \vspace{0pt}\includegraphics[width=\columnwidth]{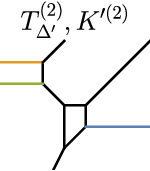}
    \end{minipage}
    \right) ~.
\end{align}

Using the technology of quiver representation theory, we have been able to map the various open string K\"{a}hler chambers of any two (pseudo-)del Pezzo theories related by a zig-zag deformation, listed in \cref{sec:RGflows} and \cref{fig:defsSummary}. For readers interested in the details, an ancillary file is available upon request.
Fixing the zig-zag $\eta$ and mapping regions $\calR(T_\Delta, K)$ and volumes $\zeta_a$ from both models, we have verified that the (resolution of) the final toric geometry is precisely the one dual to the $(p,q)$-web obtained by reversing the external leg associated to $\eta$.

%% file: Appendices/ZZflows.tex

\section{Details of the zig-zag deformations}
\label{sec:RGflows}

In this appendix we collect details of the zig-zag deformations of UV toric models and the field redefinitions needed to obtain new toric models in the IR.
We describe the deformations $\delta W$ of toric superpotentials by a (or multiple) revelant zig-zag operator $\calO_\eta$, which flow into a toric model or a marginal deformation of one.
We divide these into \cref{sec:toric-to-toric-list} and \cref{sec:toric-to-marginal-list}.
Each subsection is titled ``\emph{$\calX$ to $\calY$}''.
Futhermore, when different toric phases are involved these are also pointed out.

\subsection{Deformations to toric models}
\label{sec:toric-to-toric-list}
 
The deformations listed here include only the relevant steps, which are fully described in \cref{sec:deformation-delPezzo}.
In each subsection, describing the deformation between 2 specific geometries, we include the original starting superpotential $W$.
Additionally, each deformation is in a unique equation block with:
\begin{enumerate}[1)]
    \item The deformation $\delta W = \mu \sum_{i\in I} \calO_{\eta_i}$, associated to one or more zig-zag paths $\eta_i$.
    \item The non-trivial field redefinitions of the type in \cref{eq:field_redef}, if any. Some redefinitions include free parameters $\beta_i \in \bbC$, which do not affect the final result.
    \item The toric superpotential $W'$ representing the IR theory.
\end{enumerate}
The redefinition in item 2 is for the superpotential $W + \delta W$, after integrating out massive fields using the F-terms \eqref{eq:integrate-out-fterms}.

\subsubsection{\texorpdfstring{$\text{PdP}_{1}, \coset{\bbC^3}{\bbZ_4} \,(1,1,2)$}{PdP1, C3/Z4 (1,1,2)} to \texorpdfstring{$\mathbb{F}_0, \coset{\calC}{\bbZ_2} \,(1,1,1,1)$}{F0, Conifold/Z2 (1,1,1,1)}}
\label{sec:PdP1todP1}

\begin{align}
    \begin{aligned}
        W &= X_{13} X_{34}^1 X_{41}^2 + X_{24} X_{41}^1 X_{12}^2 + X_{31} X_{12}^1 X_{23}^2 + X_{42} X_{23}^1 X_{34}^2 \\
        & \qquad - X_{13} X_{34}^2 X_{41}^1 - X_{24} X_{41}^2 X_{12}^1 - X_{31} X_{12}^2 X_{23}^1 - X_{42} X_{23}^2 X_{34}^1
    \end{aligned}
\end{align}

\begin{align}
    \begin{aligned}
        \delta W &= \mu \left( X_{13} X_{31} - X_{24} X_{42} \right)
        \\[0.5em]
        W' &= X_{12}^1 X_{23}^2 X_{34}^2 X_{41}^1 + X_{12}^2 X_{23}^1 X_{34}^1 X_{41}^2 - X_{12}^1 X_{23}^1 X_{34}^2 X_{41}^2 - X_{12}^2 X_{23}^2 X_{34}^1 X_{41}^1 \\ 
        &= - \epsilon_{ab} \epsilon_{cd} X_{12}^a X_{23}^b X_{34}^c X_{41}^d
    \end{aligned}
    \label{eq:PdP1Def1toF0}
\end{align}

\subsubsection{\texorpdfstring{$\text{PdP}_{2}$}{PdP2} to \texorpdfstring{$\text{dP}_{2}$}{dP2}}
\label{sec:PdP2todP2}

\begin{align}
    \begin{aligned}
        W &= X_{12} X_{25}^1 X_{51}^2 + X_{14} X_{42} X_{21} + X_{53} X_{32} X_{25}^2 + X_{13} X_{34} X_{45} X_{51}^1 \\ 
        & \qquad - X_{12} X_{25}^2 X_{51}^1 - X_{13} X_{32} X_{21} - X_{14} X_{45} X_{51}^2 - X_{34} X_{42} X_{25}^1 X_{53}
    \end{aligned}
\end{align}

\begin{align}
    \begin{aligned}
        \delta W &= \mu \left( X_{12} X_{21} - X_{34} X_{45} X_{53} \right)
        \\[0.5em]
        & \quad X_{45} \mapsto +\frac{1}{\mu} X_{45} - \frac{1}{\mu} X_{42} X_{25}^1
        \qquad X_{53} \mapsto -\frac{1}{\mu} X_{53} + \frac{1}{\mu} X_{51}^1 X_{13}
        \\[0.5em]
        W' &= X_{34} X_{45} X_{53} + X_{13} X_{32} X_{25}^1 X_{51}^2 + X_{14} X_{42} X_{25}^2 X_{51}^1 \\ 
        & \qquad - X_{14} X_{45} X_{51}^2 - X_{32} X_{25}^2 X_{53} - X_{13} X_{34} X_{42} X_{25}^1 X_{51}^1        
    \end{aligned}
    \label{eq:PdP2Def1todP2}
\end{align}

\subsubsection{\texorpdfstring{$\text{PdP}_{3a}, \coset{\bbC^3}{\bbZ_6} \,(1,2,3)$}{PdP3a, C3/Z6 (1,2,3)} to \texorpdfstring{$\text{PdP}_{3c}, \coset{\text{SPP}}{\bbZ_2} \,(0,1,1,1)$}{PdP3c, SPP/Z2 (0,1,1,1)}}
\label{sec:PdP3atoPdP3c}

\subsubsection*{Phase A to Phase A}

\begin{align}
    \begin{aligned}
        W &= X_{12} X_{26} X_{61} + X_{13} X_{35} X_{51} + X_{15} X_{54} X_{41} + X_{24} X_{43} X_{32} \\ 
        & \qquad + X_{25} X_{56} X_{62} + X_{34} X_{46} X_{63} - X_{12} X_{25} X_{51} - X_{13} X_{34} X_{41} \\ 
        & \qquad - X_{15} X_{56} X_{61} - X_{24} X_{46} X_{62} - X_{26} X_{63} X_{32} - X_{35} X_{54} X_{43}
    \end{aligned}
\end{align}

\begin{align}
    \begin{aligned}
        \delta W &= \mu \left(X_{15} X_{51} - X_{34} X_{43}\right)
        \\[0.5em]
        &\quad X_{26} \mapsto -\frac{1}{\mu} X_{26} + \left( \frac{1}{\mu} - \beta_1 \right) X_{24} X_{46} + \beta_1\, X_{25} X_{56} \\ 
        &\quad X_{62} \mapsto -\frac{1}{\mu} X_{62} + \left( \frac{1}{\mu} - \beta_1 \right) X_{61} X_{12} + \beta_1\, X_{63} X_{32}
        \\[0.5em]
        W' &= X_{24} X_{46} X_{62} + X_{26} X_{63} X_{32} + X_{12} X_{25} X_{54} X_{41} + X_{13} X_{35} X_{56} X_{61} \\ 
        & \qquad - X_{12} X_{26} X_{61} - X_{25} X_{56} X_{62} - X_{13} X_{32} X_{24} X_{41} - X_{35} X_{54} X_{46} X_{63}
    \end{aligned}
    \label{eq:PdP3atoPdP3c-eta6}
\end{align}

\begin{align}
    \begin{aligned}
        \delta W &= \mu \left(X_{26} X_{62} - X_{15} X_{51}\right)
        \\[0.5em]
        &\quad X_{34} \mapsto -\frac{1}{\mu} X_{34} + \left( \frac{1}{\mu} - \beta_1 \right) X_{32} X_{24} + \beta_1\, X_{35} X_{54} \\ 
        &\quad X_{43} \mapsto -\frac{1}{\mu} X_{43} + \left( \frac{1}{\mu} - \beta_1 \right) X_{41} X_{14} + \beta_1\, X_{46} X_{63}
        \\[0.5em]
        W' &= X_{13} X_{34} X_{41} + X_{35} X_{54} X_{43} + X_{12} X_{24} X_{46} X_{61} + X_{25} X_{56} X_{63} X_{32} \\ 
        & \qquad - X_{24} X_{43} X_{32} - X_{34} X_{46} X_{63} - X_{12} X_{25} X_{54} X_{41} - X_{13} X_{35} X_{56} X_{61}
    \end{aligned}
    \label{eq:PdP3atoPdP3c-eta4}
\end{align}

\subsubsection{\texorpdfstring{$\text{PdP}_{3c}, \coset{\text{SPP}}{\bbZ_2} \,(0,1,1,1)$}{PdP3c, SPP/Z2 (0,1,1,1)} to \texorpdfstring{$\text{PdP}_{3b}$}{PdP3b}}
\label{sec:PdP3ctoPdP3b}

\subsubsection*{Phase A to Phase A}

\begin{align}
    \begin{aligned}
        W &= X_{25} X_{56} X_{62} + X_{36} X_{65} X_{53} + X_{13} X_{34} X_{45} X_{51} + X_{16} X_{64} X_{42} X_{21} \\ 
        & \qquad - X_{16} X_{65} X_{51} - X_{45} X_{56} X_{64} - X_{13} X_{36} X_{62} X_{21} - X_{25} X_{53} X_{34} X_{42}
    \end{aligned}
\end{align}

\begin{align}
    \begin{aligned}
        \delta W &= \mu \left(X_{16} X_{62} X_{21} - X_{34} X_{45} X_{53}\right)
        \\[0.5em]
        &\quad X_{16} \mapsto -\frac{1}{\mu} X_{16} + \frac{1}{\mu} X_{13} X_{36}
        \qquad X_{62} \mapsto \frac{1}{\mu} X_{62} - \frac{1}{\mu} X_{64} X_{42} \\ 
        &\quad X_{53} \mapsto -\frac{1}{\mu} X_{53} + \frac{1}{\mu} X_{51} X_{13}
        \qquad X_{45} \mapsto \frac{1}{\mu} X_{45} - \frac{1}{\mu} X_{42} X_{25}
        \\[0.5em]
        W' &= X_{16} X_{65} X_{51} + X_{25} X_{56} X_{62} + X_{34} X_{45} X_{53} + X_{13} X_{36} X_{64} X_{42} X_{21} \\ 
        & \qquad - X_{16} X_{62} X_{21} - X_{36} X_{65} X_{53} - X_{45} X_{56} X_{64} - X_{13} X_{34} X_{42} X_{25} X_{51}
    \end{aligned}
\end{align}

\subsubsection*{Phase B to Phase B}

\begin{align}
    \begin{aligned}
        W &= X_{12} X_{23} X_{31} + X_{25} X_{56} X_{62} + X_{26} X_{64} X_{42} + X_{34} X_{45} X_{53}^2 \\ 
        & \qquad + X_{15} X_{53}^1 X_{36} X_{61} - X_{12} X_{26} X_{61} - X_{15} X_{53}^2 X_{31} - X_{23} X_{36} X_{62} \\ 
        & \qquad - X_{45} X_{56} X_{64} - X_{25} X_{53}^1 X_{34} X_{42}
    \end{aligned}
\end{align}

\begin{align}
    \begin{aligned}
        \delta W &= \mu \left(X_{26} X_{62} - X_{15} X_{53}^1 X_{31}\right)
        \\[0.5em]
        &\quad X_{31} \mapsto -\frac{1}{\mu} X_{31} + \frac{1}{\mu} X_{36} X_{61}
        \qquad X_{53}^1 \mapsto \frac{1}{\mu} X_{53}^1 - \frac{1}{\mu} X_{53}^2 \\ 
        &\quad X_{45} \mapsto \frac{1}{\mu} X_{45} - \frac{1}{\mu} X_{42} X_{25}
        \\[0.5em]
        W' &= X_{15} X_{53}^1 X_{31} + X_{34} X_{45} X_{53}^2 + X_{12} X_{25} X_{56} X_{61} + X_{23} X_{36} X_{64} X_{42} \\ 
        & \qquad - X_{12} X_{23} X_{31} - X_{45} X_{56} X_{64} - X_{15} X_{53}^2 X_{36} X_{61} - X_{25} X_{53}^1 X_{34} X_{42}
    \end{aligned}
\end{align}

\subsubsection{\texorpdfstring{$\text{PdP}_{3b}$}{PdP3b} to \texorpdfstring{$\text{dP}_{3}$}{dP3}}
\label{sec:PdP3btodP3}

\subsubsection*{Phase A to Phase A}

\begin{align}
    \begin{aligned}
        W &= X_{12} X_{26} X_{61} + X_{14} X_{42} X_{21} + X_{25} X_{53} X_{32} + X_{13} X_{34} X_{46} X_{65} X_{51} \\ 
        & \qquad - X_{12} X_{25} X_{51} - X_{13} X_{32} X_{21} - X_{14} X_{46} X_{61} - X_{26} X_{65} X_{53} X_{34} X_{42}
    \end{aligned}
\end{align}

\begin{align}
    \begin{aligned}
        \delta W &= \mu \left(X_{12} X_{21} - X_{34} X_{46} X_{65} X_{53}\right)
        \\[0.5em]
        &\quad X_{53} \mapsto -\frac{1}{\mu} X_{53} + \frac{1}{\mu} X_{51} X_{13}
        \qquad X_{46} \mapsto \frac{1}{\mu} X_{46} - \frac{1}{\mu} X_{42} X_{26} 
        \\[0.5em]
        W' &= X_{13} X_{32} X_{26} X_{61} + X_{14} X_{42} X_{25} X_{51} + X_{34} X_{46} X_{65} X_{53} \\ 
        & \qquad - X_{14} X_{46} X_{61} - X_{25} X_{53} X_{32} - X_{13} X_{34} X_{42} X_{26} X_{65} X_{51}
    \end{aligned}
\end{align}

\subsubsection*{Phase B to Phase B}

\begin{align}
    \begin{aligned}
        W &= X_{53} X_{32} X_{25}^2 + X_{56} X_{62} X_{25}^1 + X_{13} X_{34} X_{45} X_{51} + X_{16} X_{64} X_{42} X_{21} \\ 
        & \qquad - X_{13} X_{32} X_{21} - X_{45} X_{56} X_{64} - X_{16} X_{62} X_{25}^2 X_{51} - X_{34} X_{42} X_{25}^1 X_{53}
    \end{aligned}
\end{align}

\begin{align}
    \begin{aligned}
        \delta W &= \mu \left(X_{12} X_{21} - X_{34} X_{46} X_{65} X_{53}\right)
        \\[0.5em]
        &\quad X_{53} \mapsto -\frac{1}{\mu} X_{53} + \frac{1}{\mu} X_{51} X_{13}
        \qquad X_{45} \mapsto \frac{1}{\mu} X_{45} - \frac{1}{\mu} X_{42} X_{25}^1 \\
        &\quad X_{21} \mapsto -\frac{1}{\mu} X_{21} + \frac{1}{\mu} X_{25}^2 X_{51}
        \qquad X_{62} \mapsto \frac{1}{\mu} X_{62} - \frac{1}{\mu} X_{64} X_{42}
        \\[0.5em]
        W' &= X_{13} X_{32} X_{21} + X_{34} X_{45} X_{53} + X_{56} X_{62} X_{25}^1 + X_{16} X_{64} X_{42} X_{25}^2 X_{51} \\ 
        & \qquad - X_{16} X_{62} X_{21} - X_{45} X_{56} X_{64} - X_{53} X_{32} X_{25}^2 - X_{13} X_{34} X_{42} X_{25}^1 X_{51}
    \end{aligned}
\end{align}

\subsubsection*{Phase C to Phase C}

\begin{align}
    \begin{aligned}
        W &= X_{13} X_{35} X_{51} + X_{16} X_{62}^2 X_{21} + X_{24} X_{43} X_{32}^2 + X_{53} X_{32}^1 X_{25}^2 \\ 
        & \qquad + X_{46} X_{62}^1 X_{25}^1 X_{54} - X_{13} X_{32}^1 X_{21} - X_{24} X_{46} X_{62}^2 - X_{35} X_{54} X_{43} \\ 
        & \qquad - X_{53} X_{32}^2 X_{25}^1 - X_{16} X_{62}^1 X_{25}^2 X_{51}
    \end{aligned}
\end{align}

\begin{align}
    \begin{aligned}
        \delta W &= \mu \left(X_{12} X_{21} - X_{34} X_{46} X_{65} X_{53}\right)
        \\[0.5em]
        &\quad X_{21} \mapsto \frac{1}{\mu} X_{21} - \frac{1}{\mu} X_{25}^2 X_{51}
        \qquad X_{62}^1 \mapsto -\frac{1}{\mu} X_{62}^1 + \frac{1}{\mu} X_{62}^2 \\ 
        &\quad X_{24} \mapsto -\frac{1}{\mu} X_{24} + \frac{1}{\mu} X_{25}^1 X_{54}
        \\[0.5em]
        W' &= X_{16} X_{62}^1 X_{21} + X_{24} X_{46} X_{62}^2 + X_{13} X_{32}^2 X_{25}^1 X_{51} + X_{43} X_{32}^1 X_{25}^2 X_{54} \\ 
        & \qquad - X_{13} X_{32}^1 X_{21} - X_{24} X_{43} X_{32}^2 - X_{16} X_{62}^2 X_{25}^2 X_{51} - X_{46} X_{62}^1 X_{25}^1 X_{54}
    \end{aligned}
\end{align}

\subsubsection{\texorpdfstring{$\text{PdP}_{4b}$}{PdP4b} to \texorpdfstring{$\text{PdP}_{4a}$}{PdP4a}}
\label{sec:PdP4btoPdP4a}

\begin{align}
    \begin{aligned}
        W &= X_{16} X_{67} X_{71} + X_{17} X_{72} X_{21} + X_{25} X_{56} X_{62} + X_{26} X_{64} X_{42} \\ 
        & \qquad + X_{37} X_{75} X_{53} + X_{13} X_{34} X_{45} X_{51} - X_{13} X_{37} X_{71} - X_{16} X_{62} X_{21} \\ 
        & \qquad - X_{17} X_{75} X_{51} - X_{26} X_{67} X_{72} - X_{45} X_{56} X_{64} - X_{25} X_{53} X_{34} X_{42}
    \end{aligned}
\end{align}

\subsubsection*{Phase A to Phase A}

\begin{align}
    \begin{aligned}
        \delta W &= \mu \left(X_{17} X_{71} - X_{26} X_{62} \right)
        \\[0.5em]
        &\quad X_{45} \mapsto -\frac{1}{\mu} X_{45} + \frac{1}{\mu} X_{42} X_{25}
        \qquad X_{53} \mapsto -\frac{1}{\mu} X_{53} + \frac{1}{\mu} X_{51} X_{13}
        \\[0.5em]
        W' &= X_{45} X_{56} X_{64} + X_{13} X_{37} X_{72} X_{21} + X_{16} X_{67} X_{75} X_{51} + X_{25} X_{53} X_{34} X_{42} \\ 
        & \qquad - X_{37} X_{75} X_{53} - X_{13} X_{34} X_{45} X_{51} - X_{16} X_{64} X_{42} X_{21} - X_{25} X_{56} X_{67} X_{72}
    \end{aligned}
\end{align}

\subsubsection*{Phase A to Phase B}

\begin{align}
    \begin{aligned}
        \delta W &= \mu \left(X_{26} X_{62} - X_{34} X_{45} X_{53} \right)
        \\[0.5em]
        &\quad X_{45} \mapsto \frac{1}{\mu} X_{45} - \frac{1}{\mu} X_{42} X_{25}
        \qquad X_{53} \mapsto -\frac{1}{\mu} X_{53} + \frac{1}{\mu} X_{51} X_{13} \\
        &\quad X_{17} \mapsto -\frac{1}{\mu} X_{17} + \left( \frac{1}{\mu} - \beta_1 \right) X_{13} X_{37} + \beta_1\, X_{16} X_{67} \\
        &\quad X_{71} \mapsto -\frac{1}{\mu} X_{71} + \left( \frac{1}{\mu} - \beta_1 \right) X_{72} X_{21} + \beta_1\, X_{75} X_{51}
        \\[0.5em]
        W' &= X_{13} X_{37} X_{71} + X_{17} X_{75} X_{51} + X_{34} X_{45} X_{53} + X_{16} X_{64} X_{42} X_{21} \\ 
        & \qquad + X_{25} X_{56} X_{67} X_{72} - X_{16} X_{67} X_{71} - X_{17} X_{72} X_{21} - X_{37} X_{75} X_{53} \\ 
        & \qquad - X_{45} X_{56} X_{64} - X_{13} X_{34} X_{42} X_{25} X_{51}
    \end{aligned}
\end{align}

\begin{align}
    \begin{aligned}
        \delta W &= \mu \left(X_{17} X_{71} - X_{34} X_{45} X_{53} \right)
        \\[0.5em]
        &\quad X_{45} \mapsto \frac{1}{\mu} X_{45} - \frac{1}{\mu} X_{42} X_{25}
        \qquad X_{53} \mapsto -\frac{1}{\mu} X_{53} + \frac{1}{\mu} X_{51} X_{13} \\
        &\quad X_{26} \mapsto \frac{1}{\mu} X_{26} + \left( - \frac{1}{\mu} - \beta_1 \right) X_{21} X_{16} + \beta_1\, X_{25} X_{56} \\
        &\quad X_{62} \mapsto \frac{1}{\mu} X_{62} + \left( - \frac{1}{\mu} - \beta_1 \right) X_{64} X_{42} + \beta_1\, X_{67} X_{72}
        \\[0.5em]
        W' &= X_{25} X_{56} X_{62} + X_{26} X_{64} X_{42} + X_{34} X_{45} X_{53} + X_{13} X_{37} X_{72} X_{21} \\ 
        & \qquad + X_{16} X_{67} X_{75} X_{51} - X_{16} X_{62} X_{21} - X_{26} X_{67} X_{72} - X_{37} X_{75} X_{53} \\ 
        & \qquad - X_{45} X_{56} X_{64} - X_{13} X_{34} X_{42} X_{25} X_{51}
    \end{aligned}
\end{align}

\subsubsection{\texorpdfstring{$\coset{\bbC^3}{(\bbZ_4 \times \bbZ_2 )} \,(1,0,3)(0,1,1)$}{C3/(Z4*Z2) (1,0,3)(0,1,1)} to \texorpdfstring{$\coset{L^{1,3,1}}{\bbZ_2} \,(0,1,1,1)$}{L131/Z2 (0,1,1,1)}}
\label{sec:C3Z4Z2toL131Z2}

\subsubsection*{Phase A to Phase A}

\begin{align}
    \begin{aligned}
        W &= X_{12} X_{28} X_{81} + X_{14} X_{43} X_{31} + X_{17} X_{72} X_{21} + X_{23} X_{34} X_{42} \\ 
        & \qquad + X_{36} X_{65} X_{53} + X_{45} X_{56} X_{64} + X_{58} X_{87} X_{75} + X_{67} X_{78} X_{86} \\ 
        & \qquad - X_{12} X_{23} X_{31} - X_{14} X_{42} X_{21} - X_{17} X_{78} X_{81} - X_{28} X_{87} X_{72} \\ 
        & \qquad - X_{34} X_{45} X_{53} - X_{36} X_{64} X_{43} - X_{56} X_{67} X_{75} - X_{58} X_{86} X_{65}
    \end{aligned}
\end{align}

\begin{align}
    \begin{aligned}
        \delta W &= \mu \left( X_{12} X_{21} - X_{34} X_{43} \right)
        \\[0.5em]
        &\quad X_{56} \mapsto \frac{1}{\mu} X_{56} + \left( - \frac{1}{\mu} - \beta_1 \right) X_{58} X_{86} + \beta_1\, X_{53} X_{36} \\
        &\quad X_{65} \mapsto \frac{1}{\mu} X_{65} + \left( - \frac{1}{\mu} - \beta_1 \right) X_{64} X_{45} + \beta_1\, X_{67} X_{75} \\
        &\quad X_{78} \mapsto \frac{1}{\mu} X_{78} + \left( - \frac{1}{\mu} - \beta_2 \right) X_{72} X_{28} + \beta_2\, X_{75} X_{58} \\
        &\quad X_{87} \mapsto \frac{1}{\mu} X_{87} + \left( - \frac{1}{\mu} - \beta_2 \right) X_{86} X_{67} + \beta_2\, X_{81} X_{17}
        \\[0.5em]
        W' &= X_{36} X_{65} X_{53} + X_{45} X_{56} X_{64} + X_{58} X_{87} X_{75} + X_{67} X_{78} X_{86} \\ 
        & \qquad + X_{14} X_{42} X_{28} X_{81} + X_{17} X_{72} X_{23} X_{31} - X_{17} X_{78} X_{81} - X_{28} X_{87} X_{72} \\ 
        & \qquad - X_{56} X_{67} X_{75} - X_{58} X_{86} X_{65} - X_{14} X_{45} X_{53} X_{31} - X_{23} X_{36} X_{64} X_{42}
    \end{aligned}
\end{align}

\begin{align}
    \begin{aligned}
        \delta W &= \mu \left(X_{34} X_{43} - X_{54} X_{43} \right)
        \\[0.5em]
        &\quad X_{12} \mapsto \frac{1}{\mu} X_{12} + \left( - \frac{1}{\mu} - \beta_1 \right) X_{17} X_{72} + \beta_1\, X_{14} X_{42} \\
        &\quad X_{21} \mapsto \frac{1}{\mu} X_{21} + \left( - \frac{1}{\mu} - \beta_1 \right) X_{23} X_{31} + \beta_1\, X_{28} X_{81} \\
        &\quad X_{78} \mapsto \frac{1}{\mu} X_{78} + \left( - \frac{1}{\mu} - \beta_2 \right) X_{75} X_{58} + \beta_2\, X_{72} X_{28} \\
        &\quad X_{87} \mapsto \frac{1}{\mu} X_{87} + \left( - \frac{1}{\mu} - \beta_2 \right) X_{81} X_{17} + \beta_2\, X_{86} X_{67}
        \\[0.5em]
        W' &= X_{12} X_{28} X_{81} + X_{17} X_{72} X_{21} + X_{58} X_{87} X_{75} + X_{67} X_{78} X_{86} \\ 
        & \qquad + X_{14} X_{45} X_{53} X_{31} + X_{23} X_{36} X_{64} X_{42} - X_{12} X_{23} X_{31} - X_{14} X_{42} X_{21} \\ 
        & \qquad - X_{17} X_{78} X_{81} - X_{28} X_{87} X_{72} - X_{36} X_{67} X_{75} X_{53} - X_{45} X_{58} X_{86} X_{64}
    \end{aligned}
\end{align}

\begin{align}
    \begin{aligned}
        \delta W &= \mu \left(X_{56} X_{65} - X_{78} X_{87} \right)
        \\[0.5em]
        &\quad X_{12} \mapsto \frac{1}{\mu} X_{12} + \left( - \frac{1}{\mu} - \beta_1 \right) X_{14} X_{42} + \beta_1\, X_{17} X_{72} \\
        &\quad X_{21} \mapsto \frac{1}{\mu} X_{21} + \left( - \frac{1}{\mu} - \beta_1 \right) X_{28} X_{81} + \beta_1\, X_{23} X_{31} \\
        &\quad X_{34} \mapsto \frac{1}{\mu} X_{34} + \left( - \frac{1}{\mu} - \beta_2 \right) X_{31} X_{14} + \beta_2\, X_{36} X_{64} \\
        &\quad X_{43} \mapsto \frac{1}{\mu} X_{43} + \left( - \frac{1}{\mu} - \beta_2 \right) X_{45} X_{53} + \beta_2\, X_{42} X_{23}
        \\[0.5em]
        W' &= X_{12} X_{28} X_{81} + X_{14} X_{43} X_{31} + X_{17} X_{72} X_{21} + X_{23} X_{34} X_{42} \\ 
        & \qquad + X_{36} X_{67} X_{75} X_{53} + X_{45} X_{58} X_{86} X_{64} - X_{12} X_{23} X_{31} - X_{14} X_{42} X_{21} \\ 
        & \qquad - X_{34} X_{45} X_{53} - X_{36} X_{64} X_{43} - X_{17} X_{75} X_{58} X_{81} - X_{28} X_{86} X_{67} X_{72}
    \end{aligned}
\end{align}

\begin{align}
    \begin{aligned}
        \delta W &= \mu \left( X_{78} X_{87} - X_{12} X_{21} \right)
        \\[0.5em]
        &\quad X_{34} \mapsto \frac{1}{\mu} X_{34} + \left( - \frac{1}{\mu} - \beta_1 \right) X_{36} X_{64} + \beta_1\, X_{31} X_{14} \\
        &\quad X_{43} \mapsto \frac{1}{\mu} X_{43} + \left( - \frac{1}{\mu} - \beta_1 \right) X_{42} X_{23} + \beta_1\, X_{45} X_{53} \\
        &\quad X_{56} \mapsto \frac{1}{\mu} X_{56} + \left( - \frac{1}{\mu} - \beta_2 \right) X_{53} X_{36} + \beta_2\, X_{58} X_{86} \\
        &\quad X_{65} \mapsto \frac{1}{\mu} X_{65} + \left( - \frac{1}{\mu} - \beta_2 \right) X_{67} X_{75} + \beta_2\, X_{64} X_{45}
        \\[0.5em]
        W' &= X_{14} X_{43} X_{31} + X_{23} X_{34} X_{42} + X_{36} X_{65} X_{53} + X_{45} X_{56} X_{64} \\ 
        & \qquad + X_{17} X_{75} X_{58} X_{81} + X_{28} X_{86} X_{67} X_{72} - X_{34} X_{45} X_{53} - X_{36} X_{64} X_{43} \\ 
        & \qquad - X_{56} X_{67} X_{75} - X_{58} X_{86} X_{65} - X_{14} X_{42} X_{28} X_{81} - X_{17} X_{72} X_{23} X_{31}
    \end{aligned}
\end{align}

\subsubsection{\texorpdfstring{$\coset{\bbC^3}{(\bbZ_4 \times \bbZ_2 )} \,(1,0,3)(0,1,1)$}{C3/(Z4*Z2) (1,0,3)(0,1,1)} to \texorpdfstring{$\text{PdP}_{5}, \coset{\calC}{\bbZ_2 \times \bbZ_2} \,(1,0,0,1)(0,1,1,0)$}{PdP5, Conifold/(Z2*Z2) (1,0,0,1)(0,1,1,0)}}
\label{sec:C3Z4Z2toPdP5}

\begin{align}
    \begin{aligned}
        W &= X_{12} X_{28} X_{81} + X_{14} X_{43} X_{31} + X_{17} X_{72} X_{21} + X_{23} X_{34} X_{42} \\ 
        & \qquad + X_{36} X_{65} X_{53} + X_{45} X_{56} X_{64} + X_{58} X_{87} X_{75} + X_{67} X_{78} X_{86} \\ 
        & \qquad - X_{12} X_{23} X_{31} - X_{14} X_{42} X_{21} - X_{17} X_{78} X_{81} - X_{28} X_{87} X_{72} \\ 
        & \qquad - X_{34} X_{45} X_{53} - X_{36} X_{64} X_{43} - X_{56} X_{67} X_{75} - X_{58} X_{86} X_{65}
    \end{aligned}
\end{align}

\subsubsection*{Phase A to Phase A}

\begin{align}
    \begin{aligned}
        \delta W &= \mu \left(X_{12} X_{21} - X_{34} X_{43} + X_{56} X_{65} - X_{78} X_{87} \right)
        \\[0.5em]
        W' &= X_{14} X_{42} X_{28} X_{81} + X_{17} X_{72} X_{23} X_{31} + X_{36} X_{67} X_{75} X_{53} \\ 
        & \qquad + X_{45} X_{58} X_{86} X_{64} - X_{14} X_{45} X_{53} X_{31} - X_{17} X_{75} X_{58} X_{81} \\ 
        & \qquad - X_{23} X_{36} X_{64} X_{42} - X_{28} X_{86} X_{67} X_{72}
    \end{aligned}
\end{align}

\subsubsection*{Phase A to Phase B}

\begin{align}
    \begin{aligned}
        \delta W &= \mu \left(X_{12} X_{21} - X_{56} X_{65}\right)
        \\[0.5em]
        &\quad X_{34} \mapsto -\frac{1}{\mu} X_{34} + \left(\frac{1}{\mu} - \beta_1 \right) X_{31} X_{14} + \beta_1\, X_{36} X_{64} \\
        &\quad X_{43} \mapsto -\frac{1}{\mu} X_{43} + \left(\frac{1}{\mu} - \beta_1 \right) X_{45} X_{53} + \beta_1\, X_{42} X_{23} \\
        &\quad X_{78} \mapsto \frac{1}{\mu} X_{78} + \left( - \frac{1}{\mu} - \beta_2 \right) X_{72} X_{28} + \beta_2\, X_{75} X_{58} \\
        &\quad X_{87} \mapsto \frac{1}{\mu} X_{87} + \left( - \frac{1}{\mu} - \beta_2 \right) X_{86} X_{67} + \beta_2\, X_{81} X_{17}
        \\[0.5em]
        W' &= X_{34} X_{45} X_{53} + X_{36} X_{64} X_{43} + X_{58} X_{87} X_{75} + X_{67} X_{78} X_{86} \\ 
        & \qquad + X_{14} X_{42} X_{28} X_{81} + X_{17} X_{72} X_{23} X_{31} - X_{14} X_{43} X_{31} - X_{17} X_{78} X_{81} \\ 
        & \qquad - X_{23} X_{34} X_{42} - X_{28} X_{87} X_{72} - X_{36} X_{67} X_{75} X_{53} - X_{45} X_{58} X_{86} X_{64}
    \end{aligned}
\end{align}

\begin{align}
    \begin{aligned}
        \delta W &= \mu \left(X_{34} X_{43} - X_{78} X_{87}\right)
        \\[0.5em]
        &\quad X_{12} \mapsto \frac{1}{\mu} X_{12} + \left( -\frac{1}{\mu} - \beta_1 \right) X_{14} X_{42} + \beta_1\, X_{17} X_{72} \\
        &\quad X_{21} \mapsto \frac{1}{\mu} X_{21} + \left( -\frac{1}{\mu} - \beta_1 \right) X_{28} X_{81} + \beta_1\, X_{23} X_{31} \\
        &\quad X_{56} \mapsto -\frac{1}{\mu} X_{56} + \left( \frac{1}{\mu} - \beta_2 \right) X_{53} X_{36} + \beta_2\, X_{58} X_{86} \\
        &\quad X_{65} \mapsto -\frac{1}{\mu} X_{65} + \left( \frac{1}{\mu} - \beta_2 \right) X_{67} X_{75} + \beta_2\, X_{64} X_{45}
        \\[0.5em]
        W' &= X_{12} X_{28} X_{81} + X_{17} X_{72} X_{21} + X_{56} X_{67} X_{75} + X_{58} X_{86} X_{65} \\ 
        & \qquad + X_{14} X_{45} X_{53} X_{31} + X_{23} X_{36} X_{64} X_{42} - X_{12} X_{23} X_{31} - X_{14} X_{42} X_{21} \\ 
        & \qquad - X_{36} X_{65} X_{53} - X_{45} X_{56} X_{64} - X_{17} X_{75} X_{58} X_{81} - X_{28} X_{86} X_{67} X_{72}
    \end{aligned}
\end{align}

\subsection{Deformations to marginal deformations of toric models}
\label{sec:toric-to-marginal-list}

Listed here are relevant zig-zag deformations of toric models for which the endpoint of the RG flow is not toric.
Instead, the IR models in this section are described by a toric superpotential plus an exactly marginal zig-zag deformation.
In each subsection, we present the original toric superpotential $W$.
Each deformation is in a unique equation block with:
\begin{enumerate}[1)]
    \item The deformation $\delta W = \mu \calO_{\eta}$, associated to a zig-zag path $\eta$.
    \item The field redefinitions of the type in \cref{eq:field_redef}, if any.
    \item The superpotential $W'$ representing the IR theory, which is of the form $W_{\mathrm{toric}}' + \frac{1}{\mu} \calO_{\eta'}'$, where $R_{sc}\big[\calO_{\eta'}' \big] = 2$. The superpotential $W_{\mathrm{toric}}'$ defines a brane tiling model, and $\calO_{\eta'}'$ corresponds to the zig-zag operator of $W_{\mathrm{toric}}'$ for the reversed zig-zag path $\eta'$. 
\end{enumerate}

\subsubsection{\texorpdfstring{$\coset{L^{1,3,1}}{\bbZ_2} \,(0,1,1,1)$}{L131/Z2 (0,1,1,1)} to \texorpdfstring{$\text{PdP}_{5}, \coset{\calC}{\bbZ_2 \times \bbZ_2} \,(1,0,0,1)(0,1,1,0)$}{PdP5, Conifold/(Z2*Z2) (1,0,0,1)(0,1,1,0)}}
\label{sec:L131Z2toPdP5}

\subsubsection*{Phase A to Phase A}

\begin{align}
    \begin{aligned}
        W &= X_{17} X_{78} X_{81} + X_{18} X_{83} X_{31} + X_{27} X_{73} X_{32} + X_{37} X_{75} X_{53} \\ 
        & \qquad + X_{14} X_{45} X_{56} X_{61} + X_{24} X_{48} X_{86} X_{62} - X_{14} X_{48} X_{81} - X_{17} X_{73} X_{31} \\ 
        & \qquad - X_{18} X_{86} X_{61} - X_{37} X_{78} X_{83} - X_{24} X_{45} X_{53} X_{32} - X_{27} X_{75} X_{56} X_{62}
    \end{aligned}
\end{align}

\begin{align}
    \begin{aligned}
        \delta W &= \mu \left(X_{37} X_{73} - X_{18} X_{81}\right)
        \\[0.5em]
        &\quad X_{17} \mapsto \mu X_{17} \\
        &\quad X_{83} \mapsto \mu X_{83}
        \\[0.5em]
        W' &= X_{14} X_{45} X_{56} X_{61} + X_{17} X_{75} X_{53} X_{31} + X_{24} X_{48} X_{86} X_{62} + X_{27} X_{78} X_{83} X_{32} \\ 
        & \quad - X_{14} X_{48} X_{83} X_{31} - X_{17} X_{78} X_{86} X_{61} - X_{24} X_{45} X_{53} X_{32} - X_{27} X_{75} X_{56} X_{62} \\
        & \qquad +\frac{1}{\mu} \left( X_{14} X_{48} X_{86} X_{61} - X_{27} X_{75} X_{53} X_{32} \right)
    \end{aligned}
\end{align}

\subsubsection*{Phase A to Phase B}

\begin{align}
    \begin{aligned}
        W &= X_{17} X_{78} X_{81} + X_{18} X_{83} X_{31} + X_{27} X_{73} X_{32} + X_{37} X_{75} X_{53} \\ 
        & \qquad + X_{14} X_{45} X_{56} X_{61} + X_{24} X_{48} X_{86} X_{62} - X_{14} X_{48} X_{81} - X_{17} X_{73} X_{31} \\ 
        & \qquad - X_{18} X_{86} X_{61} - X_{37} X_{78} X_{83} - X_{24} X_{45} X_{53} X_{32} - X_{27} X_{75} X_{56} X_{62}
    \end{aligned}
\end{align}

\begin{align}
    \begin{aligned}
        \delta W &= \mu \left( X_{18} X_{81} - X_{24} X_{45} X_{56} X_{62} \right) 
        \\[0.5em]
        &\quad X_{61} \mapsto \mu X_{61} \\
        &\quad X_{48} \mapsto \mu X_{48}
        \\[0.5em]
        W' &= X_{24} X_{46} X_{62} + X_{27} X_{73} X_{32} + X_{37} X_{75} X_{53} + X_{45} X_{56} X_{64} \\ 
        & \qquad + X_{14} X_{48} X_{83} X_{31} + X_{17} X_{78} X_{86} X_{61} - X_{14} X_{46} X_{61} - X_{17} X_{73} X_{31} \\ 
        & \qquad - X_{37} X_{78} X_{83} - X_{48} X_{86} X_{64} - X_{24} X_{45} X_{53} X_{32} - X_{27} X_{75} X_{56} X_{62} \\
        & \qquad +\frac{1}{\mu} \left( X_{46} X_{64} - X_{17} X_{78} X_{83} X_{31} \right)
    \end{aligned}
\end{align}

\subsubsection*{Phase B to Phase C}

\begin{align}
    \begin{aligned}
        W &= X_{17} X_{78} X_{81} + X_{18} X_{83} X_{31} + X_{23} X_{34} X_{42} + X_{26} X_{67} X_{72} + X_{37} X_{75} X_{53} \\ 
        & \quad + X_{48} X_{86} X_{64} + X_{14} X_{45} X_{56} X_{61} - X_{14} X_{48} X_{81} - X_{18} X_{86} X_{61} - X_{26} X_{64} X_{42} \\ 
        & \quad - X_{34} X_{45} X_{53} - X_{37} X_{78} X_{83} - X_{56} X_{67} X_{75} - X_{17} X_{72} X_{23} X_{31}
    \end{aligned}
\end{align}

\begin{align}
    \begin{aligned}
        \delta W &= \mu \left( X_{18} X_{81} - X_{45} X_{56} X_{64} \right)
        \\[0.5em]
        &\quad X_{14} \mapsto \mu X_{14} \\
        &\quad X_{86} \mapsto \mu X_{86} \\
        &\quad X_{64} \mapsto \frac{1}{\mu} X_{64} + X_{61} X_{14} 
        \\[0.5em]
        W' &= X_{23} X_{34} X_{42} + X_{26} X_{67} X_{72} + X_{37} X_{75} X_{53} + X_{45} X_{56} X_{64} \\ 
        & \qquad + X_{14} X_{48} X_{83} X_{31} + X_{17} X_{78} X_{86} X_{61} - X_{34} X_{45} X_{53} - X_{37} X_{78} X_{83} \\ 
        & \qquad - X_{48} X_{86} X_{64} - X_{56} X_{67} X_{75} - X_{14} X_{42} X_{26} X_{61} - X_{17} X_{72} X_{23} X_{31} \\
        & \qquad +\frac{1}{\mu} \left( X_{26} X_{64} X_{42} - X_{17} X_{78} X_{83} X_{31} \right)
    \end{aligned}
\end{align}

%% file: Appendices/QuiverReps.tex

\section{Moduli spaces from quiver representations}
\label{sec:quiver-reps-moduli}

\subsection{Quiver representations and path algebras}

A quiver $\calQ$ \cite{Soibelman:2019lect} is a multidigraph, a directed graph where multiple edges between two vertices and loops are allowed.
The data defining the quiver $\calQ$ consists of a set of vertices $\calV$, a set of edges $\calE$, and two mappings $s, t : \calE \to \cal V$ defined as the ``source'' and ``tail'' of an edge (defined in \cref{sec:brane-tiling-dict}).
Formally, $\calQ$ is the ordered tuple $(\calV, \calE, s, t)$.

A representation $V$ of a quiver $\calQ$ is a collection of $\bbC$-vector spaces $\set{V_i}_{i\in\calV}$ and a collection of linear maps $\set{\phi_e : V_{t(e)} \to V_{s(e)}}_{e\in\calE}$. 
A finite-dimensional representation $V$ has a dimension vector $\alpha \in (\bbZ_{n\ge 0})^G$, where $\alpha_i = \dim V_i$.
We denote the set of representations of $\calQ$ with dimension vector $\alpha$ by $\Rep(\calQ; \alpha)$.

Many of the notions of linear algebra can be extended to quiver representations.
In particular, given two representations $V = (\set{V_i}_{i\in\calV}, \set{\phi_e}_{e\in\calE})$ and $W = (\set{W_i}_{i\in\calV}, \set{\psi_e}_{e\in\calE})$, a morphism between quiver representations $f: S \to R$ is a collection of linear maps $f_i: W_i \to V_i$ such that $f_{s(e)} \,\psi_e = \phi_e \,f_{t(e)}$ for all edges $e\in\calE$.
Two representations $V$ and $W$ are isomorphic if $f$ is bijective.
A \emph{subrepresentation} $W \subset V$ of a representation of $\calQ$ consists of subspaces $W_i \subset V_i$ and arrows $\set{\psi_e}_{e\in\calE}$ such that $\phi_e|_{W_{t(e)}}  = \psi_e$. 
In this case, we can find an injective morphism $f$ between representations.

For 4d $\calN=1$ quiver gauge theories, we can view $\Rep(\calQ; \alpha)$ as the $\bbC$-valued affine space ($V_i \cong \bbC^{\alpha_i}$), 
\begin{align}
    \Rep(\calQ, \alpha) \cong \prod_{e\in\calE} \Hom\big(\bbC^{\alpha_{t(e)}}, \bbC^{\alpha_{s(e)}}\big) ~,
\end{align}
where points represent VEVs of the chiral bifundamentals $\set{X_e}_{e\in\calE}$.
The complexified gauge group $G(\alpha) = \prod_{i\in\calV} GL(\alpha_i, \bbC)$ produces a natural group action on $V \in \Rep(\calQ, \alpha)$,
\begin{align}
    g \cdot V = \setst{ g_{t(e)} \phi_e \mkern2mu g_{s(e)}^{-1} }{e\in\calE} ~,
\end{align}
for $g = (g_i) \in G(\alpha)$.
Orbits of $G(\alpha)$ are isomorphism classes of quiver representations. 

To include the F-term equations that describe the singular geometry of $\calY$ it is simpler to look at quiver path algebra and its subalgebras.
The quiver path algebra $\bbC\calQ$ is an associative $\bbC$-algebra with elements $u_{p}$ for all paths $p = e_1 e_2 \dots e_\ell$ in the quiver $\calQ$ and a multiplication rule
\begin{align}
    u_{p} u_{q} =
    \begin{cases}
        u_{p q} & t(p) = s(q) \\
        0       & \text{otherwise}
    \end{cases}
    ~ .
\end{align}
For each vertex $i\in\calV$ there is a trivial path (no arrow) denoted by $p_i$ such that $s(p_i) = t(p_i) = i$, with algebra element $u_{i} \equiv u_{p_i}$. 
The identity element of the path algebra is $\mathbbl{1}_{\calQ} = \sum_{i\in\calV} u_{i}$.

The abelian category of representations of $\calQ$ is the same as the category of modules over the path algebra, $\bbC \calQ$-Mod, i.e. the category of finite dimensional representations of $\bbC \calQ$.
For a quiver representation $V$, a path $p = e_1 \dots e_{\ell}$ defines a linear map
$\phi_{p} : V_{t(e)} \to V_{s(e)}$, given by $\phi_{p} = \phi_{e_1} \,\phi_{e_2} \dots \phi_{e_\ell}$.
Thus, we can trivially define a $\bbC\calQ$-module structure on $\bigoplus_{i\in\calV} V_i$, meaning
\begin{align}
    u_p v =
    \begin{cases}
        \phi_p(v) & v \in V_{t(p)} \\
        0         & \text{otherwise}
    \end{cases}
    ~ .
\end{align}
On the other hand, if $V$ is a $\bbC\calQ$-module, we can define the vector subspaces $V_i = u_i V$, since the trivial path elements $u_i$ are idempotent and act as a projection.
Additionally, for each edge $e\in\calE$ we can define field expectations $\phi_e : V_{t(e)} \to V_{s(e)}$ as $\phi_e(v) = u_e v$.
Therefore, $(\set{V_i}, \set{\phi_e})$ is a representation of $\calQ$.

The F-term equations come into play in the form of the path subalgebra
\begin{align}
    \calA = \coset{\bbC\calQ}{ \angl*{\partial_X W|_{\set{u_e} }} } ~,
\end{align}
\emph{i.e.} we quotient the path algebra $\bbC \calQ$ by the (Jacobian) ideal generated by open paths relations in the quiver of the form $u_{p_i} = u_{q_i}$ for two open paths $p_i, q_i$ in the quiver, provided by the F-terms.
From the same equivalency established above, the representations in $\Rep(\calQ; \alpha)$ associated to $\calA$-modules form a closed $G(\alpha)$-invariant subvariety that we can denote by $\calZ_\calA(\alpha)$.

\subsection{Kempf–Ness theorem and the Proj GIT quotient}

We would like to construct the moduli space as a quotient of a $\calG$-action, that relates to the K\"{a}hler reduction for a nonzero level $\xi$ \cite{Soibelman:2019lect,Closset:2012ep}.
For that we need to use the \emph{Proj GIT quotient} description of a variety, which can be physically interpreted as the VEVs of chiral superfields that obey the F-term conditions, modded out by the action of the complexified gauge group.
We consider the abelian $U(1)^G$ quiver, so we take the subspaces dimensions as $\alpha_i = 1$ for all $1 \le i\le G$ in the previous description. The reductive group $G(\alpha)$ acting on an affine algebraic variety $\calZ \equiv \calZ_\calA(\alpha)$ is the complexified gauge group $\calG_\bbC = (\bbC^\times)^G$.
We can extend the group action on the line bundle $\calZ \times \bbC$ using a
character $\chi_\theta : \calG_\bbC \mapsto \bbC^\times$ by
\begin{align}
    \begin{aligned}
        \calG_\bbC \times (\calZ \times \bbC) &\to \calZ \times \bbC \\
        g \cdot (z, t) &\mapsto (g \cdot z, \,\chi_\theta(g) \,t)
    \end{aligned}
    \quad,\quad\text{with}\quad
    \chi_\theta(g) = \prod_{i=1}^G (g_i)^{-\theta_i}
    ~ .
\end{align}
Note that $\theta\in\bbZ^G \cong \Hom\left((\bbC^\times)^G, \bbC^\times\right)$.
In the algebra of polynomial functions of the line bundle, $\bbC[\calZ\times\bbC]$, the group acts 
as $g \cdot f(z, t) = f(g^{-1} \cdot z, \,\chi_\theta(g)^{-1}\, t)$, which can be rewritten as
\begin{align}
    g \cdot f(z, t) = g \cdot \sum_{n \ge 0} f_n(z) t^n = \sum_{n \ge 0} f_n(g^{-1} \cdot z) \,\chi_\theta(g)^{-n}\, t^n ~.
\end{align}
Thus, for $f$ to be $\calG_\bbC$-invariant, we must have $g \cdot f_n$ = $\chi_\theta(g)^n \,f_n$ for all $n$. 
We call such functions $f_n\in\bbC[\calZ]$ as $\chi_\theta^n$-semi-invariants, which defines a $\bbZ_{n\ge 0}$-grading on the ring of polynomials of the fiber bundle
\begin{align}
    R = \bbC[\calZ \times \bbC]^{\calG_\bbC(\theta)} = \bigoplus_{n \ge 0} \bbC[\calZ]^{\chi_\theta^n} ~,
\end{align}
where $\bbC[\calZ]^{\chi_\theta^n}$ is a ring of semi-invariants.

The moduli space of quiver representations $\calM(Q,W; \theta)_\text{GIT}$ is then defined as the Proj GIT quotient
\begin{align}
    \ccoset{\calZ}{\chi_\theta}{\calG_\bbC} = \Proj \bbC[\calZ \times \bbC]^{\calG_\bbC(\theta)}
\end{align}
The closed points of $\Proj R$, for a graded algebra $R = \bigoplus_{n\ge 0} R_n$, are naturally in correspondence with the homogeneous (under the induced grading) maximal ideals that do not contain the \emph{irrelevant ideal} $R_{+} = \bigoplus_{n>0} R_n$.
This quotient is connected with the previously defined K\"{a}hler quotient (reduction) through the \emph{Kempf–Ness theorem}
\begin{align}
    \calM(Q,W; \xi)_\text{K} = \calM(Q,W; \theta)_\text{GIT} ~, \quad\text{for}\quad \xi = \theta \in \bbZ^{G} ~.
\end{align}
Note that, while the K\"{a}hler quotient is defined for real levels of the moment map, in the GIT quotient the space of characters that produces the $\bbZ_{n\ge 0}$-grading in the ring of regular functions in the line bundle of quiver representations is equivalent to $\bbZ^{G}$.

\subsection{Stability of moduli of quiver representations}
\label{sec:quiver-rep-moduli-stability}

There is a way to understand the closed points of $\ccoset{\calZ}{\chi_\theta}{\calG_\bbC}$ instead of looking at ideals of $\bbC[\calZ \times \bbC]^{\calG_\bbC(\theta)}$, using the idea of $\chi_\theta$-semistability \cite{Soibelman:2019lect,King:1994}.
A point $z \in \calZ$ is \emph{$\chi_\theta$-semistable} if for some $n>0$ there exists a nonvanishing semi-invariant $f \in \bbC[\calZ]^{\chi_\theta^n}$ such that $z \in \calZ_f \equiv \setst{w \in \calZ}{f(w) \ne 0}$.
Thus, the set of all $\chi_\theta$-semistable points are defined as
\begin{align}
    \calZ^\text{ss}_{\chi_\theta} = \bigcup_{f\in R_{+}} \setst{z \in \calZ}{f(z) \ne 0} ~.
\end{align}
If, in addition, $\calZ_f$ is closed under the action of $\calG_\bbC$ and the stabilizer $(\calG_\bbC)_z$  is finite, then we say that $z$ is \emph{$\chi_\theta$-stable}.

Two points in $z_1, z_2 \in \calZ^\text{ss}_{\chi_\theta}$ are \emph{S-equivalent} if and only if the the intersection of the closure of the $\calG_\bbC$-orbits is nonempty in the set of semistable points, meaning that $\overline{\calG_\bbC \cdot z_1} \cap \overline{\calG_\bbC \cdot z_2} \cap \calZ^\text{ss}_{\chi_\theta} \ne \emptyset$.
For each point in $z\in \calZ$, we can define a maximal ideal $I_z = \setst{f \in R}{f(z) = 0}$, and in particular if $z$ is $\chi_\theta$-semistable then $I_z$ is also homogeneous and does not contain the irrelevant ideal $R_{+}$.
Moreover, we have that $I_{z_1} = I_{z_2}$ for two S-equivalent points $z_1, z_2$.
Therefore, the GIT quotient $\ccoset{\calZ}{\chi_\theta}{\calG_\bbC}$ is a variety whose points are in natural bijection with the S-equivalence classes of $\chi_\theta$-semistable points of $\calZ$.

We would like to better understand the characters and one-parameter subgroups of $G(\alpha)$.
A character $\chi$ is an element of $\Hom(G(\alpha), \bbC^\times)$, while one-parameter subgroups $\lambda \in \Hom(\bbC^\times, G(\alpha))$.
We can define the inner-product
\begin{align}
    \angl{ \cdot, \cdot } : \Hom(\bbC^\times, G(\alpha)) \times \Hom(G(\alpha), \bbC^\times) \to \bbZ
\end{align}
This is possible since the composition of the two elements $\chi\circ\lambda \in \Hom(\bbC^\times, \bbC^\times)$, which is group-isomorphic to $\bbZ$ via the map $t^n \mapsto n$.
We can use this to simplify the criterion for semistability, which is where the Hilbert-Mumford criterion comes in.

Let $G(\alpha)$ be the reductive group acting on the affine variety $\calX$.
A point $x \in \calX$ is $\chi$-semistable if and only if for any one-parameter subgroup $\lambda$ such that $\lim_{t\to\infty} \lambda(t)\cdot x$ exists we have $\angl{ \lambda, \chi } \ge 0$.
If the inequality is strict for any nontrivial such $\lambda$, then $x$ is $\chi$-stable.
For unframed quivers like the ones we focus on, each point of $\Rep(\calQ, \alpha)$ has a nontrivial subgroup $\Gamma\cong \bbC^\times$ contained in its stabilizer, corresponding to the complexification of the trivially acting diagonal gauge $U(1)$. Therefore, we must include the condition that $\chi(\Gamma) = 1$ in the semistability condition.

Since $G(\alpha) = \prod_{i\in\calV} GL(\alpha_i, \bbC)$, a character $\chi$ can be written as a product $\prod_{i\in\calV} \chi_i$, where $\chi_i = \det^{-\theta_i}$ is a character of $GL(\alpha_i, \bbC)$ for some $\theta_i \in \bbZ$.
For simplicity, we fixed $\alpha_i = 1$, meaning that we can write $\chi(t_1, \dots, t_G) = \prod_{i=1}^{G} (t_i)^{-\theta_i}$.
Similarly, a one-parameter subgroup must be of the form $\lambda(t) = (t^{\beta_1}, \dots, t^{\beta_G})$, for $\beta \in \bbZ^G$.
So the \emph{Hilbert-Mumford criterion} can be simply be written as
\begin{align}
    \angl{ \lambda, \chi } \ge 0 \quad\Leftrightarrow\quad \theta \cdot \beta \le 0 ~.
\end{align}
Note that $\chi(\Gamma) = 1$ requires $\chi$ to be a character of $\bbP G(\alpha)$. This  is equivalent to $\theta \cdot \alpha = 0$.

Given the Hilbert-Mumford criterion, we can define $\theta$-(semi)stability. 
Given $\theta \in \bbR^G$, a representation $V$ of the quiver $\calQ$ with (nonzero) dimension vector $\alpha$ is called $\theta$-semistable if $\theta \cdot \alpha = 0$ and for any subrepresentation $W \subset V$ with dimension vector $\beta$ we have $\theta \cdot \beta \le 0$.
We say that V is $\theta$-stable if under the previous assumptions $\theta \cdot \beta < 0$ for any nontrivial proper subrepresentation $W \subset V$ with dimension vector $\beta$.

These results can be tied together with a theorem by King \cite{King:1994}.
Let $\calQ$ be a quiver, and let $\theta \in \bbZ$. Let $\alpha \in \bbZ_{\ge 0}$ be a dimension vector such that $\theta \cdot \alpha = 0$. Then, any $V \in \Rep(\calQ, \alpha)$ is $\chi_\theta$-semistable (resp. $\chi_\theta$-stable) if and only if $V$ is $\theta$-semistable (resp. $\theta$-stable).